\documentclass[12pt]{article}

\usepackage{amsthm, amsmath, amssymb}
\usepackage{latexsym,amsmath,amsfonts,amsthm,amssymb}

\newtheorem{theorem}{Theorem}[section]
\usepackage{tikz}
\usepackage[dvips]{epsfig}
\usepackage{enumerate}
\usepackage{bbm}
\usepackage{graphics}
\usepackage[font={footnotesize},margin=1cm]{caption}
\usepackage[breaklinks=false,pdfpagemode=None,pdfview=FitH,pdfstartview=FitH,citebordercolor={0 0 1},linkbordercolor={0 0 1},urlbordercolor={0 0 1},pagebordercolor={0 0 1},pdfborder={0 0 1}]{hyperref}

\usepackage[normalem]{ulem}

\usetikzlibrary{arrows,shapes}
\usetikzlibrary{decorations.pathmorphing}

\newcommand{\vphi}{\varphi}
\newcommand{\veps}{\varepsilon}

\newcommand{\calC}{{\mathcal{C}}}

\newcommand{\calI}{{\mathcal{I}}}

\newcommand{\calN}{{\mathcal{N}}}

\newcommand{\calR}{{\mathcal{R}}}
\newcommand{\calS}{{\mathcal{S}}}

\newcommand{\calU}{{\mathcal{U}}}

\numberwithin{equation}{section}

\newtheorem{corollary}[theorem]{Corollary}
\newtheorem{lemma}[theorem]{Lemma}
\newtheorem{definition}[theorem]{Definition}
\newtheorem{proposition}[theorem]{Proposition}
\title  {
        Localization and eigenvalue statistics for the lattice Anderson model with discrete disorder
                 }

\author{
John Z. Imbrie
\footnote{This work was supported by the Simons Foundation, \#638557.}
\\Department of Mathematics,
University of Virginia \\
Charlottesville, VA 22904-4137, USA
\\ {\tt imbrie@virginia.edu}
}
\date{}
\begin{document}
\maketitle
\begin{abstract}
We prove localization and probabilistic bounds on the minimum level spacing for the Anderson tight-binding model on the lattice in any dimension, with single-site potential having a discrete distribution taking $N$ values, with $N$ large. These results hold for all energies under an assumption of weak hopping.

 \end{abstract}
\tableofcontents
\section{Introduction}\label{sec:introduction}
\subsection{Background}\label{sec:background}

There are a wealth of results on the phenomenon of localization for Schr\"odinger operators with random potentials. The Anderson model \cite{Anderson1958} describes a quantum particle hopping in a random potential. Localization occurs when the particle cannot escape to infinity; this is the case, for example, when an eigenfunction correlator $\sum_\beta \lvert \vphi_\beta(x)\vphi_\beta(y) \rvert $ exhibits rapid decay in a suitable sense. Mathematically, this problem has been analyzed using multi-scale analysis (for example \cite{Frohlich1983}) or fractional-moment bounds (for example \cite{Aizenman1993}). However, these methods demand a degree of regularity of the distribution of the random potential, in order to obtain control over the density of states via some variant of the Wegner argument \cite{Wegner1981}. 

It is natural to consider the case of a discrete disorder distribution. The Anderson-Bernoulli model is a particularly appealing example wherein the potential takes two values only. These might reflect, for example, the presence or absence of an impurity. Localization has been proven for the Anderson-Bernoulli model on the lattice in one dimension \cite{carmona1987anderson,shubin1998some}.
Further results include improved regularity of the density of states for weak disorder \cite{bourgain2012furstenbergmeasure,bourgain2014application}. 
Results on localization have been obtained for the Anderson-Bernoulli model in the continuum, both in one dimension \cite{damanik2002localization}, and in higher dimensions \cite{bourgain2005localization}. In the latter work, localization was established near the bottom of the spectrum, using a quantitative form of the unique continuation principle to ensure that eigenfunctions do not decay too rapidly. This idea was implemented in  a number of generalizations and other cases involving singular potentials \cite{germinet2007cantor,germinet2007localization,germinet2013comprehensive,klein2016quantitative}. 
New work on unique continuation in the lattice \cite{Buhovsky2017} led to proofs of Anderson localization near the bottom of the spectrum in two and three dimensions \cite{Ding2019,Li2019}.
The higher-dimensional case remains open.

In this work, we demonstrate localization at all energies for the Anderson model on the lattice, with a discrete disorder distribution taking $N$ values, with $N \gg 1$. The case $N=2$ remains open. However, the method introduced here may be useful in working toward that goal. Results such as  \cite{Frohlich1983,Aizenman1993} are valid in the case of large disorder or extreme energies, the latter being technically more demanding. 
The situation here is somewhat analogous, in that localization should be expected for large $N$ or at extreme energies.

Bounds on the density of states are interesting in their own right. When an a priori bound on the density of states is not available, it becomes necessary to prove such bounds in parallel with spatial decay estimates. For example, log-H\"older continuity of the integrated density of states is proven in \cite{germinet2007localization}, with an exponent $p<\tfrac{3}{8}d$, where $d$ is the dimension. For the lattice model considered here, we obtain log-H\"older continuity with exponent $p$, which may be chosen arbitrarily large, provided $N$ is sufficiently large and the hopping is sufficiently weak (Theorem \ref{thm:1.1}).
In principle, localization should go hand-in-hand with a log-H\"older exponent $p>d$ since in that case the density of regions resonant to some $E$ to within $\delta$ would be of order $\lvert \log \delta\rvert^{-p}$, the typical separation would be of order 
$\lvert\log \delta\lvert^{p/d}$, and then an exponentially decaying interaction would be much smaller than the energy denominator $\delta$. One might be able to push our method down to this threshold, but in the present work we take $p$ to be fairly large.

For the eigenfunction correlator, we establish decay in mean as a large power of the distance, as well as exponential bounds with probability tending to 1 (Theorem \ref{thm:1.2}).

We also prove estimates on eigenvalue separation in parallel with decay and density of states bounds. This becomes necessary because the random potential produces a rank one perturbation to a local version of 
the Hamiltonian, 
and if there is more than one eigenvalue in play, this is insufficient for moving the spectrum out of the interval.
We prove that the probability of a near-degeneracy of size $\delta$ in the spectrum decays as a large power of $\lvert\log \delta\rvert$ (Theorem \ref{thm:1.3}).
A minimum level-spacing condition also arises as an assumption under which many-body localization could be proven \cite{Imbrie2016a}. 
One can obtain a level-spacing condition from a Minami estimate \cite{Minami1996} only if the disorder distribution is sufficiently regular
\cite{Klein2006}.

We take as a starting point the method of iterated Schur complements 
that was introduced in \cite{Imbrie2016c} (which in turn uses ideas from \cite{Frohlich1983}). The dimension of the Hilbert space of active modes is systematically reduced as the energy window is narrowed, until only a single mode is in play. Resonant regions connect via a multiscale percolation process whose connectivity function decays as a large power of the distance. This procedure provides a systematic way of producing successive local approximations to the eigenvalues and eigenfunctions of the Hamiltonian. The $k^{\text{th}}$ 
approximation brings in the effect of the random potential in a neighborhood of size $L_k \sim 2^k$ of a localization center. Changes are exponentially small in $L_k$, with probability $1-L_k^{-p}$ for some large $p$. Thus one may say that the eigenvalues and eigenfunctions are quasilocal functions of the random potentials. This is also a feature of the constructions in \cite{Imbrie2016,Imbrie2016a,Imbrie2016c}.

In order to exploit the randomness in each new annular neighborhood, we find a particular site that is most influential for a group of eigenfunctions. Eigenfunctions in $\mathbb{Z}^d$ cannot grow faster than exponentially. (See, for example \cite{craig1983log}.
Our methods would not work, say, for the Kagome lattice, where this property fails.) This is used to demonstrate that the influence of this site is no smaller than an exponential in the distance from the localization center. 
With some care, one can show that cancellations do not nullify the effect of this site. It turns out that the randomness at one site per annulus is sufficient to drive a gradual breakdown of nearly degenerate situations, and eventually, to move spectrum out of a narrow energy window.

\subsection{Model and Main Results}\label{sec:model}

We consider the Anderson model on a rectangle $\Lambda \subset \mathbb{Z}^d$. The Hamiltonian is
\begin{equation}\label{(1)}
H = H_{\Lambda}= -\gamma \Delta + v,
\end{equation}
where $\Delta$ is the restriction of the lattice Laplacian to $\Lambda$,  $0<\gamma \ll 1$, and $v$ is multiplication by the lattice potential $v_x$, $x\in \Lambda$. We take $\mathbf{v} = \{v_x\}_{x\in \Lambda}$ to be a collection of iid random variables, each with a uniform distribution on $\{0, \frac{1}{N-1}, \frac{2}{N-1}, \ldots , 1\}$, with $N$ an integer greater than 1. Thus we have a generalization of the Anderson-Bernouilli model, which corresponds to the case $N=2$. We may write
\begin{equation}\label{(2)}
H = H_0-\gamma J,
\end{equation}
where
\begin{equation}\label{(3)}
H_0 = \text{diag}\big(\{2d\gamma + v_x\}_{x\in \Lambda}\big)
\end{equation}
and
\begin{equation}\label{(4)}
J_{xy} = \begin{cases}
1, &\text{if } \lvert x-y \rvert = 1; \\
0, &\text{otherwise.}
\end{cases}
\end{equation}
We use the metric $\lvert x \rvert = \sum_{i=1}^d \lvert x_i\rvert$ on $\mathbb{Z}^d$, and $\text{diam}(X)$ is the corresponding diameter for subsets of $\Lambda$. However, it will be convenient to write $\text{Diam}(\Lambda)$ for the sup-norm diameter of $\Lambda$.
Note that the spectrum of $H$ is confined to the interval $[0, 1+4d\gamma]$.

For $\gamma$ small, we have a largely diagonally dominant matrix. However, the problem of resonances is particularly acute when the potential has a discrete distribution, because the probability that $v_x$ lies in an interval of width $\delta$ does not go to zero with $\delta$.

In the following results, we use a parameter $p$, which corresponds to the exponent for log-H\"older continuity in our bounds for the density of states (Theorem \ref{thm:1.1}). It also determines the exponent for power-law decay of probabilities (Theorem \ref{thm:1.2}).
In the course of the proofs, we will require $p > d$ to be a sufficiently large constant. Then we require $N$ to be sufficiently large, depending on the chosen value of $p$. Finally, we require $\gamma$ to be sufficiently small, depending on the chosen value of $N$. Specifically, we take $\gamma \le \veps^{20}$, where $\veps \equiv \tfrac{1}{N-1}$. 
Thus, the choice of parameters is made in the order $p$, $N$, $\gamma$. 

We introduce some notation. Let $I_\delta(E)$ denote the interval $[E-\delta, E+\delta]$, and let $\calN(I)$ denote the number of eigenvalues of $H$ in $I$. Let $\{E_\beta, \varphi_\beta\}_{\beta = 1,\ldots,\lvert \Lambda \rvert}$ denote the eigenvalues and associated normalized eigenvectors of $H$. In view of Theorem \ref{thm:1.3} below, the probability of an exact degeneracy decreases as a power of $\text{Diam}(\Lambda)$. If necessary, a basis can be chosen for an eigenspace of multiplicity greater than 1. All bounds are independent of the choice.

We establish log-H\"older continuity of the density of states, with exponent $p$.
\begin{theorem}
\label{thm:1.1}
Choose a sufficiently large $p$. Then for $N$ sufficiently large (depending on $p$) and $\gamma$ sufficiently small (depending on $N$),
\begin{equation}\label{(5)}
\mathbb{E}\,\calN\big(I_\delta(E)\big) \le \lvert \Lambda \rvert (\log_\gamma \delta)^{-p}.
\end{equation}
for any rectangle $\Lambda$ and any $\delta \in [\gamma^{\mathrm{Diam}(\Lambda)/2},1]$.
\end{theorem}
Next, we prove bounds on the eigenfunction correlator, establishing localization and exponential decay of the eigenfunctions.
 \begin{theorem}
\label{thm:1.2} \label{(6)}
Choose $p$ sufficiently large, then $N$ sufficiently large (depending on $p$), and $\gamma$ sufficiently small (depending on $N$). For any rectangle $\Lambda$, the eigenfunction correlator satisfies
\begin{equation}\label{(6')}
\mathbb{E}\, \sum_\beta \lvert \varphi_\beta(x)\varphi_\beta(y) \rvert  \le \big(\lvert x-y \rvert  \vee 1\big)^{-(p/2 - d - 1)}.
\end{equation}
Furthermore, 
the following bound holds for all $x \in \Lambda$, $R \ge 4$:
\begin{equation}\label{(7)}
P\bigg(\max_{y:\,\lvert y-x \rvert  \ge R} \sum_\beta \lvert \varphi_\beta(x)\varphi_\beta(y) \rvert  \gamma^{-\lvert x-y \rvert /5} > 1\bigg) \le R^{-(p/2-4d-1)}.
\end{equation}
\end{theorem}
Lastly, we establish probabilistic estimates on the minimum eigenvalue spacing.
\begin{theorem}
\label{thm:1.3} 
Choose a sufficiently large $p$. Then for $N$ sufficiently large (depending on $p$) and $\gamma$ sufficiently small (depending on $N$),
\begin{equation}\label{(8)}
P\Big(\min_{\beta \ne \tilde{\beta}} \lvert E_\beta  - E_{\tilde{\beta}} \rvert  < \delta\Big) \le \lvert \Lambda \rvert^2 (\log_\gamma \delta)^{-(p/2 - 1)},
\end{equation}
for any rectangle $\lvert \Lambda \rvert$ and any $\delta \in [\gamma^{\mathrm{Diam}(\Lambda)},1]$.
\end{theorem}

\subsection{A Lemma on Schur Complements}\label{ssec:fl}
The following lemma from \cite{Imbrie2016c} will be used throughout as a way of reducing the analysis to an equivalent lower-dimensional problem focusing only on those eigenvalues (or approximate eigenvalues) in a small interval of energy. 
\begin{lemma}
\label{fundamental}
Let $K$ be a $(p+q)\times(p+q)$ symmetric matrix in block form, $K = \left( \begin{smallmatrix}
A&B \\C&D \end{smallmatrix}\right)$, with $A$ a $p\times p$ matrix, $D$ a 
$q\times q$ matrix, and $C = B^T$. Assume that $\|(D-E)^{-1}\| \le \tilde{\veps}^{-1},
\|B\| \le \tilde{\gamma}, \|C\| \le \tilde{\gamma}$. Define the Schur complement with respect to 
$\lambda$:
\begin{equation}\label{(1.6)}
F_{\lambda} \equiv A - B(D-\lambda)^{-1}C.
\end{equation}
Let $\tilde{\veps}$ and $\tilde{\gamma} / \tilde{\veps}$ be small, and $\lvert \lambda-E \rvert  \le \tilde{\veps} /2$. Then
\begin{enumerate}[(i)]
\item{
If $\varphi$ is an eigenvector for $F_\lambda$ with eigenvalue $\lambda$, then
$(\varphi, -(D-\lambda)^{-1}C\varphi)$ is an eigenvector for $K$ with eigenvalue $\lambda$, and all eigenvectors of $K$ with eigenvalue $\lambda$ are of this form.
}
\item{
\begin{equation}\label{(10)}
\|F_\lambda - F_E\| \le 2\Big(\frac{\tilde{\gamma}}{\tilde{\veps}}\Big)^2 \lvert \lambda - E \rvert .
\end{equation}
}
\item{
The spectrum of $K$ in $[E-\tilde{\veps}/2,E+\tilde{\veps}/2]$ is in close agreement with that of $F_E$ in the following sense. If $\lambda_1 \le \lambda_2 \le \ldots \le \lambda_m$ are the eigenvalues of $K$ in $[E-\tilde{\veps}/2,E+\tilde{\veps}/2]$, then there are corresponding eigenvalues $\tilde{\lambda}_1 \le \tilde{\lambda}_2 \le \ldots \le \tilde{\lambda}_m$ of $F_E$, and 
\begin{equation}\label{evmove}
\lvert\lambda_i - \tilde{\lambda}_i\rvert \le 2(\tilde{\gamma} /\tilde{\veps})^2\lvert\lambda_i - E\rvert.
\end{equation}
}
\end{enumerate}
\end{lemma}
Observe that the lemma actually provides an algorithm for finding the eigenvalues of $K$ near $E$. Weyl's inequality and (\ref{(10)}) show that the eigenvalues of $F_\lambda$ can be taken as Lipschitz continuous functions of $\lambda$, with a small Lipschitz constant. Hence we can determine the eigenvalues of $K$ near $E$ by a fixed point argument, effectively solving the condition $\lambda \in \text{spec}\,F_\lambda$.

\textit{Proof.}
(i) We have that $\big(\left( \begin{smallmatrix}
A&B \\C&D \end{smallmatrix}\right)-\lambda\big) \left( \begin{smallmatrix}
\vphi \\\tilde{\vphi}  \end{smallmatrix}\right)
 = 0$ if and only if
$C\vphi + (D-\lambda)\tilde{\vphi}= 0$ (\textit{i.e.} $\tilde{\vphi} = -(D-\lambda)^{-1}C\vphi$) and $(F_\lambda - \lambda)\vphi = 0$. 
%Corrected this equation by adding ^{-1}
Thus we have a 1-1 mapping between the
$\lambda$-eigenspaces of $K$ and of $F_\lambda$.
(ii) We write
\begin{equation}\label{(10')}
F_E-F_\lambda  = B(D-E)^{-1}(\lambda-E)(D-\lambda)^{-1}C.
\end{equation}
Since $\lvert \lambda-E \rvert  \le \tilde{\veps}/2$ and $\mathrm{dist(spec\,}D,E) \ge \tilde{\veps}$, we have that $\|(D-\lambda)^{-1}\| \le 2/\tilde{\veps}$, and then (\ref{(10)}) follows by inserting the assumed bounds for each operator.
(iii) By Weyl's inequality, the eigenvalues of $F_{\lambda_i}$ and $F_E$ differ by no more than $2(\tilde{\gamma} /\tilde{\veps})^2\lvert\lambda_i - E\lvert$ when shifting from $F_{\lambda_i}$ to $F_E = F_{\lambda_i} + (F_E-F_{\lambda_i})$.\qed

In what follows, we will be iterating this argument on a sequence of length scales $L_k = L_02^k$ and spectral window widths  
\begin{equation}\label{window}
\veps_1 \equiv \tfrac{1}{3(N-1)},\quad\veps_k = \gamma^{1.6L_{k}} \text{ for }k>1. 
\end{equation}
Using a local approximation to $F_\lambda^{(k)}$ (the $k^{\text{th}}$ Schur complement of $H$), we may identify resonant sites where spectrum should be within $\veps_k$ of $E$, and these determine the subspace for the next Schur complement. Clusters of resonant sites become farther apart as $k$ grows, ensuring that the off-diagonal blocks $B_\lambda^{(k)}$ and $C_\lambda^{(k)}$ tend rapidly to 0 with $k$. Eventually, the window width is $\sim \delta$, and then we will determine how many eigenvalues are present. The construction produces as well the associated eigenfunction, demonstrating exponential decay with high probability.
Note that we are taking the Schur complement of a $\lambda$-dependent $K$, but this does not affect
(i). As long as we have a Lipschitz condition on $F_\lambda^{(k)}$ 
(see Theorem \ref{thm:3} below), we will have a corresponding statement on its spectrum as in (iii).

\section{Iterated Schur Complements and Random Walk Expansions}\label{sec:2}
\subsection{First Step}\label{ssec:first}
The first Schur complement will be organized so as to examine spectrum near some energy $E \in [0,1+4d\gamma]$. 
The allowed values of $v_x$ are multiples of $\frac{1}{N-1}$ in $[0,1]$.
Let $\veps \equiv \tfrac{1}{N-1}$ and $\veps_1 \equiv \veps/3  $.
In the first step, we say a site $x$ is resonant to $E$ if $v_x+2d\gamma \in I_{\veps_1}(E)$, \textit{i.e.} if 
\begin{equation}\label{(2.1n)}
\lvert v_x +2d\gamma - E\rvert \le \veps_1.
\end{equation}
Then the probability that $x$ is resonant to $E$ is bounded by $\veps$. We see that the set of resonant sites will typically be a very dilute set. Define
\begin{equation}\label{(12)}
R^{(1)} = \{x \in \Lambda\,\text{:  }  x \text{ is resonant to }E \}.
\end{equation}

The box $\Lambda$ is divided into resonant sites $R^{(1)}$ and nonresonant sites
$R^{(1)\text{c}} = \Lambda \setminus R^{(1)}$. The associated index sets determine the block form of the Hamiltonian:
\begin{equation}\label{(14)}
H = \begin{pmatrix}
A^{(1)} & B^{(1)}\\  C^{(1)} & D^{(1)} \end{pmatrix},
\end{equation}
with $A^{(1)}$ denoting the restriction of $H$ to the subspace with indices in $R^{(1)}$, and $D^{(1)}$ denoting the restriction to the subspace with indices in $R^{(1)\text{c}}$. This allows us to write down the Schur complement
 \begin{equation}\label{(15)}
F_{\lambda}^{(1)} \equiv A^{(1)} - B^{(1)}(D^{(1)}-\lambda)^{-1}C^{(1)}.
\end{equation}

Let us decompose
\begin{equation}\label{(16)}
D^{(1)} = W^{(1)} - V^{(1)},
\end{equation}
where
\begin{align}
W_{xy}^{(1)} &= (2d\gamma +v_x)\delta_{xy}, \label{17} \\ \label{(18)}
V_{xy}^{(1)} &= \gamma J_{xy} = \begin{cases}
\gamma, &\text{if } \lvert x-y \rvert  = 1; \\
0, &\text{otherwise.}
\end{cases}
\end{align}
Let us assume that $\lambda - E \le \veps_1/2$, so that
$\|(W^{(1)} - \lambda)^{-1}\| \le 2/\veps_1$. Note that $\|V^{(1)}\| \le 2d\gamma$. Hence for $\gamma$ small, the Neumann series
\begin{equation}\label{(19)}
(D^{(1)} - \lambda)^{-1} = (W^{(1)} - \lambda)^{-1} + (W^{(1)} - \lambda)^{-1}V^{(1)}(W^{(1)} - \lambda)^{-1} + \ldots
\end{equation}
converges, and we obtain a random-walk expansion
\begin{equation}\label{(20)}
\left[B^{(1)}(D^{(1)} - \lambda)^{-1}C^{(1)}\right]_{xy} = \sum_{g_1:x\rightarrow y} \,\prod_{i=1}^m \frac{1}{2d\gamma + v_{x_i} - \lambda}\,\prod_{j=0}^{m} V^{(1)}_{x_jx_{j+1}}.
\end{equation}
Here $g_1 = \{x=x_0, x_1, \ldots, x_m, x_{m+1} = y\}$ is a random walk with $m+1$ nearest-neighbor steps, $m \ge 1$; return visits are allowed. Note that $x$, $y$ are in $R^{(1)}$, while $x_1,\ldots,x_m$ are in $R^{(1)\text{c}}$.
It should be clear that $[B^{(1)}(D^{(1)}-\lambda)^{-1}C^{(1)}]_{xy}$ decays exponentially in $\lvert x-y \rvert $, as each additional step in the walk brings a factor $\gamma$ from the interaction $V^{(1)}$ and a factor $\le 2/\veps_1 = 6/\veps$ from
$(2d\gamma+v_{x_i} -\lambda)^{-1}$; recall that $\gamma \le \veps^{20}$. A similar decay holds for the eigenfunction-generating kernel $-(D^{(1)}-\lambda)^{-1}C^{(1)}$, see Lemma \ref{fundamental}(i). Precise bounds will be stated below in Theorem \ref{thm:2}.
Note that $A^{(1)}$ does not connect different components of $R^{(1)}$, but $B^{(1)}(D^{(1)}-\lambda)^{-1}C^{(1)}$ produces a long-range (but exponentially decaying) effective interaction between components.

\subsection{Isolated Blocks}\label{ssec:isolated}
We need to define a set of isolated blocks that are candidates for elimination from the resonant set. 
Let us make the needed definitions here both for the first step and for the general step. 
The set $R^{(1)}$ can be broken into connected components, where we declare $x$ and $y$ to be connected if $\lvert x-y \rvert  \le L_1^\alpha$. 
Here $\alpha = \tfrac{3}{2}$ is a fixed power that sets the scale for isolation. We are using the first in a sequence of length scales,
\begin{equation}\label{(13)}
L_k = L_02^k, k = 1, 2, \ldots.
\end{equation}
We take $L_0$ to be a large integer, whose choice will depend on the value of $p$. Thus our parameters will be fixed in the following order: $p$, $L_0$, $N$, $\gamma$, with each choice depending on the size of the previous parameter.
Let $B_1$ denote a connected component of $R^{(1)}$, based on 
connections with range $L_1^\alpha$.
 Then let $\bar{B}_1$ denote the set of lattice points within a distance $2L_1$ of $B_1$. In the $k^{\text{th}}$ step, we will have a resonant set $R^{(k)}$; the sequence satisfies $R^{(k)} \subseteq R^{(k-1)}$. 
 We declare that two sites of $R^{(k)}$ are connected if they are within a distance $L_k^\alpha$. This leads to a decomposition of $R^{(k)}$ into a set of components
 $\{B_{k,\beta}\}_{\beta =1,\ldots,m}$.
 For simplicity, we will drop the subscript $\beta$ when discussing a single component $B_k$.

\begin{definition}\label{def:2.1}
Let $B_{k-1}$ be a component of $R^{(k-1)}$ on scale $k$ with $k\ge2$. We say that $B_{k-1}$ is  \textbf{isolated in step $k$}
  if
\begin{equation}\label{isolated}
\mathrm{diam}(B_{k-1})\le L_{k-1}.
\end{equation}
\end{definition}

\textit{Remark.} This  condition  on the diameter ensures that the distance from $B_{k-1}$ to other components is much larger than $\text{diam}(B_{k-1})$. The distance conditions and some other constructions introduced below should be familiar to readers of \cite{Frohlich1983}.

 For each isolated component $B_1$ of $R^{(1)}$, we define a localized version of $F^{(1)}_\lambda$ by restricting the random-walk expansion to $\bar{B}_1$:
\begin{equation}\label{(21)}
[\tilde{F}^{(1)}_{\lambda}(B_1)]_{xy} \equiv A^{(1)}_{xy} - \sum_{g_1:x\rightarrow y \text{, }g_1 \subseteq \bar{B}_1} \,\prod_{i=1}^m \frac{1}{2d\gamma + v_{x_i} - \lambda}\,\prod_{j=0}^{m} V^{(1)}_{x_jx_{j+1}}.
\end{equation}
Here $x$, $y$ are restricted to $B_1$; $g_1 \subseteq \bar{B}_1$ means that each of the sites visited by $g_1$ lie in $\bar{B}_1$. Note that the 
separation between components is greater than $L_1^\alpha$, which is much greater than $2L_1$, the width of the collar $\bar{B}_1 \setminus B_1$.
Hence, the expanded blocks $\bar{B}_1$ do not have any sites in common. In view of the smallness of the terms dropped in this definition, $\tilde{F}_\lambda^{(1)}(B_1)$ may be used to determine whether the block $B_1$ remains resonant in the next step.

\textit{Remark.} It is easy to see that (\ref{(21)}) is the same as what would be obtained by replacing $H=H_{\Lambda}$ with $H_{\bar{B}_1}$ in (\ref{(14)})-(\ref{(20)}), in which case the resolvent $(D^{(1)}-\lambda)^{-1}$ becomes the resolvent in the region $\bar{B}_1 \setminus B_1$.

\subsection{Resonant Blocks}\label{ssec:resonant}

Let us give the condition for resonance in the general step. We will need the flexibility to shift the energy $E$ from step to step in our procedure. Thus we allow for a sequence of energies $E_k$ with $E_1 = E$ and $\lvert E_k - E_{k-1} \rvert  \le \veps_{k-1}/3$. Here
\begin{equation}\label{(22)}
\veps_1 \equiv \veps/3 = \tfrac{1}{3(N-1)}, \,\, \veps_k \equiv \gamma^{1.6L_{k}} \text{ for } k > 1
\end{equation}
are the energy windows for each step. One possibility would be to put $E_k = E$ for all $k$ (fixed energy procedure), in order to investigate spectrum in small windows about $E$. Another possibility would be to put $E_k$ close to a solution to $\lambda \in \text{spec}\,
\tilde{F}^{(k-1)}_{\lambda}(B_{k-1})$ (energy-following procedure), in order to obtain a convergent sequence of approximate eigenvalues.
(The definition of $\tilde{F}^{(k-1)}_{\lambda}(B_{k-1})$ for $k>2$ will be given in the next subsection, generalizing (\ref{(21)}), once the random-walk expansion is defined for the general step.)

\begin{definition}\label{def:resonant}
Let $B_{k-1}$ be a component of $R^{(k-1)}$ with $k\ge2$. We say that $B_{k-1}$ is \textbf{resonant in step} $k$
  if it is isolated in step $k$ and if 
  \begin{equation}  \label{(23)}
  \mathrm{dist}\big(\mathrm{spec}\,\tilde{F}_{E_k}^{(k-1)}(B_{k-1}),E_k\big) \le \veps_k.
  \end{equation}
\end{definition}
We define the new resonant set $R^{(k)}$ by deleting from $R^{(k-1)}$ all of its components that are isolated but not resonant in step $k$. Thus
\begin{equation}\label{(24)}
R^{(k)}=R^{(k-1)} \setminus  \bigcup_{\beta:\,B_{k-1,\beta} \text{ is  isolated  but not resonant in step }k }B_{k-1,\beta}.
\end{equation}
This set of sites is then used to determine the block decomposition
\begin{equation}\label{(25)}
F_\lambda^{(k-1)} = \begin{pmatrix}
A^{(k)} & B^{(k)}\\  C^{(k)} & D^{(k)} \end{pmatrix},
\end{equation}
where the blocks are determined by the decomposition of $R^{(k-1)}$ into $R^{(k)}$ (upper-left block) and $R^{(k-1)}\setminus R^{(k)}$ (lower-right block).
(We do not make the $\lambda$-dependence explicit for the matrices $A^{(k)}$, $B^{(k)}$, $C^{(k)}$, $D^{(k)}$.)
Note that the blocks $B_{k-1}$ that were taken out of the resonant set in (\ref{(24)}) are nonresonant; this ensures the invertibility of $D^{(k)}-\lambda$, for $\lvert \lambda - E_k \rvert \le \veps_k/2$ -- see the estimates below on the random-walk expansion. Thus we may define
\begin{equation}\label{(26)}
F_{\lambda}^{(k)} = A^{(k)} - B^{(k)}(D^{(k)}-\lambda)^{-1}C^{(k)}.
\end{equation}

The overall plan of this construction is very similar to the one in \cite{Frohlich1983}. With each new step in the procedure, isolated, nonresonant components are removed from $R^{(k-1)}$ to form $R^{(k)}$, and the connectivity distance is increased to $L_k^{\alpha}$.
The condition (\ref{(23)}) for resonance of a component $B_{k-1}$ is based on a localized version of the construction using $H_{\bar{B}_{k-1}}$, for a suitable neighborhood $\bar{B}_{k-1}$ of $B_{k-1}$.
(The definition of $\bar{B}_{k}$ will be given in the next subsection, but the key point is to draw the boundary of $\bar{B}_{k}$ at a distance $2L_{k}$ from $B_{k}$, but deformed so as to be sufficiently far from the isolated components $B_j$ for $j<k$.)
Locality of the condition (\ref{(23)}) is important for the probability estimates of Section \ref{sec:3}, which are performed via a sequence of conditionings that determine the resonant sets $R^{(1)}$, $R^{(2)}$,\ldots. In this context, the event $\{B_{k-1} \text{ is resonant in step }k\}$ is determined completely by the potentials in $\bar{B}_{k-1}$, as it is based on the spectrum of $H_{\bar{B}_{k-1}}$.

\subsection{Random Walk Expansion}\label{ssec:random}

To complete our constructions in the general step, we need to describe the collared blocks $\bar{B}_k$, give the random-walk expansion for $F^{(k)}_\lambda$, and use these to define $\tilde{F}^{(k)}_\lambda(B_k)$. These inductive definitions depend on earlier incarnations of the objects being defined.

We need a construction that forces the boundary of $\bar{B}_k$ to go around blocks from earlier scales that are no longer part of $R^{(k)}$. The
blocks $B_j$, $j<k$ are isolated and nonresonant in step $j+1$; thus
$\text{diam}(B_j) \le L_j$. The block $B_k$, on the other hand, is a component of 
$R^{(k)}$, and it is not necessarily isolated; there is no limitation on its diameter.

Connectivity in $R^{(j)}$ is defined so that each $B_j$ is at least a distance $L_j^\alpha$ from the rest of $R^{(j)}$. This implies that $B_j$ is similarly distant from any $B_i$
that is formed out of $R^{(i)}$ for $i \ge j$.
We give here an inductive construction of a set of collared 
blocks $\bar{B}_k$ and $\bar{\bar{B}}_k$. 
Assume $\bar{B}_j$ has been constructed for $j<k$. 
Then let $\bar{\bar{B}}_j$ denote an $L_j^{\sqrt{\alpha}}$-neighborhood of $\bar{B}_j$. 
(Neighborhoods will be taken within $\Lambda$ throughout.)
Write $\calU_{k-1}$ for the union of all $\bar{\bar{B}}_j$ for $j<k$. 
Then define $\bar{B}_k$ by taking a $2L_k$-neighborhood of $B_k$ and combining it with any connected component of $\calU_{k-1}$ that intersects it.
Thus $\bar{B}_k$ depends on previous scale collared blocks $\bar{\bar{B}}_j$, $j < k$. 
The boundary of $\bar{B}_k$  skirts around nearby $\bar{B}_j$, $j < k$ at a distance $L_j^{\sqrt{\alpha}}$. 
A similar construction was done in \cite{Frohlich1983}. This definition automatically produces 
collections of collared blocks $\{\bar{B}_{j,\beta}\}_{j\le k}$ such that any pair
of distinct blocks
$\{\bar{B}_{i,\beta},\bar{B}_{j,\beta'}\}$ with $i \le j$ satisfies
$\text{dist}(\bar{B}_{i,\beta},\bar{B}_{j,\beta'}) > L_i^{\sqrt{\alpha}}$ or 
$\text{dist}(\bar{B}_{i,\beta},\bar{B}_{j,\beta'}^{\text{c}}) > L_i^{\sqrt{\alpha}}$. In other words, 
$\bar{B}_{i,\beta}$ is either well separated from $\bar{B}_{j,\beta^\prime}$ or well inside of it.

Let $U_k$ be one of the connected components of $\calU_k$. We prove the following estimates by induction on $k$, assuming $L_0$ is sufficiently large (see Appendix D of \cite{Frohlich1983}, which has similar arguments). Here we assume that $B_k$ is isolated, \textit{i.e.} $\text{diam}(B_k) \le L_k$.
\begin{equation}\label{(27)}
\text{diam}({\bar{B}_k}) \le 5.1L_k;\quad \text{diam}({U_k}) \le 2.1L_k^{\sqrt{\alpha}}.
\end{equation}
These bounds hold for $k=1$ because $\calU_0$ is empty, and so $\text{diam}(\bar{B}_1) \le 5L_1$.
For $k>1$, observe that 
\begin{equation}\label{(27a)}
\text{diam}(U_k) 
\le 5.1L_k+2L_k^{\sqrt{\alpha}}+2\cdot 2.1L_j^{\sqrt{\alpha}} 
\le 2.1L_k^{\sqrt{\alpha}},
\end{equation}
where the first two terms bound the diameter of the $L_k^{\sqrt{\alpha}}$-neighborhood of $\bar{B_k}$, and the third bounds the expansion due to components of $\calU_{k-1}$ that intersect it, with $j$ being the maximum scale index for such components. 
The last inequality holds because $\text{dist}(B_j,B_k) \ge L_j^\alpha$, and so $L_j^\alpha \le 5.1L_k+L_k^{\sqrt{\alpha}}+2.1L_j^{\sqrt{\alpha}}$, which implies that $.9 L_j^\alpha \le 1.1L_k^{\sqrt{\alpha}}$, and hence $L_j^{\sqrt{\alpha}} \le 1.3 L_k$.
Similarly, we may argue that
\begin{equation}\label{(27b)}
\text{diam}(\bar{B}_k) 
\le 5 L_k+2\cdot2.1L_j^{\sqrt{\alpha}} 
\le 5.1L_k.
\end{equation}
For the last inequality, we have used the separation condition to obtain
$L_j^{\alpha} \le 2L_k + 2.1L_j^{\sqrt{\alpha}}$, so that $.9L_j^\alpha \le 2L_k$, and then $4.2L_j^{\sqrt{\alpha}} \le .1 L_k$.
Later, we will use the fact -- implicit in (\ref{(27b)}) -- that $\bar{B}_k$ is 
contained within a $2.05L_k$-neighborhood of $B_k$.
Note that $L_j \ll L_k$ in both cases, which means that two blocks on the same scale never combine in $\bar{B}_k$ or $U_k$.

In order to generate the random-walk expansion in the $k^{\text{th}}$ step,
we need to restrict to the neighborhood $\lvert \lambda - E_k \rvert  \le \veps_k/2$.
Then we write
\begin{equation}\label{(28)}
D^{(k)}=W^{(k)}-V^{(k)},
\end{equation}
where $W^{(k)}$ is block diagonal, each block being $\tilde{F}_\lambda^{(k-1)}(B_{k-1})$ for some $B_{k-1}$. The matrix $\tilde{F}_\lambda^{(k-1)}(B_{k-1})$ will be constructed by restricting the set of graphs that define $F^{(k-1)}_\lambda$ to those that start and end in $B_{k-1}$ and remain within $\bar{B}_{k-1}$. 
This means that $V^{(k)}$ consists of the long graphs not included in $\tilde{F}_\lambda^{(k-1)}(B_{k-1})$. It generates matrix elements both within blocks and between blocks.

We show below in Theorem \ref{thm:3} that 
\begin{equation}\label{(1.47b)}
\|\tilde{F}_\lambda^{(k-1)}(B_{k-1})-\tilde{F}_{E_k}^{(k-1)}(B_{k-1})\| \le \gamma\lvert \lambda - E_k \rvert ,
\end{equation}
which is less than $\veps_k/6$, because $\lvert \lambda - E_k \rvert \le \veps_k/2$. 
Since all the blocks of $R^{(k-1)}\setminus R^{(k)}$ are nonresonant,
\begin{equation}\label{(1.48)}
\text{dist}\big(\text{spec}\,\tilde{F}_{E_k}^{(k-1)}(B_{k-1}),E_k\big) \ge \veps_k ,
\end{equation}
and so
\begin{equation}\label{(1.49)}
\|(W^{(k)}-\lambda)^{-1}\| \le 3\veps_k^{-1}.
\end{equation}
Hence, as in the first step, we may expand $(D^{(k)}-\lambda)^{-1}$ in a Neumann series, and then after expanding out the matrix products, we obtain the random-walk expansion:
\begin{equation}\label{(1.50)}
[B^{(k)}(D^{(k)}-\lambda)^{-1}C^{(k)}]_{xy} = \sum_{g_k:x \rightarrow y}\, B^{(k)}_{xx_1}\prod_{i=1}^m \,[(W^{(k)}-\lambda)^{-1}]_{x_i\tilde{x}_i} \prod_{j=1}^{m-1} V^{(k)}_{\tilde{x}_jx_{j+1}}C^{(k)}_{\tilde{x}_my}. 
\end{equation}
Here $g_k = \{x=x_0,x_1,\tilde{x}_1,x_2,\tilde{x}_2,\ldots,x_m,\tilde{x}_m,x_{m+1}=y\}$, with each $x_i,\tilde{x}_i$ in the same block $B_{k-1}$
for $i = 1,\ldots,m$ and $x, y$ in $R^{(k)}$. 
Note that $V^{(k)}_{x,y}$ is given by a sum of graphs contributing to $F_\lambda^{(k-1)} - \oplus_{B_{k-1}} \tilde{F}_\lambda^{(k-1)} (B_{k-1})$, where
\begin{equation}\label{(1.51)}
F_\lambda^{(k-1)} = 
A^{(k-1)} -B^{(k-1)} (D^{(k-1)}-\lambda)^{-1}  C^{(k-1)}.
\end{equation}
Also, $B^{(k)}$, $C^{(k)}$ are blocks of $F_\lambda^{(k-1)}$. Thus we see that each step of $g_k$ is either a matrix element of $(W^{(k)}-\lambda)^{-1}$ or a sum of graphs $g_{k-1}$ that contribute to $F_\lambda^{(k-1)}$. We obtain inductively-defined, nested walk structures that we term \textit{multigraphs}. We may expand these structures down to the first random-walk expansion. Then one may visualize multigraphs as ordinary walks with nearest-neighbor steps, except that upon reaching a block $B_{j-1}$, there is a matrix element of $(W^{(j)}-\lambda)^{-1}$
that produces an intra-block jump.

We use the multigraph expansion to define 
$\tilde{F}_\lambda^{(k)}(B_{k})$ by restricting the multigraphs for $F_\lambda^{(k)}$ to those that remain within $\bar{B}_k$. 
Equivalently, $\tilde{F}_\lambda^{(k)}(B_{k})$ may be defined as
$F_\lambda^{(k)}$ computed in volume $\bar{B_k}$ instead of $\Lambda$. Then the support restriction for multigraphs is automatically satisfied.
This equivalence depends on the fact (explained at the start of this section) that all blocks from scales $j < k$ are either completely contained in $\bar{B}_k$ or completely outside of $\bar{B}_k$. 
%This should be evident from the way $\bar{B}_k$ was defined. 

We will also need multigraph expansions for the matrices that generate the eigenfunctions. Recall from Lemma \ref{fundamental} 
that if $\varphi^{(k)}$ is an eigenvector of $F_\lambda^{(k)}$ with eigenvalue $\lambda$, then 
\begin{equation}\label{(1.52)}
\varphi^{(k-1)}=
\begin{pmatrix}
\varphi^{(k)}\\ -(D^{(k)}-\lambda)^{-1}  C^{(k)}\varphi^{(k)}
\end{pmatrix}
\end{equation}
is an eigenvector of $F_\lambda^{(k-1)}$ with the same eigenvalue. This process may be repeated to extend the eigenvector $\varphi^{(k)}$ all the way down to the original lattice $\Lambda$, that is, to produce $\varphi^{(0)}$, an eigenvector of $H$. Let us write
\begin{equation}\label{(1.53)}
\varphi^{(0)}=G_\lambda^{(k)}\varphi^{(k)},
\end{equation}
and then we may give a multigraph expansion for $G_\lambda^{(k)}$ in the same manner as was just described for $F_\lambda^{(k)}$. Indeed, the same operators $C^{(k)},D^{(k)}$ appear when unwrapping (\ref{(1.52)}). 
Note that $G_\lambda^{(k)}$ has one index in $R^{(k)}$ and the other in $\Lambda$. In contrast, $F_\lambda^{(k)}$ has both its indices in $R^{(k)}$.

We now state our main theorem on graphical bounds. Let $\calS^{(k)}_{x,z,y}$ denote the sum of the absolute values of all multigraphs for $B^{(k)} (D^{(k)}-\lambda)^{-1}  C^{(k)}$ that go from $x$ to $y$ and that contain $z$. Here, $x,y$ are in $R^{(k)}$, and $z$ is  in $\Lambda \setminus R^{(k)}$. We say that a multigraph contains $z$ if any of the sites or blocks that it passes through contain $z$.
\begin{theorem}
\label{thm:2}
Let $L_0$ be sufficiently large. Take $\veps = \tfrac{1}{N-1}$ to be sufficiently small, depending on $L_0$, and take $\gamma \le \veps^{20}$. 
Assume that $\lvert \lambda - E_k \rvert  \le \veps_k/2$. Put $r_1=.9$, $r_k = r_{k-1}(1-6L_{k-1}^{1-\alpha})$ for $k \ge 2$. Then for all $k$, $r_k \ge r_\infty = .85$, and
\begin{align}
\calS^{(1)}_{x,z,y} &\le \gamma^{ r_1 [(\lvert x-z \rvert +\lvert z-y \rvert )\vee 2]}\cdot2^{-1} ,\label{(1.54)}
\\
\calS^{(k)}_{x,z,y} &\le \gamma^{r_k  [(\lvert x-z \rvert +\lvert z-y \rvert )\vee L_{k-1}^\alpha] }\cdot2^{-k}, \text{ for }k \ge 2, \label{(1.55)}
\\
\sum_{j \le k}\calS^{(j)}_{x,z,y} &\le \gamma^{ r_k [(\lvert x-z \rvert +\lvert z-y \rvert )\vee 2]}.
\label{(1.55a)}
\end{align}
\end{theorem}
\textit{Proof}. For $k=1$, we have the random-walk expansion (\ref{(20)}). For each of the $m+1$ steps of $g_1$, we have a factor $\gamma$. At each of the intermediate sites $x_1,\ldots,x_m$, we have
factors $\lvert(2d\gamma+v_{x_i}-\lambda)^{-1}\rvert\le 2/\veps_1 = 6/\veps$.
With a combinatoric factor
\footnote{Any sum 
$\sum_\rho \lvert T_\rho \rvert$ can be bounded by $\sup_\rho\lvert T_\rho\rvert c_\rho$ provided $\sum_\rho c_\rho^{-1} \le 1$. Then we call $c_\rho$ the combinatoric factor.}
$2^m(2d)^{m+1}$, we can replace the sum over walks with a supremum. Since $m+1 \ge (\lvert x-z \rvert +\lvert z-y \rvert )\vee 2$, we have a bound
\begin{equation}\label{(R1)}
(2d\gamma)^{m+1}\cdot \big(\tfrac{12}{\veps}\big)^m \le
\big(\tfrac{24d\gamma}{\veps}\big)^{[(\lvert x-z \rvert +\lvert z-y \rvert )\vee 2]} \le
\gamma^{r_1 [(\lvert x-z \rvert +\lvert z-y \rvert )\vee 2]}\cdot 2^{-1}.
\end{equation}
For the second inequality, we have taken $r_1 = .9$ and used the fact that $\gamma \le \veps^{20}$ is small.

\begin{figure}[h]
\centering
\includegraphics[width=.85\textwidth]{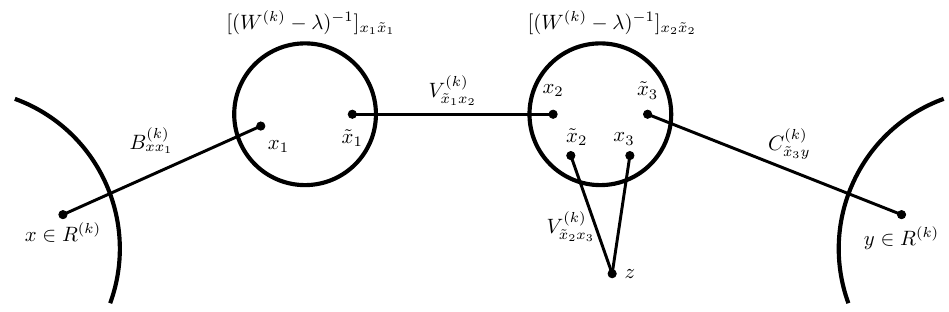}
\caption{A graph from $x$ to $y$ in the step $k$ random walk expansion. Intermediate blocks are components of $R^{(k-1)}\setminus R^{(k)}$. In this example, a type II step contains $z$.} \label{fig:2}
\end{figure}

For $k>1$ we have the random-walk expansion (\ref{(1.50)}).
The walk from $x$ to $y$ has the following structure. See Fig. \ref{fig:2}. The points $x,y$ are in $R^{(k)}$. Each of these matrices is given by a sum of graphs contributing to $B^{(j)}(D^{(j)}-\lambda)^{-1}C^{(j)}$ for $j<k$, so we may work inductively. Each of the blocks $B_{k-1}$ traversed by $g_k$ satisfies $\text{diam}(B_{k-1}) \le L_{k-1}$. Furthermore, they are separated from each other and from $R^{(k)}$ by at least a distance 
$L_{k-1}^\alpha$, with $\alpha = \tfrac{3}{2}$.
The steps $V^{(k)}_{\tilde{x}_jx_{j+1}}$ are of two types.
Type I steps move between different blocks $B_{k-1}$, while type II steps have $\tilde{x}_j,x_{j+1}$ in the same block $B_{k-1}$. Type II steps necessarily involve multigraphs containing a point $z_j \notin \bar{B}_{k-1}$; hence the inductive bound involves a total distance
$w_j \equiv \lvert\tilde{x}_j-z_j\rvert + \lvert z_j-x_{j+1}\rvert \ge 4L_{k-1} = 2L_k$. For type I steps we have $w_j \ge L^\alpha_{k-1}$.

We work inward toward $z$ from $x,y$, summing successively 
$x_1, \tilde{x_1},z_1, x_2, \tilde{x}_2, z_2, \ldots$. A combinatoric factor
 $(2L_{k-1}+1)^d$ controls the sum over $\tilde{x}_{j}$; a factor $(c_dw_{j}^{d+1})^2$ suffices to control the sums over $z_j$, $x_{j+1}$. Noting that $(c_dw_{j}^{d+1})^2(2L_{k-1}+1)^d \le \exp(w_jL_k^{1-\alpha})$ for $w_j \ge 2L_k$ and $L_0$ large, we have bounds for type II steps such as
\begin{align}\label{(R2)}
\sum_{x_j,\tilde{x}_j} \left| V^{(k)}_{\tilde{x}_{j-1}x_j }[(W^{(k)}-\lambda)^{-1}]_{x_j\tilde{x}_j} \right| &\le \sup_{w_{j} \ge L_k}\exp(w_{j}L_k^{1-\alpha})\gamma^{r_{k-1}w_{j}}\cdot 3\veps_k^{-1} 
\nonumber\\ 
&\le 3\exp(L_k^{2-\alpha})\gamma^{(2r_{k-1}-1.6)L_{k}}
\ll 1 .
\end{align}
for $\gamma$ small. Here we use the floor $r_{k-1} \ge .85$ for all $k$.
The bound (\ref{(R2)}) works because the graphs neglected in the truncation $F^{(k-1)}_\lambda \rightarrow \tilde{F}^{(k-1)}_\lambda (B_{k-1})$ are much smaller than $\veps_k$, the width of the spectral window.
Using (\ref{(R2)}), the sums over type II steps may be bounded by a factor 2 per type I step.

The type I steps span the entire distance $\lvert x-z \rvert +\lvert z-y \rvert $, except for gaps at blocks (we may need to include one type II step to cover $z$). The
minimum inter-block distance is $L_{k-1}^\alpha$ and the maximum block diameter is $L_{k-1}$. Hence the ratio between block diameter and inter-block distance is $\le L_{k-1}^{1-\alpha}$. Thus when converting sums into suprema for these steps, we have a constraint that $\sum_j w_j$ is at least
$[(\lvert x-z \rvert +\lvert z-y \rvert )\vee L_{k-1}^\alpha](1+L_{k-1}^{1-\alpha})^{-1}$.
In a manner similar to (\ref{(R2)}) we may bound
\begin{equation}\label{(R3)}
\calS^{(k)}_{x,z,y} \le \sup_{\{w_j\}} \, \prod_j \left[2^2\exp(w_jL_k^{1-\alpha})\gamma^{r_{k-1}w_j-3.2L_{k-1}}\right],
\end{equation}
where a second factor of two per step is included to control the sum over the number of steps. Since $w_j\ge L_{k-1}^\alpha$, we have that
\begin{equation}\label{(R4)}
3.2L_{k-1} + \lvert\log \gamma\rvert^{-1}(w_jL_k^{1-\alpha} + 2\ln 2)
\le 3.3w_jL_{k-1}^{1-\alpha} \le 4r_{k-1}w_jL_{k-1}^{1-\alpha}.
\end{equation}
Therefore,
\begin{equation}\label{(R5)}
\calS^{(k)}_{x,z,y} \le \sup_{\{w_j\}} 
\, \prod_j \gamma^{r_{k-1}w_j (1-4L_{k-1}^{1-\alpha}) } 
\le \gamma^{r_k[(\lvert x-z \rvert +\lvert z-y \rvert )\vee L_{k-1}^{\alpha}]}\cdot 2^{-k}.
\end{equation}
The decay rate has been adjusted downward to $r_k \equiv r_{k-1}(1-6L_{k-1}^{1-\alpha})$, with the difference between
$r_k$ and $r_{k-1}(1-4L_{k-1}^{1-\alpha})(1+L_{k-1}^{1-\alpha})^{-1}$ leading to some additional smallness and the factor $2^{-k}$ in this bound.
%an extra factor $\le \gamma^{r_{k-1}L_{k-1}^{1-\alpha}\cdot L_{k-1}^{\alpha}/2} \le 2^{-k}$. 
The above arguments use a uniform floor $r_k\ge .85$. This holds because $\sum_jL_{j-1}^{1-\alpha}$ is a convergent series, so for $L_0$ large enough,
the product $\prod_j(1-6L_{j-1}^{1-\alpha})$ can be made as close to 1 as required. Theorem \ref{thm:2} follows immediately from (\ref{(R1)}),(\ref{(R5)}). \qed

An immediate consequence of Theorem \ref{thm:2} is the following bound, which ensures that the terms neglected in truncating $F^{(k)}_\lambda$ to its block diagonal approximation are much smaller than the energy window used in the next step.
\begin{corollary} \label{cor:2'}
Under the same assumptions as Theorem \ref{thm:2},
\begin{equation}\label{(1.59a)}
\|F_\lambda^{(k)}  - \oplus_{\beta} \tilde{F}_\lambda^{(k)}(B_{k,\beta})\| \le \gamma^{3.3L_k}\ll \veps_{k+1}.
\end{equation}
\end{corollary}
\textit{Proof.} Graphs contributing to the difference go from $x$ to $y$ via a point $z$ such that $\lvert x-z \rvert \ge 2L_{k}$, $\lvert y-z \rvert \ge 2L_{k}$. We may bound the norm by estimating the maximum absolute row sum of the matrix. This means fixing $x$ and taking the sum over $z$ and $y$ of (\ref{(1.55a)}). Theorem \ref{thm:2} establishes decay at rate $r_\infty = .85$ over a distance $\lvert x-z \rvert +\lvert z-y \rvert  \ge 4L_k$. With a small decrease in rate to control the sum, we obtain (\ref{(1.59a)}).\qed

We will need bounds on the eigenfunction-generating kernel 
$G_\lambda^{(k)}$, which maps a function on $R^{(k)}$ to a function on $\Lambda$, see (\ref{(1.52)}),(\ref{(1.53)}).
\begin{theorem}
\label{thm:G}
Under the same assumptions as Theorem \ref{thm:2}, take $y \in R^{(k)}$. Then
\begin{equation}\label{(R6)}
G_{\lambda,xy}^{(k)} = \delta_{xy} + G_{\lambda,xy}^{\prime(k)},
\end{equation}
with $G_{\lambda,xy}^{\prime(k)}$ nonzero only for $x \in \Lambda \setminus R^{(k)}$, and
\begin{equation}\label{(R7)}
\lvert G_{\lambda,xy}^{\prime(k)}\rvert \le \gamma^{r_k\lvert x-y \rvert }.
\end{equation}
\end{theorem}
\textit{Proof.} The recursion (\ref{(1.52)}) can be written as
\begin{equation}\label{(R8)}
G_{\lambda,xy}^{(k)} = G_{\lambda,xy}^{(k-1)} 
\begin{pmatrix}
I \\- (D^{(k)}-\lambda)^{-1}C^{(k)} 
\end{pmatrix} ,
\end{equation}
where $I$ is the identity matrix for $R^{(k)}$ and $(D^{(k)}-\lambda)^{-1}C^{(k)}$ takes functions on $R^{(k)}$ to functions on $R^{(k-1)}\setminus R^{(k)}$. 
The bound (\ref{(R7)}) holds for $k=1$ as in the proof of (\ref{(1.54)}).
Working inductively, we have a setup similar to that of Theorem \ref{thm:2}, if we replace $B^{(k)}$ with $G_\lambda^{(k)}$ -- see Fig. \ref{fig:2}.
Arguing as in the previous proof, we find that only a summable and small fraction of decay is lost in step $k$. \qed

Recall that we defined $\tilde{F}^{(k)}_\lambda(B_k)$ by reducing the working volume from $\Lambda$ to $\bar{B}_k$. We do the same for the eigenfunction-generating kernel, writing $\tilde{G}_\lambda^{(k)}$ for the eigenfunction-generating kernel in $\bar{B}_k$. (We drop the $B_k$ argument, which will be understood from context.) As in Theorem \ref{thm:G}, we have that 
\begin{equation}\label{(R6prime)}
\tilde{G}_{\lambda,xy}^{(k)} = \delta_{xy} + \tilde{G}_{\lambda,xy}^{\prime(k)},
\end{equation}
where $\delta_{xy}$ is the identity matrix on $B_k$, and the off-diagonal part satisfies $\tilde{G}_{\lambda,xy}^{\prime(k)} \le \gamma^{r_k\lvert x-y \rvert }$.

We also need to control the difference $\tilde{F}_\lambda^{(k)}(B_k)-\tilde{F}_{E_k}^{(k)}(B_k)$ in norm, so that when isolated blocks are defined via the condition $\text{dist}\big(\text{spec}\,\tilde{F}_{E_k}^{(k-1)}(B_{k-1}),{E_k}\big) > \veps_k$, it is still safe to build the random walk expansion for the Schur complement with respect to $\lambda$.
\begin{theorem}
\label{thm:3}
Under the same assumptions as Theorem \ref{thm:2}, let $\lambda$, $\mu$ be in $I_{\veps_k/2}(E_k)$. Then
\begin{equation}\label{(1.60)}
\|\tilde{F}_\lambda^{(k)}(B)  - \tilde{F}_{\mu}^{(k)}(B)\| \le \gamma\lvert\lambda-{\mu}\rvert.
\end{equation}
\end{theorem}
\textit{Proof}. 
We have $F_\lambda^{(k)} = A^{(k)} - B^{(k)}(D^{(k)}-\lambda)^{-1}C^{(k)}$. In addition to the explicit appearance of $\lambda$, the matrices $A^{(k)},B^{(k)},C^{(k)},D^{(k)}$ depend on $\lambda$ for $k \ge 2$.  We already have control of the graphs contributing to these expressions by Theorem \ref{thm:2}. Similar arguments will allow us to control differences when we change $\lambda$ to $\mu$.

We begin by proving an analog of Theorem \ref{thm:2} to control the sum of differences of graphs, \textit{i.e.} each graph is evaluated at $\lambda$ and at $\mu$ and the difference taken. Let $\tilde\calS^{(k)}_{x,y}$ denote the sum of the absolute values of all difference multigraphs that contribute to $[B^{(k)}(D^{(k)}-\lambda)^{-1}C^{(k)}]_{xy}$.
We claim that
\begin{align}
\tilde{\calS}^{(1)}_{x,y} &\le \frac{1}{\veps_1}\gamma^{  r_1(\lvert x-y \rvert \vee 2)}\cdot2^{-1} \lvert \lambda - \mu \rvert ,
\label{(1.60a)} \\
\tilde{\calS}^{(k)}_{x,y} &\le \frac{1}{\veps_k}\gamma^{ r_k (\lvert x-y \rvert \vee L_{k-1}^\alpha) }\cdot2^{-k} \lvert \lambda - \mu \rvert , \text{ for } k\ge 2;
\label{(1.61)}
\end{align}
and hence that
\begin{equation}\label{(1.61a)}
\sum_{j \le k}\tilde{\calS}^{(j)}_{x,y} \le \frac{1}{\veps_1}\gamma^{r_1  (\lvert x-y \rvert \vee 2)}\lvert \lambda-\mu \rvert .
\end{equation}
Consider the case $k=1$. Redoing the proof of (\ref{(1.54)}) for differences, we obtain a sum of graphs wherein a difference
\begin{equation}\label{(1.62)}
[(W^{(1)}-\lambda)^{-1} - (W^{(1)}-\mu)^{-1}]_{x_i\tilde{x}_i} = (\lambda-\mu)[(W^{(1)}-\lambda)^{-1} (W^{(1)}-\mu)^{-1}]_{x_i\tilde{x}_i}
\end{equation}
appears in place of the corresponding matrix element of $(W^{(1)}-\lambda)^{-1}$ or $(W^{(1)}-\mu)^{-1}$. 
In the bound, this leads to
an extra factor $3/\veps_1$ from the additional $(W^{(1)}-\mu)^{-1}$. Estimating as in  (\ref{(R1)}), we obtain (\ref{(1.60a)}).

For step $k \ge 2$, we apply the difference operation to each factor in (\ref{(1.50)}).
Each matrix 
$W^{(k)},V^{(k)},B^{(k)},C^{(k)}$ is covered by (\ref{(1.61a)}), by induction, and this leads to an incremental factor of $\veps_1^{-1}\lvert \lambda-\mu \rvert $, compared to before. When we difference the explicit factors of $\lambda$ in (\ref{(1.50)}), we obtain as in (\ref{(1.62)}) a new factor of $(W^{(k)}-\mu)^{-1} \lvert \lambda - \mu \rvert $. This leads to an incremental factor of $3\veps_k^{-1}\lvert \lambda-\mu \rvert $, compared to before, coming from the bound (\ref{(1.49)}). Thus in all cases, we get no worse than an extra factor $3\veps_k^{-1}\lvert \lambda-\mu \rvert $. This completes the proof of (\ref{(1.61)}). The minimum decay length $L_{k-1}^\alpha$ is much greater than $1.6L_{k}$, so the factor $\veps_k^{-1}$ can be absorbed with a small change in the rate $r_k$. Then (\ref{(1.61a)}) follows immediately.

Note that  (\ref{(1.61a)}) provides an estimate on the matrix elements of $F_\lambda^{(k)} - F_\mu^{(k)}$, so that
\begin{equation}\label{(1.64)}
\sum_y \left|\big[F_\lambda^{(k)} - F_\mu^{(k)}\big]_{xy}\right| \le \gamma\lvert \lambda-\mu \rvert ,
\end{equation}
and hence
\begin{equation}
\|F_\lambda^{(k)} - F_\mu^{(k)}\| \le \gamma\lvert \lambda-\mu \rvert ,
\label{(1.65)}
\end{equation}
The same bound applies to $\|\tilde{F}_\lambda^{(k)}(B_k)- \tilde{F}_\mu^{(k)}(B_k)\|$, since in this case we are just looking at a subset of the collection of multigraphs (the ones that remain within $\bar{B}_k$).\qed

\section{Probability Bounds}\label{sec:3}
\subsection{Movement of Eigenvalues}\label{ssec:3.1}

Here we demonstrate that when we transition from $\bar{B}_{k-1}$ to 
$\bar{B}_{k}$, the new approximate eigenvalues depart the spectral window $I_{\veps_{k+1}}(E_{k+1})$ with probability at least $1-\tfrac{1}{N-1}$.

The first step is to establish the existence of sites with significant influence on the relevant eigenvalues. Let $H_X$ be the matrix obtained by restricting both indices of $H$ to $X\subseteq \Lambda$.

\begin{lemma}\label{lem:3.1}
Let $B_{k-1}$ be isolated in step $k$, \textit{i.e.} $\mathrm{diam}(B_{k-1}) \le L_{k-1}$.
Let $\psi = \tilde{G}_\lambda^{(k-1)}\vphi$, with $\big(\tilde{F}^{(k-1)}_\lambda(B_{k-1})-\lambda \big)\vphi = 0$, $\|\vphi \| =1$, and
$\lambda \in I_{\veps_k/2}(E_k)$. 
Recall that $\tilde{G}_\lambda^{(k-1)}$ is defined in the domain $\bar{B}_{k-1}$,  
so that
$\big(H_{\bar{B}_{k-1}}-\lambda\big)\psi = 0$.
For any $y$ with $\mathrm{dist}(y,\bar{B}_{k-1}) = 1$, define the influence of $y$ as
\begin{equation}\label{(P1)}
\calI_\psi(y) =  \Bigg|\sum_{x\in\bar{B}_{k-1},\, \lvert x-y \rvert =1} \psi(x) \Bigg|.
\end{equation}
If $\mathrm{diam}(\bar{B}_{k-1}) < \mathrm{Diam}(\Lambda)$, 
then for $\gamma$ small, there exists at least one $y \in \Lambda 
\setminus \bar{B}_{k-1}$ with $\calI_\psi(y) \ge \gamma^{3.1L_{k-1}}$.
\end{lemma}
\textit{Proof.} 
Choose coordinates in the rectangle $\Lambda$ so that: (1) The origin is at a point $\bar{x}$ of $B_{k-1}$ such that $\lvert \vphi(\bar{x}) \rvert  \ge \lvert B_{k-1}\rvert^{-1/2}$ -- such a point exists because $\vphi$ is normalized;
(2) the $z$-coordinate runs toward a boundary face of $\Lambda$ that
contains no points of $\bar{B}_{k-1}$ -- such a face exists because
$\text{diam}(\bar{B}_{k-1}) < \text{Diam}(\Lambda)$.
Our definition of $\bar{B}_{k-1}$ ensures that it extends no further than 
a distance $2.05L_{k-1}$ from $B_{k-1}$. Therefore, $z_{\text{max}} \le 3.05L_{k-1}$, where $z_{\text{max}}$ denotes the maximal $z$-coordinate for points in $\bar{B}_{k-1}$.

We give a proof by contradiction.
Suppose there is no site $y \in \Lambda \setminus \bar{B}_{k-1}$ 
with $\calI_\psi(y) \ge \gamma^{3.1L_{k-1}}$. Then for each $x_0$ in the top layer at $z=z_{\text{max}}$ we have $\lvert\psi(x_0)\rvert < \gamma^{3.1L_{k-1}}$. (Each site $y$ with $d^{\text{th}}$ coordinate
$z_{\text{max}}+1$ is in $\Lambda$ and is adjacent to no more than one site of $\bar{B}_{k-1}$, so the sum in (\ref{(P1)}) reduces to a single term.)

Let $x$ be a site of $\bar{B}_{k-1}$ that is immediately below a top-layer site $x_0$ of $\bar{B}_{k-1}$. Let $y_1,\ldots,y_{2d-1}$ denote the other neighbors of $x_0$. Then
\begin{equation}\label{(P2)}
-\gamma \bigg(\psi(x) + \sum_{i-1}^{2d-1} \psi(y_i) \bigg) + (2d\gamma + v_{x_0} -\lambda )\psi(x_0) = 0,
\end{equation}
where we put $\psi(x) = 0$ for $x \notin \bar{B}_{k-1}$. Observing that $\lambda \in [0, 1+4d\gamma]$, $v_{x_0} \in [0,1]$, we have that
\begin{equation}\label{(P3)}
\lvert 2d\gamma + v_{x_0} - \lambda\rvert \le 1+2d\gamma.
\end{equation}
Hence
\begin{equation}\label{(P4)}
\lvert \psi(x) \rvert  \le [\tfrac{1}{\gamma}(1+2d\gamma) + 2d-1]\gamma^{3.1L_{k-1}} = (\tfrac{1}{\gamma}+4d-1)\gamma^{3.1L_{k-1}} \le \tfrac{2}{\gamma}\gamma^{3.1L_{k-1}}.
\end{equation}
The remaining sites $x$ in the second layer lie below a site
$y\notin\bar{B}_{k-1}$, which then must satisfy $\calI_\psi(y)< \gamma^{3.1L_{k-1}}$.
We have already established that the other neighbors of $y$ satisfy 
$\lvert\psi(y_i)\rvert\le \gamma^{3.1L_{k-1}}$. Therefore, $\lvert \psi(x) \rvert \le2d\gamma^{3.1L_{k-1}}$ (otherwise, even after a cancellation with the other neighbors, $\calI_\psi(y)$ would be too large). Thus (\ref{(P4)}) holds for all sites in the second layer (for $\gamma$ small).

We continue this argument on successive layers, obtaining a bound
\begin{equation}\label{(P5)}
\lvert \psi(x) \rvert \le \big(\tfrac{2}{\gamma}\big)^{z_{\text{max}}-z}\gamma^{3.1L_{k-1}} \le (2\gamma)^{.05L_{k-1}}
\end{equation}
for the layer with $d$-coordinate $z
\ge 0$. 
Thus we learn that $\lvert\psi(\bar{x})\rvert = \lvert\vphi(\bar{x})\rvert \le (2\gamma)^{.05L_{k-1}}$.
This contradicts the 
condition $\lvert \vphi(\bar{x}) \rvert  \ge \lvert B_{k-1} \rvert ^{-1/2} \ge (2L_{k-1}+1)^{-d/2}$.
Hence there must be at least one influential site 
$y\in \Lambda \setminus \bar{B}_{k-1}$ satisfying $\calI_\psi(y) \ge \gamma^{3.1L_{k-1}}$.
\qed

We will need to follow the behavior of the number of eigenvalues in small windows around various energies. Define for each $k$ and each component of $R^{(k)}$
\begin{equation}\label{(P6)}
\hat{n}_{k}(B_{k}) = \text{the number of eigenvalues of }\tilde{F}^{(k)}_{E_{k+1}}(B_{k}) \text{ in } I_{\veps_{k+1}}(E_{k+1}).
\end{equation}
Here we count eigenvalues with multiplicity.
We will see that this is a non-increasing function of $k$. 
Under the right circumstances, we can show that $\hat{n}_{k}(B_{k})<\hat{n}_{k-1}(B_{k-1})$.
To this end, we consider the implications of Lemma \ref{lem:3.1} for randomness-driven 
movement of the eigenvalues. We work in a specific situation, where
a block $B_{k-1}$ of $R^{(k-1)}$ is isolated and resonant in step $k$ with respect to an energy $E_k$. 
We assume $B_{k-1}$ remains the same in the next step (\textit{i.e.} $B_{k-1} = B_k$, because it does not combine with other components of $R^{(k-1)}$ when forming components of $R^{(k)}$).
Since $B_{k-1}$ is isolated in step $k$, $\text{diam}(B_{k-1})\le L_{k-1}$, and so
$\text{diam}(B_k)\le L_{k-1}$ as well; hence
$B_k$ is isolated in step $k+1$.
We have an energy $E_{k+1} = E_k = E$ (fixed energy procedure) or else choose an energy $E_{k+1} \in I_{\veps_k/3}(E_k)$ (energy-following procedure).
The central questions that we need to address are the following. How likely is it that $B_k$ is resonant in step $k+1$? If it is resonant, how many eigenvalues of $\tilde{F}^{(k)}_{E_{k+1}}(B_k)$ are in $I_{\veps_{k+1}}(E_{k+1})$? 
The plan is to identify a site $\bar{y} \in \bar{B}_k\setminus\bar{B}_{k-1}$, and control the shift in spectrum as we transition from $\tilde{F}^{(k-1)}_{E_k}(B_{k-1})$ to  $\tilde{F}^{(k)}_{E_{k+1}}(B_{k})$, as a function of $v_{\bar{y}}$,
with all other potentials fixed. We show that $\hat{n}_{k}(B_{k}) \le \hat{n}_{k-1}(B_{k-1}) $ and that $\hat{n}_{k}(B_{k}) = \hat{n}_{k-1}(B_{k-1})$ for at most one value of $v_{\bar{y}}$.

\begin{proposition}\label{prop:3.3}
Let $L_0$ be sufficiently large. Take $\veps = \tfrac{1}{N-1}$ to be sufficiently small, depending on $L_0$, and take $\gamma \le \veps^{20}$. 
Assume that $B_{k-1}$ is isolated and resonant in step $k$ with respect to energy $E_k$, and that $\mathrm{diam}(\bar{B}_{k-1}) < \mathrm{Diam}(\Lambda)$. 
Assume that $B_{k-1}$ remains isolated in step $k+1$, so that $B_{k} = B_{k-1}$. Fix all $v_{y}$ for $y \in \bar{B}_{k-1}$. 
Then there is a choice of $\bar{y}$, adjacent to $\bar{B}_{k-1}$ and depending only on $v_y$ for $y \in \bar{B}_{k-1}$, such that the following statements hold. 
Fix all remaining $v_y \in \bar{B}_k$, $y \ne \bar{y}$. Then
\begin{equation}\label{equals}
\hat{n}_k(B_k) \le \hat{n}_{k-1}(B_{k-1}). 
\end{equation}
Furthermore, consider  two cases:
\begin{enumerate}
\item $E_{k+1} = E_k = E$ (fixed energy procedure);
\item $E_{k+1} = E_{k+1}(v_{\bar{y}})$ is chosen in $I_{\veps_k/3}(E_k)$ (energy-following procedure), and $\hat{n}_{k-1}(B_{k-1}) > 1$.
\end{enumerate}
In either case, we have that
\begin{equation}\label{less}
\hat{n}_k(B_k) < \hat{n}_{k-1}(B_{k-1})
\end{equation}
for all but one value of $v_{\bar{y}}$.
\end{proposition}

\textit{Preliminaries.} Let $\lambda_0$ be the closest eigenvalue of $\tilde{F}^{(k)}_{E_{k+1}}(B_{k})$ to $E_{k+1}$. Note that $\lambda_0$ depends on $E_{k+1}$ and hence on $v_{\bar{y}}$, in case 2. We will assume going forward that $\lambda_0 \in I_{\veps_{k+1}}(E_{k+1})$, because otherwise definition (\ref{(P6)}) would imply that 
$\hat{n}(B_k) = 0$, and then (\ref{less}) is automatically true. As $\veps_{k+1}\ll \gamma\veps_k$ and  $\lvert E_k-E_{k+1} \rvert  \le \veps_k/3$, we have that $\lvert \lambda_0-E_k\rvert \le (\tfrac{1}{3} + \gamma)\veps_k$.
 There must be an eigenvalue $\lambda_1$ of $\tilde{F}^{(k-1)}_{E_k}(B_{k-1})$ within $\gamma\veps_k$ of $\lambda_0$.
(Here we use (i) $\lvert E_k-E_{k+1} \rvert  \le \veps_k/3$, so by Theorem \ref{thm:3} the shift in spectrum in the transition $\tilde{F}^{(k-1)}_{E_k}(B_{k-1})
\rightarrow \tilde{F}^{(k-1)}_{E_{k+1}}(B_{k-1})$ is $\le \gamma \veps_k/3$; and (ii) as explained earlier, the shift in spectrum due to graphs extending to $\bar{B}_k\setminus\bar{B}_{k-1}$ is $\le \gamma^{3.3L_{k-1}} \ll \veps_k$.) 
%Then we have $\lvert \lambda_0 - E_k \rvert  \le (\tfrac{1}{3}+\tfrac{1}{20})\veps_k $. 
Consequently, $\lvert\lambda_1 - E_{k}\rvert \le (\tfrac{1}{3} + 2\gamma)\veps_k$. 
By Theorem \ref{thm:3} and a fixed point argument, there is a  solution to 
$\mu \in \text{spec}\,\tilde{F}^{(k-1)}_{\mu}(B_{k-1})$ satisfying
$\lvert \mu - \lambda_1 \rvert  \le 2\gamma\lvert\lambda_1 - E_{k}\rvert \le \gamma\veps_k$. 
Hence  
$\lvert\mu - E_{k}\rvert \le (\tfrac{1}{3}+3\gamma)\veps_k < \tfrac{2}{5}\veps_k$. 
Let $\lambda$ be the solution to $\mu \in \text{spec}\,
\tilde{F}^{(k-1)}_{\mu}(B_{k-1})$ that is closest to $E_k$. 
Clearly, $\lambda$ also lies in $I_{2\veps_k/5}(E_k)$.
As $\lambda$ is the closest eigenvalue of $H_{\bar{B}_{k-1}}$ to $E_k$, it evidently depends only on the potentials in $\bar{B}_{k-1}$. We will use $\lambda$ as the starting point for much of the analysis to follow.

Define
\begin{equation}\label{(P7)}
\hat{n}= \text{the number of eigenvalues of }\tilde{F}^{(k-1)}_{\lambda}(B_{k-1}) \text{ in } I_{\veps_{k}/2}(E_{k}) .
\end{equation}
Then we have that 
\begin{equation}\label{(P8)}
\hat{n} \le \hat{n}_{k-1}(B_{k-1}),
\end{equation}
because by Theorem \ref{thm:3}, the change in the spectrum is 
$\le \gamma\lvert \lambda - E_k \rvert  \le \gamma\veps_k/2$; this means that eigenvalues of $\tilde{F}^{(k-1)}_{E_k}(B_{k-1}) $ outside of $I_{\veps_k}(E_k)$ cannot migrate into $I_{\veps_k/2}(E_{k})$.

Let us write
\begin{equation}\label{starstar}
\Delta F_\lambda = \tilde{F}^{(k-1)}_{\lambda}(B_{k-1}) -\tilde{F}^{(k)}_{\lambda}(B_{k}).
\end{equation}
Recall that we are assuming $B_k = B_{k-1}$. Recall also that
$\tilde{F}^{(k)}_{\lambda}(B_{k})$ is defined by restricting the multigraphs in (\ref{(1.50)}) to $\bar{B}_k$, while for $\tilde{F}^{(k-1)}_{\lambda}(B_{k-1})$ they must remain within $\bar{B}_{k-1}$.
By expanding the set of multigraphs, we gain access to the randomness in $\bar{B}_k\setminus\bar{B}_{k-1}$; this will be used to demonstrate eigenvalue movement.

The expansion of the domain from $\bar{B}_{k-1}$ to $\bar{B}_k$ leads to a useful representation for $\Delta F_\lambda$. Let us put
\begin{align}\label{(P9)}
d &\equiv H_{\bar{B}_k\setminus B_{k-1}} - \lambda, \quad 
d_1 \equiv \big(H_{\bar{B}_{k-1}\setminus B_{k-1}}-\lambda\big)\oplus
\big(H_{\bar{B}_{k}\setminus \bar{B}_{k-1}}-\lambda\big),
\\
\Gamma_{xy} &= \begin{cases}
\gamma, &\text{if } \lvert x-y \rvert  = 1 \text{ with one in }\bar{B}_{k-1} \text{ and the other in }\bar{B}_k\setminus\bar{B}_{k-1}; \\
0, &\text{otherwise.}
\end{cases}
\end{align}
Then by the second resolvent identity,
\begin{equation}\label{(P10)}
d^{-1} = d_1^{-1} + d_1^{-1}\Gamma d^{-1} = d_1^{-1} + d_1^{-1}\Gamma d_1^{-1}+ d_1^{-1}\Gamma d^{-1}\Gamma d_1^{-1}. 
\end{equation}
The second term vanishes if both indices are taken in $\bar{B}_{k-1}$. We may write
\begin{equation}\label{(P11)}
H_{\bar{B}_{k-1}} = \begin{pmatrix}
a &b \\c & d_1+\lambda  \end{pmatrix} , \quad
H_{\bar{B}_{k}} = \begin{pmatrix}
a &b \\c & d +\lambda \end{pmatrix} ,
\end{equation}
where $a = H_{B_{k-1}}$, $d_1$ is restricted to $\bar{B}_{k-1}\setminus B_{k-1}$, and $b$, $c$ contain the nearest-neighbor interactions connecting $B_{k-1}$ to $\bar{B}_{k-1}\setminus B_{k-1}$. Then
\begin{equation}\label{(P111a)}
\tilde{F}^{(k-1)}_{\lambda}(B_{k-1}) = a-bd_1^{-1}c,\quad
 \tilde{F}^{(k)}_{\lambda}(B_{k}) = a-bd^{-1}c.
\end{equation}
Thus we see that
\begin{equation}\label{(star)}
\Delta F_\lambda = bd_1^{-1}\Gamma d^{-1} \Gamma d_1^{-1}c
=\tilde{G}_\lambda^{\prime(k-1)\text{tr}}\Gamma\big(H_{\bar{B}_{k}\setminus 
B_{k-1}}-\lambda\big)^{-1}\Gamma\tilde{G}_\lambda^{\prime(k-1)}.
\end{equation}
For the second equality we recognize $-d_1^{-1}c$ as the eigenfunction-generating kernel $\tilde{G}_\lambda^{\prime(k-1)}$ from (\ref{(R6prime)});
likewise $-bd_1^{-1}$ is its transpose.
We have not emphasized the connection with resolvents, but they provide a useful perspective here. When a region $X$ is divided into $R^{(k)}$ and $X\setminus R^{(k)}$, for example, the resolvent in $X$ has an expression as a Neumann series involving the resolvent in $X\setminus R^{(k)}$
via the formula for the inverse of a block matrix. Continuing to $X\setminus R^{(k)} \setminus R^{(k-1)}$, etc., we obtain the full multigraph expansion. This is closely connected to the procedure in \cite{Frohlich1983}, the main difference being that we use the full Neumann expansion rather than a finite iteration of the resolvent identity.

Nevertheless, when it comes to estimates we find it convenient to work directly with the multigraph expansions. In fact the representation (\ref{(star)}) for $\Delta F_\lambda$ can be seen directly at the level of multigraph expansions.
Consider a multigraph contributing to the difference 
$(\Delta F_\lambda)_{x\tilde{x}}$.
It begins at $x$, then departs $\bar{B}_{k-1}$ through a link $\langle x',x'' \rangle$ with $x'' \in \bar{B}_k\setminus\bar{B}_{k-1}$, $x''$ adjacent to $x'\in\bar{B}_{k-1}$. It returns to $\bar{B}_{k-1}$ for the last time via a link $\langle \tilde{x}'',\tilde{x}' \rangle$ with $\tilde{x}'\in \bar{B}_{k-1}$, $\tilde{x}'$ adjacent to $x''\in\bar{B}_{k}\setminus\bar{B}_{k-1}$. The sum over multigraphs leading from $x$ to $x'$ leads to the eigenfunction-generating kernel $\tilde{G}_{\lambda,xx'}^{\prime(k-1)\text{tr}}$ in $\bar{B}_{k-1}$, the sum over multigraphs leading from $\tilde{x}'$ to $\tilde{x}$ leads to a kernel $\tilde{G}_{\lambda,\tilde{x}'\tilde{x}}^{\prime(k-1)}$.
In between, there are two factors of $\gamma$ from the steps 
$\langle x',x'' \rangle$, $\langle \tilde{x}'',\tilde{x}' \rangle$,
and a sum of multigraphs going from $x''$ to $\tilde{x}''$ in $\bar{B}_k\setminus B_{k-1}$. As explained above, the latter may be identified with
$\big(H_{\bar{B}_{k}\setminus B_{k-1}}-\lambda\big)^{-1}$ as in (\ref{(star)}).

We may use Theorem \ref{thm:G} to control the sum of multigraphs contributing to $\tilde{G}_\lambda^{\prime(k-1)}$.
As it stands, Theorem \ref{thm:2} does not apply to the graphical expansion for $[\Gamma\big(H_{\bar{B}_{k}\setminus B_{k-1}}-\lambda\big)^{-1}\Gamma]_{xy}$. However, the only substantive difference with the situation considered there is the fact that the minimum distance from a block $B_j$ to $\{x,y\}$ is $L_j^{\sqrt{\alpha}}$, instead of $L_j^{\alpha}$. This does not affect the proof of (\ref{(1.55a)}), as the requirement $\alpha > 1$ is satisfied also for $\sqrt{\alpha}$.

We switch to a basis of normalized eigenvectors $\{\vphi_1,
\ldots,\vphi_{\hat{n}},\vphi_{\hat{n}+1},\ldots,\vphi_n\}$  corresponding to eigenvalues 
$\lambda_1,
\ldots,\lambda_{\hat{n}},\lambda_{\hat{n}+1},\ldots,\lambda_n$.
of $\tilde{F}_\lambda^{(k-1)}(B_{k-1})$. Here $\lambda = \lambda_1,\lambda_2,\ldots,\lambda_{\hat{n}}$ are the eigenvalues in 
$I_{\veps_k/2}(E_{k})$, and $n = \lvert B_{k-1} \rvert $ is the number of sites in $B_{k-1}$. In this basis,
\begin{equation}\label{(P12)}
\Delta F_{\lambda,\beta\tilde{\beta}} = \langle\vphi_\beta,\tilde{G}_\lambda^{\prime(k-1)\text{tr}}\Gamma\big(H_{\bar{B}_{k}\setminus B_{k-1}}-\lambda\big)^{-1}\Gamma\tilde{G}_\lambda^{\prime(k-1)}\vphi_{\tilde{\beta}}\rangle.
\end{equation}
Note that $\psi\equiv \tilde{G}_\lambda^{(k-1)}\vphi_1=(\vphi_1,\tilde{G}_\lambda^{\prime(k-1)}\vphi_1)$ is an eigenvector of 
$H_{\bar{B}_{k-1}}$ with eigenvalue $\lambda$.
In view of the decay of $\tilde{G}_\lambda^{\prime(k-1)}$ that was 
established in
Theorem \ref{thm:G}, $\|\tilde{G}_\lambda^{\prime(k-1)}\| \le \gamma^{4/5}$, 
so $1 \le \|\psi\| \le 1+\gamma^{4/5}$. 

Define, for any $y\in \bar{B}_k\setminus \bar{B}_{k-1}$ with $\text{dist}(\bar{y},\bar{B}_{k-1})=1$,
\begin{equation}\label{(P13)}
\chi_y(x) = \begin{cases} 1, \text{ if } \lvert x-y \rvert  = 1 \text{ and } x\in \bar{B}_{k-1};
\\ 0, \text{ otherwise.} \end{cases}
\end{equation}
This is the indicator function for the set of sites in $\bar{B}_{k-1}$ that are adjacent to $y$. Then put
\begin{equation}\label{(P14)}
a_\beta(y) \equiv \langle \chi_y, \tilde{G}_\lambda^{\prime(k-1)}\vphi_\beta\rangle.
\end{equation}
The vector $\mathbf{a}^{(\text{r})}(y) = (a_1(y),\ldots,a_{\hat{n}}(y))$
 -- in particular its length-squared $\lvert \mathbf{a}^{(\text{r})}(y) \rvert ^2 = \sum_{\beta = 1}^{\hat{n}}\lvert a_\beta(y)\rvert^2$ -- is a measure of the influence of $v_y$ on the family of eigenvalues $\{\lambda_1,\ldots,\lambda_{\hat{n}}\}$ -- the ones resonant with $E_{k}$ to within $\veps_k/2$.
Let us choose $\bar{y}\in\Lambda$ with $\text{dist}(\bar{y},\bar{B}_{k-1})=1$ 
to be a site that maximizes $\lvert \mathbf{a}^{(\text{r})}(y) \rvert $ from amongst all neighbors of $\bar{B}_{k-1}$.
Lemma \ref{lem:3.1} implies that $\lvert a_1(y) \rvert  \ge \gamma^{3.1 L_{k-1}}$
for at least one $y$ adjacent to $\bar{B}_{k-1}$. Hence
$\lvert\mathbf{a}^{(\text{r})}(\bar{y})\rvert \ge \gamma^{3.1 L_{k-1}}$.
All of these definitions are based on $H_{\bar{B}_{k-1}}$ and its Schur complement $\tilde{F}_\lambda^{(k-1)}(B_{k-1})$, so $\bar{y}$ depends only on the potentials in $\bar{B}_{k-1}$.

Let us write, for $x,y  \in \bar{B}_k\setminus B_{k-1}$ and adjacent
to $\bar{B}_{k-1}$,
\begin{equation}\label{(a)}
K(x,y) \equiv \gamma^2\big(H_{\bar{B}_k\setminus \bar{B}_{k-1}}-\lambda\big)^{-1}_{xy} = K_0(x,y) + K_1(x,y) + K_2(x,y).
\end{equation}
For $K_0$, the sum of multigraphs for $\big(H_{\bar{B}_k\setminus \bar{B}_{k-1}}-\lambda\big)^{-1}$ is restricted to those that do not include the site $\bar{y}$ (which means that $K_0$ is independent of $v_{\bar{y}}$).
For $K_1$, we include only the trivial multigraph of at $\bar{y}$; thus
\begin{equation}\label{(b)}
K_1(x,y) = \frac{\gamma^2\delta_{x\bar{y}}\delta_{\bar{y}y}}{v_{\bar{y}} + 2d\gamma - \lambda}.
\end{equation}
The remaining graphs make up $K_2$; they must contain $\bar{y}$ and have at least one additional step. As explained above, a variant of Theorem \ref{thm:3} implies that
\begin{align}
\lvert K_0(x,y) \rvert  &\le \gamma^{.85(\lvert x-y \rvert \vee 2)},
\label{(c0)}
\\
\lvert K_2(x,y)\rvert &\le \gamma^{.85[(\lvert x-\bar{y}\rvert+\lvert \bar{y}-y\rvert)\vee 3)]}.
\label{(c')}
\end{align}
With these definitions, we may write
\begin{equation}\label{(d')}
\Delta F_{\lambda,\beta\tilde{\beta}} = \sum_{xy} a_\beta(x)K(x,y)a_{\tilde{\beta}}(y).
\end{equation}
In order to obtain precise control over the behavior of eigenvalues
in $I_{\veps_k/2}(E_{k})$ as we make the perturbation
 (\ref{starstar}), we work with another Schur complement. Write
\begin{equation}\label{asterisk}
\tilde{F}^{(k-1)}_{\lambda}(B_{k-1})= \begin{pmatrix}
q &r \\s & t  \end{pmatrix} ,\quad\tilde{F}^{(k)}_{\lambda}(B_{k})=
\begin{pmatrix}
\tilde{q} &\tilde{r} \\ \tilde{s} & \tilde{t}  \end{pmatrix},
\end{equation}
where $q$, $\tilde{q}$ are the restrictions of $\tilde{F}^{(k-1)}_{\lambda}(B_{k-1})$, $\tilde{F}^{(k)}_{\lambda}(B_{k})$ to the subspace spanned by $\{\vphi_1,\ldots,\vphi_{\hat{n}}\}$, and $r,s,t,\tilde{r},\tilde{s},\tilde{t}$ fill out the remainder of the block decomposition of these matrices. Note that $r=s=0$, because $\tilde{F}^{(k-1)}_{\lambda}(B_{k-1})$ is diagonal in the basis of eigenvectors. Define the Schur complements
\begin{align}
f_\lambda^{(k-1)} &= q-r(t-\lambda)^{-1}s = q,
 \label{starA}
 \\
f_\lambda^{(k)} &= \tilde{q}-\tilde{r}(\tilde{t}-\lambda)^{-1}\tilde{s}.
\label{starB}
\end{align}
Note that the operators $\tilde{q},\tilde{r},\tilde{s},\tilde{t}$ depend on $\lambda$ through the operators $\tilde{G}_\lambda^{\prime(k-1)}$, $\big(H_{\bar{B}_k\setminus B_{k-1}}-\lambda\big)^{-1}$ used in the constructions above.
Then if we change the energy $\lambda \rightarrow E_{k+1}$, (\ref{starB}) serves to define $f_{E_{k+1}}^{(k)}$ as well.
(However, the basis $\{\vphi_1,\ldots,\vphi_{\hat{n}},\vphi_{\hat{n}+1},\ldots,\vphi_n\}$ that defines the block decomposition (\ref{asterisk}) is kept fixed.) 

A final preliminary definition is needed in order to identify the constant ($v_{\bar{y}}$-independent) part of $\tilde{F}_\lambda^{(k)}(B_k)$. (It might be much bigger than the estimates we can manage on the $v_{\bar{y}}$-dependent part.) To do this, we define $\tilde{F}_\lambda^{(k)\text{const}}$ by deleting $\bar{y}$ from the working region $\bar{B}_k$ in the definition of $\tilde{F}_\lambda^{(k)}(B_k)$. Equivalently, we may write as in (\ref{(star)}) 
\begin{equation}\label{const}
\tilde{F}_\mu^{(k)\text{const}} = 
\tilde{G}_\mu^{\prime(k-1)\text{tr}}\Gamma\big(H_{\bar{B}_{k}\setminus 
B_{k-1}\setminus\{\bar{y}\}}-\mu\big)^{-1}\Gamma\tilde{G}_\mu^{\prime(k-1)},
\end{equation}
using now $\mu$ for the running spectral parameter.
In fact, $\gamma^2\big(H_{\bar{B}_{k}\setminus 
B_{k-1}\setminus\{\bar{y}\}}-\mu\big)^{-1}$ is exactly $K_0$, which was given a graphical definition after (\ref{(a)}). Now let us define $\tilde{\lambda}$ to be the closest solution to $\lambda$ of the relation
$\mu \in \text{spec}\,\tilde{F}_\mu^{(k)\text{const}}$. Then it is evident that $\tilde{\lambda}$ does not depend on $v_{\bar{y}}$.
Note that $\lambda \in \text{spec}\,\tilde{F}_\lambda^{(k-1)}(B_{k-1})$ and $\tilde{\lambda} \in \text{spec}\,\tilde{F}_{\tilde{\lambda}}^{(k)\text{const}}$, and since the difference $\tilde{F}_\lambda^{(k-1)}(B_{k-1})-\tilde{F}_{\lambda}^{(k)\text{const}}$ involves graphs going 
outside of $\bar{B}_{k-1}$, we have as in Corollary \ref{cor:2'} that the norm of the difference is $\ll \veps_k$. Therefore,
\begin{equation}\label{star1A}
\lvert \lambda -\tilde{\lambda}\rvert \le \gamma\veps_k.
\end{equation}
We have that both $\lambda$ and $\tilde{\lambda}$ lie in $I_{2\veps_k/5}$, since the argument above actually placed $\lambda$ in a slightly smaller interval.
We will also need the second Schur complement for $\tilde{F}_\mu^{(k)\text{const}}$, which we denote by $f_\mu^{(k)\text{c}}$.

The following lemma gives the estimates we need to control the behavior of the spectrum in $I_{\veps_k/2}(E_{k})$. There are two cases. If $E_{k+1}$ is close to $\tilde{\lambda}$, then $v_{\bar{y}}$ creates spread in the spectrum by varying the coefficient of a rank 1 operator. If $E_{k+1}$ is far from $\tilde{\lambda}$, then the presence of spectrum near $E_{k+1}$ creates spread.
 
\begin{lemma}\label{prop:3.2}
Consider the situation in Proposition \ref{prop:3.3} (cases 1 and 2), and assume (as explained above) that $\lambda_0 \in  I_{\veps_k/20}(E_{k+1})$. If $\lvert\tilde{\lambda}-E_{k+1}\rvert \le \gamma^2
\lvert\mathbf{a}^{(\text{r})}(\bar{y})\rvert^2\cdot 10N$, then 
there is a decomposition
\begin{equation}\label{psi}
\big(f_\lambda^{(k-1)} - f_{E_{k+1}}^{(k)}\big)_{\beta\tilde{\beta}} = \frac{\gamma^2}{v_{\bar{y}}+2d\gamma-\lambda}
a^{(\text{r})}_\beta(\bar{y})a^{(\text{r})}_{\tilde{\beta}}(\bar{y}) + \calC_{\beta\tilde{\beta}}
+\calR(v_{\bar{y}})_{\beta\tilde{\beta}},
\end{equation}
where $\calC$ is independent of $v_{\bar{y}}$, and
\begin{align}
\|\calC\| &\le \gamma\veps_k,
\label{chi} 
\\
\|\calR(v_{\bar{y}})\| &\le \gamma^{2.5}\lvert\mathbf{a}^{(\text{r})}(\bar{y})\rvert^2.
\label{chiprime}
\end{align}
Furthermore, the following bound holds without limitation on $\lvert\tilde{\lambda}-E_{k+1}\rvert$:
\begin{equation}\label{3.29a}
\|f_{\tilde{\lambda}}^{(k)\mathrm{c}} - f_{E_{k+1}}^{(k)}\| \le
 \gamma^2
\lvert\mathbf{a}^{(\mathrm{r})}(\bar{y})\rvert^2\cdot 5N + 2\gamma\lvert\tilde{\lambda} - E_{k+1}\rvert.
\end{equation}
\end{lemma}
\textit{Proof.} We write
\begin{align}\label{mu}
f_\lambda^{(k-1)} - f_{E_{k+1}}^{(k)} &= \big(f_\lambda^{(k-1)} - f_{\lambda}^{(k)}\big)+\big(f_\lambda^{(k)} - f_{E_{k+1}}^{(k)}\big) \nonumber\\
&=(q-\tilde{q}) + \tilde{r}(\tilde{t}-\lambda)^{-1}\tilde{s} + \big(f_\lambda^{(k)} - f_{E_{k+1}}^{(k)}\big).
\end{align}
Consider the first term in (\ref{mu}). By (\ref{(a)})-(\ref{starB}), we have that
\begin{equation}\label{muprime}
(q-\tilde{q})_{\beta\tilde{\beta}} = \frac{\gamma^2}{v_{\bar{y}}+2d\gamma-\lambda}
a^{(\text{r})}_\beta(\bar{y})a^{(\text{r})}_{\tilde{\beta}}(\bar{y}) + \calC^{(1)}_{\beta\tilde{\beta}}
+\calR^{(1)}(v_{\bar{y}})_{\beta\tilde{\beta}},
\end{equation}
with
\begin{align} \label{nu}
 \calC^{(1)}_{\beta\tilde{\beta}}&= \sum_{xy} a^{(\text{r})}_\beta(x)K_0(x,y)a^{(\text{r})}_{\tilde{\beta}}(y) 
 =
 \langle \vphi_\beta, \tilde{G}^{\prime(k-1)\text{tr}}_\lambda \tfrac{\Gamma}{\gamma}K_0\tfrac{\Gamma}{\gamma}\tilde{G}_\lambda^{\prime(k-1)}\vphi_{\tilde{\beta}}\rangle,
 \\
 \calR^{(1)}_{\beta\tilde{\beta}}(v_{\bar{y}})&= \sum_{xy} a^{(\text{r})}_\beta(x)K_2(x,y)a^{(\text{r})}_{\tilde{\beta}}(y).
 \label{omega}
\end{align}

By Theorem \ref{thm:G} and the minimum distance $2L_{k-1}$ from $B_{k-1}$ to the boundary of $\bar{B}_{k-1}$, we have that
$\lvert a_\beta(y) \rvert  = \lvert\langle \chi_y,\tilde{G}^{\prime(k-1)}_\lambda \vphi_\beta\rangle\rvert \le 2dn\gamma^{1.7L_{k-1}}$.
For a crude estimate, we may take the supremum over $xy$ in (\ref{nu}) by adding a factor $(2d\lvert \bar{B}_{k-1} \rvert )^2$, and then using (\ref{(c0)}), we obtain that $\lvert \calC^{(1)}_{\beta\tilde{\beta}} \rvert  \le (2d\lvert \bar{B}_{k-1} \rvert )^2(2dn\gamma^{1.7L_{k-1}})^2$.
We may estimate $\|\calC^{(1)}\|$ by $\hat{n}\,\text{max}_{\beta\tilde{\beta}}\lvert \calC^{(1)}_{\beta\tilde{\beta}} \rvert $, and then since 
$\hat{n}\le n\le \lvert \bar{B}_{k-1} \rvert $, we obtain a bound 
\begin{equation}\label{omega1}
\|\calC^{(1)}\| \le  \lvert \bar{B}_{k-1} \rvert ^5(2d)^4\gamma^{3.4L_{k-1}}  \le  (5.1L_{k-1}+1)^{5d}(2d)^4\gamma^{3.4L_{k-1}}\le \gamma\veps_k/2; 
\end{equation}
recall that $\veps_k \equiv \gamma^{3.2L_{k-1}}$, $L_0$ is large, and $\gamma$ is small.
To estimate $\calR^{(1)}(v_{\bar{y}})$, recall that $\bar{y}$ is defined as the site that maximizes $\lvert \mathbf{a}^{(\text{r})}(y) \rvert $, so
$\lvert \mathbf{a}^{(\text{r})}(y) \rvert \le\lvert \mathbf{a}^{(\text{r})}(\bar{y}) \rvert $
for all $y$ adjacent to $\bar{B}_{k-1}$. Using (\ref{(c')}), we obtain
\begin{equation}\label{delta}
\|\calR^{(1)}(v_{\bar{y}})\| \le \sum_{xy}
\|\mathbf{a}^{(\text{r})}(x)\mathbf{a}^{(\text{r})}(y)^{\text{tr}}\|
\lvert K_2(x,y)\rvert
\le \tfrac{1}{4}\gamma^{2.5}\lvert \mathbf{a}^{(\text{r})}(\bar{y}) \rvert ^2.
\end{equation}
We have used the fact that the norm of an outer product matrix $\mathbf{uw}^{\text{tr}}$ is bounded by $\lvert \mathbf{u}\rvert\lvert\mathbf{w}\rvert$. The key point here is that all graphs that contribute to $\calR^{(1)}(v_{\bar{y}})$ are sub-leading (at least three steps) and can be estimated in terms of $\lvert \mathbf{a}^{(\text{r})}(\bar{y}) \rvert $.

The second term of (\ref{mu}) is another remainder term
$\calR^{(2)}(v_{\bar{y}}) \equiv \tilde{r}(\tilde{t}-\lambda)^{-1}\tilde{s}$, 
which can be written as
\begin{equation}\label{epsilon}
\calR^{(2)}_{\beta\tilde{\beta}}(v_{\bar{y}}) = \sum_{\beta',\tilde{\beta}' \in [\hat{n}+1,n]}
\sum_{xx'y'y}
a^{(\text{r})}_\beta(x)K(x,x')a^{(\text{r})}_{\beta'}(x')
(\tilde{t}-\lambda)^{-1}_{\beta'\tilde{\beta}'}a^{(\text{r})}_{\tilde{\beta}'}(y')K(y',y)a^{(\text{r})}_{\tilde{\beta}}(y). 
\end{equation}
The eigenvalues of $t$ are outside of $I_{\veps_k/2}(E_{k})$, by construction, and $\lambda \in I_{2\veps_k/5}(E_{k})$. Thus the gap from $\lambda$ is at least $\veps_k/10$. Furthermore, 
$\|\tilde{t}-t\|$ can be estimated as in (\ref{omega1}) by $\gamma\veps_{k}$. Hence $\|(\tilde{t}-\lambda)^{-1}\|\le 20/\veps_k$. Then using the abovementioned estimate on $\lvert a_\beta(y) \rvert $, we can bound
\begin{equation}\label{epsilonprime}
\sum_{\beta',\tilde{\beta}' \in [\hat{n}+1,n]}
\lvert a^{(\text{r})}_{\beta'}(x')
(\tilde{t}-\lambda)^{-1}_{\beta'\tilde{\beta}'}a^{(\text{r})}_{\tilde{\beta}'}(y')\rvert \le (n-\hat{n})^2(2dn)^2\gamma^{3.4L_{k-1}}(20/\veps_k)
\le \gamma^{.1L_{k-1}}.
\end{equation}
The kernels $K$ control the sums over $x',y'$; the sums over $x,y$ lead to factors of $2d\lvert \bar{B}_{k-1} \rvert $, and $\lvert a^{(\text{r})}_\beta(x)a^{(\text{r})}_{\tilde{\beta}}(y) \rvert  \le \lvert \mathbf{a}^{(\text{r})}(\bar{y}) \rvert ^2$; thus
\begin{equation}\label{epsilonprimeprime}
\calR^{(2)}_{\beta\tilde{\beta}}(v_{\bar{y}}) \le (2d)^2(5.1L_{k-1}+1)^{2d} \gamma^{.1L_{k-1}} \lvert \mathbf{a}^{(\text{r})}(\bar{y}) \rvert ^2.
\end{equation}
Another factor of $\hat{n} \le n$ converts this into a norm bound, which leads to an estimate $\|\calR^{(2)}(v_{\bar{y}})\| \le \tfrac{1}{4}\gamma^{2.5}\lvert \mathbf{a}^{(\text{r})}(\bar{y}) \rvert ^2$.

Now consider the third term of (\ref{mu}), and write it as
\begin{equation}\label{star3a}
f_\lambda^{(k)}-f_{E_{k+1}}^{(k)}
= (f_\lambda^{(k)}-f_{\tilde{\lambda}}^{(k)}) + (f_{\tilde{\lambda}}^{(k)}-f_{E_{k+1}}^{(k)}) = \calC^{(2)} + \calR^{(3)}(v_{\bar{y}})+ \calR^{(4)}(v_{\bar{y}}),
\end{equation}
where
\begin{align}
\calC^{(2)} &= f_\lambda^{(k-1)}-f_{\tilde{\lambda}}^{(k-1)} \label{star3b},
\\
\calR^{(3)}(v_{\bar{y}}) &=   (f_\lambda^{(k)}-f_{\tilde{\lambda}}^{(k)}) -(f_\lambda^{(k-1)}-f_{\tilde{\lambda}}^{(k-1)})
=  (f_\lambda^{(k)}-f_{\lambda}^{(k-1)}) -(f_{\tilde{\lambda}}^{(k)}-f_{\tilde{\lambda}}^{(k-1)}) \label{star3c},
\\
\calR^{(4)}(v_{\bar{y}}) &= (f_{\tilde{\lambda}}^{(k)}-f_{E_{k+1}}^{(k)}) .\label{star3d}
\end{align}

To estimate $\calC^{(2)}$, note that by Theorem \ref{thm:3} 
\begin{equation}\label{star5}
\|\tilde{F}_\lambda^{(k-1)}(B_{k-1}) - \tilde{F}_\mu^{(k-1)}(B_{k-1})\| \le \gamma\lvert\lambda -\mu\rvert ,
\end{equation}
for $\lambda$, $\mu$ in $I_{\veps_k/2}(E_k)$.
Note that $\lambda$ is in $I_{2\veps_k/5}(E_k)$.
We claim that if $\mu \in I_{2\veps_k/5}(E_k)$, then
\begin{equation}\label{star5a}
\|f_\lambda^{(k-1)}-f_{\mu}^{(k-1)}\| \le 2\gamma\lvert \lambda-\mu \rvert.
\end{equation}
Note that $q$, $r$, $s$, $t$ are sub-matrices of $\tilde{F}_\mu^{(k-1}(B_{k-1})$, so their change when shifting $\lambda \rightarrow \mu$ are bounded in norm by $\gamma\lvert\lambda-\mu\rvert$ as well. Since $r = s = 0$ at $\lambda$, they are bounded in norm at $\mu$ by $\gamma\lvert\lambda-\mu\rvert \le \gamma\veps_k$. (Here we use the fact that both $\lambda$ and $\mu$ are in $I_{2\veps_k/5}$.) 
Again, because $r = s = 0$ at $\lambda$, we have that 
\begin{equation}\label{star5b}
f_\lambda^{(k-1)}-f_{\mu}^{(k-1)} = (q_\lambda - q_\mu) + r_\mu(t_\mu - \mu)^{-1}s_\mu,
\end{equation}
where we have introduced subscripts for clarity.
As explained earlier,  $\text{dist}(\text{spec}\,t_\lambda, E_k)\ge \veps_k/2$, by construction. 
Therefore $\text{dist}(\text{spec}\,t_\mu, E_k)\ge 9\veps_k/20$, and hence 
$\|(t_\mu-\mu)^{-1}\| \le 20/\veps_k$.
Combining bounds, we have that $\|r_\mu(t_\mu - \mu)^{-1}s_\mu\| \le 20\gamma^2\lvert\lambda-\mu\rvert$, and (\ref{star5a}) follows immediately. In particular, if we insert $\tilde{\lambda}$ for $\mu$ and apply (\ref{star1A}), we obtain that $\|\calC^{(2)}\| \le 2\gamma^2\veps_k \le \gamma\veps_k/2$.

To estimate $\calR^{(3)}(v_{\bar{y}})$, observe that $\Delta F_\lambda \equiv \tilde{F}_\lambda^{(k-1)}(B_{k-1})-\tilde{F}_\lambda^{(k)}(B_{k})$ involves a sum of graphs extending from $B_{k-1}$ to $\bar{B}_k \setminus \bar{B}_{k-1}$ and back.
Hence if we take the difference $\Delta F_\lambda - \Delta F_{\tilde{\lambda}}$, we obtain a sum of differenced graphs, each with length $\ge 4L_{k-1}$. The sum can be estimated as in the proof of Theorem \ref{thm:3}.
Allowing for a small decrease in decay rate to handle the factors $\veps_j^{-1}$ and using (\ref{star1A}), we obtain
\begin{align}
\|\Delta F_\lambda - \Delta F_{\tilde{\lambda}}\| &\le \gamma^{3.3L_{k-1}}\lvert \lambda-\tilde{\lambda} \rvert  \le \gamma^{3.3L_{k-1}}\gamma\veps_k. \nonumber \\
&\le \gamma^{(3.3+3.2)L_{k-1}} = \gamma^{3.25L_k} \ll \veps_{k+1}
\ll \tfrac{1}{4}\gamma^{2.5} \lvert \mathbf{a}^{(\text{r})}(\bar{y}) \rvert ^2.
\label{(X)}
\end{align}
Here we recall that $\veps_{k+1} = \gamma^{3.2L_k}$ and use
$\lvert \mathbf{a}^{(\text{r})}(\bar{y}) \rvert ^2 \ge (\gamma^{3.1 L_{k-1}})^2 = \gamma^{3.1 L_{k}}$.
If we extend the double-difference operation to $q(t-\lambda)^{-1}r$,
we obtain a sum of terms with two differences. As in (\ref{(X)}), 
the difference $\lambda \rightarrow \tilde{\lambda}$ leads to a factor $\gamma\veps_k/4$, and the difference $\bar{B}_{k-1} \rightarrow \bar{B}_k$ leads to a factor $\gamma^{3.3L_{k-1}}$. 
As in the arguments above, neither correction affects the bound on $\|(t-\lambda)^{-1}\|$ by more than a factor of 2, since $\lambda$, $\tilde{\lambda}$ are in $I_{2\veps_k/5}(E_k)$. Thus the bound (\ref{(X)}) extends to $\calR^{(3)}(v_{\bar{y}})$. 

Finally, we consider $\calR^{(4)}(v_{\bar{y}})$. We claim that
\begin{equation}\label{star4}
\|f_{\tilde{\lambda}}^{(k)}-f_{E_{k+1}}^{(k)}\| \le 2\gamma\lvert \tilde{\lambda}-E_{k+1} \rvert.
\end{equation}
As in the proof of (\ref{star5b}), we have that $\tilde{\lambda}$, $E_{k+1}$ are in $I_{2\veps_k/5}(E_k)$ and Theorem \ref{thm:3} implies that
$\|\tilde{F}_{\tilde{\lambda}}^{(k)}(B_{k}) - \tilde{F}_{E_{k+1}}^{(k)}(B_{k})\| \le \gamma\lvert\tilde{\lambda} -E_{k+1}\rvert$. Then if we apply the difference $\tilde{\lambda}\rightarrow E_{k+1}$ to (\ref{starB}), there are five terms, corresponding to differencing of
$\tilde{q}$, $\tilde{r}$, $\tilde{s}$, $\tilde{t}$, or of $\lambda$ itself.
Note that $\|\tilde{r}\|$ is bounded by $2\gamma\veps_k$, based on (1) the shift $r \rightarrow \tilde{r}$ involves graphs exiting $\bar{B}_{k-1}$, which are $\ll \veps_k$ and (2) the shift in spectral parameter from $\lambda$ to $\tilde{\lambda}$ or $E_{k+1}$ makes a change $\le \gamma\veps_k$ (recall that $r_\lambda = 0$). The same bound holds for $\|\tilde{s}\|$. We have as before bounds $\|(\tilde{t}-\tilde{\lambda})^{-1}\| \le 20/\veps_k$, $\|(\tilde{t}-E_{k+1})^{-1}\| \le 20/\veps_k$, since the corresponding shifts for $t$ preserve the gap to the spectrum in  $I_{2\veps_k/5}(E_k)$. These large factors can always be matched up against $\tilde{r}$ or $\tilde{s}$, the result being small.
If the difference applies to $\tilde{r}$, for example, we have a bound 
$\gamma\lvert\tilde{\lambda} -E_{k+1}\rvert \cdot 40\gamma$. This completes the proof of (\ref{star4}).
Now, in Lemma \ref{prop:3.2} it is assumed that $\lvert\tilde{\lambda}-E_{k+1}\rvert \le \gamma^2
\lvert\mathbf{a}^{(\text{r})}(\bar{y})\rvert^2\cdot 10N$. Hence (\ref{star4}) implies that $\|\calR^{(4)}(v_{\bar{y}})\| \le 2\gamma^3
\lvert\mathbf{a}^{(\text{r})}(\bar{y})\rvert^2\cdot 10N \le \tfrac{1}{4}\gamma^{2.5}\lvert\mathbf{a}^{(\text{r})}(\bar{y})\rvert^2$.

Putting
$\calC = \calC^{(1)} + \calC^{(2)}$, $\calR(v_{\bar{y}}) = \calR^{(1)}(v_{\bar{y}}) + \calR^{(2)}(v_{\bar{y}})+ \calR^{(3)}(v_{\bar{y}})+\calR^{(4)}(v_{\bar{y}})$, and combining the bounds proven above, we obtain (\ref{chi}), (\ref{chiprime}). 

It remains for us to prove (\ref{3.29a}). This statement is meant to cover the situation where $E_{k+1}$ is far from $\lambda$, $\tilde{\lambda}$.
The arguments are variations on ones given above. 
Write
\begin{align}\label{new3.53}
f_{\tilde{\lambda}}^{(k)\text{c}} - f_{E_{k+1}}^{(k)}&= 
\big(f_\lambda^{(k)\text{c}} - f_{\lambda}^{(k)}\big)
+
\big[\big(f_{\tilde{\lambda}}^{(k)\text{c}} - f_{\tilde{\lambda}}^{(k)}\big) 
-\big(f_{{\lambda}}^{(k)\text{c}} - f_{{\lambda}}^{(k)}\big)\big]
 + \big(f_{\tilde{\lambda}}^{(k)} - f_{E_{k+1}}^{(k)}\big)
 \nonumber
 \\ 
 &=\big(f_\lambda^{(k)\text{c}} - f_{\lambda}^{(k)}\big)
 + \hat{\calR}^{(3)} + \hat{\calR}^{(4)} .
\end{align}
The first term involves the expansion of the region from $\bar{B}_k \setminus \{\bar{y}\}$ to $\bar{B}_k$, so it can be handled in a similar manner as (\ref{muprime})-(\ref{epsilonprimeprime}), which involve the expansion from $\bar{B}_{k-1}$ to $\bar{B}_k$.
We write
\begin{equation}\label{new3.54}
f_{\lambda}^{(k)\text{c}} - f_\lambda^{(k)}= 
\big(q^{\text{c}} - \tilde{q}\big) + \big[ \tilde{r}(\tilde{t}-\lambda)^{-1}\tilde{s} - r^{\text{c}}(t^{\text{c}}-\lambda)^{-1}s^{\text{c}}\big] = \hat{\calR}^{(1)} +  \hat{\calR}^{(2)},
\end{equation}
where 
$\tilde{F}^{(k)\text{const}}_{\lambda}= \left( \begin{smallmatrix}q^{\text{c}}&r^{\text{c}} \\s^{\text{c}}&t^{\text{c}} \end{smallmatrix}\right)$.
Observe that $\hat{\calR}^{(1)}$ is given by the sum of graphs that touch $\bar{y}$, so 
\begin{equation}\label{new3.55}
\hat{\calR}^{(1)}_{\beta\tilde{\beta}}= \sum_{xy} a^{(\text{r})}_\beta(x)\big(K_1(x,y)+K_2(x,y)\big)a^{(\text{r})}_{\tilde{\beta}}(y).
\end{equation}
Note the absence of $K_0$ means there is no $\calC$ term, which is an important aspect of (\ref{3.29a}).
The $K_2$ term is $\calR^{(1)}(v_{\bar{y}})$, and so we can use (\ref{delta}) to bound it by $\tfrac{1}{4}\gamma^{2.5}\lvert \mathbf{a}^{(\text{r})}(\bar{y}) \rvert ^2$.
The $K_1$ term is the first term of (\ref{muprime}) and so is bounded by 
$\gamma^2 \lvert \mathbf{a}^{(\text{r})}(\bar{y}) \rvert ^2 \cdot 5(N-1)$.
Here we use the following two facts: (1) Since $E_{j+1} \in I_{\veps_j/3}(E_j)$, the intervals $I_{2\veps_j/5}$ are nested, and so $\lambda \in I_{2\veps_1/5}$. (2) 
As $v_{\bar{y}}$ is nonresonant in step 1, it satisfies
$\lvert v_{\bar{y}} + 2d\gamma - E_1\rvert \ge \veps_1$, and so  $\lvert v_{\bar{y}} + 2d\gamma - \lambda\rvert \ge 3\veps_1/5 = \tfrac{1}{5(N-1)}$. Combining the $K_1$ and $K_2$ bounds, we obtain that $\|\hat{\calR}^{(1)}\| \le \gamma^2 \lvert \mathbf{a}^{(\text{r})}(\bar{y}) \rvert ^2\cdot (5N-5+ \gamma^{1/2}/4)$.

We may handle $\hat{\calR}^{(2)}$ as in (\ref{epsilon})-(\ref{epsilonprimeprime}). In fact $\tilde{r}(\tilde{t}-\lambda)^{-1}\tilde{s}$ is $\calR^{(2)}(v_{\bar{y}})$, and $r^{\text{c}}(t^{\text{c}}-\lambda)^{-1}s^{\text{c}}$ can be bounded in exactly the same way. Thus 
$\|\hat{\calR}^{(2)}\| \le \tfrac{1}{2}\gamma^{2.5} \lvert \mathbf{a}^{(\text{r})}(\bar{y}) \rvert ^2$.
For $\hat{\calR}^{(3)}$, we have a double difference, and it can be estimated exactly as we did for $\calR^{(3)}(v_{\bar{y}})$, in the paragraph of (\ref{(X)}). Finally, $\hat{\calR}^{(4)}$ is bounded already in (\ref{star4}). Combining these estimates, we obtain (\ref{3.29a}).

\qed

\textit{Proof of Proposition \ref{prop:3.3}.} 
First, we establish statements analogous to (\ref{equals}),(\ref{less}) for the second Schur complement matrices. Let 
\begin{equation}
\hat{n}_f = \text{ the number of eigenvalues of } f^{(k)}_{E_{k+1}} \text{ in } I_{2\veps_{k+1}}(E_{k+1}).
\label{nf}
\end{equation}
We work on demonstrating that $\hat{n}_f \le \hat{n}$ and that $\hat{n}_f < \hat{n}$ for all but one value of $v_{\bar{y}}$. Note that
 $\hat{n}_f \le \hat{n}$ is true by construction, as $\hat{n}$ is the dimension of the matrix $f^{(k)}_{E_{k+1}}$.

With $\lambda \in [0,1+4d\gamma]$, take any $v_1$, $v_2$ in the set of allowed potential values $\{0,\tfrac{1}{N-1},\\ \tfrac{2}{N-1}, \ldots,1\}$.
Then $\lvert v_i+2d\gamma-\lambda\rvert \le 1+2d\gamma$ for $i=1,2$, and so
\begin{align}
\left| \frac{1}{v_1+2d\gamma-\lambda} -  \frac{1}{v_2+2d\gamma-\lambda}\right| &= \frac{\lvert v_2-v_1\rvert}{\lvert v_1+2d\gamma-\lambda\rvert \lvert v_2+2d\gamma-\lambda\rvert}
\nonumber \\
&\ge\frac{1}{(N-1)(1+2d\gamma)^2} \ge \frac{1}{N},
\label{kappa}
\end{align}
as we are taking $\gamma$ small, depending on $N$. Thus we see that there are $N$ distinct values of $(v+2d\gamma-\lambda)^{-1}$ as $v$ varies over $\{0,\tfrac{1}{N-1}, \tfrac{2}{N-1}, \ldots,1\}$, and the minimum gap between these values is $\tfrac{1}{N}$.

Let us work first assuming that $\lvert\tilde{\lambda}-E_{k+1}\rvert \le \gamma^2
\lvert\mathbf{a}^{(\text{r})}(\bar{y})\rvert^2\cdot 10N$, so that (\ref{psi})-(\ref{chiprime}) are true (the alternative will be handled below).

Consider what happens when $\hat{n} = 1$. 
We may remove case 2 of Proposition \ref{prop:3.3} from consideration, because then $\hat{n}_{k-1}(B_{k-1})>1$, and so if $\hat{n}=1$ then $\hat{n}_f \le \hat{n} <\hat{n}_{k-1}(B_{k-1})$. (We show below that $\hat{n}_k(B_k) \le \hat{n}_f$.)
So consider case 1 of Proposition \ref{prop:3.3} with $E_{k+1}$ fixed.
Then the matrix
$a^{(\text{r})}_\beta(\bar{y})a^{(\text{r})}_{\tilde{\beta}}(\bar{y})$
reduces to a number 
$\lvert a^{(\text{r})}_1(\bar{y})\rvert^2 \ge \gamma^{3.1L_{k}}$.
Likewise $f^{(k)}_{E_{k+1}}$ and $f^{(k-1)}_{\lambda}$ are numbers,
with the latter independent of $v_{\bar{y}}$.
It is evident from Lemma \ref{prop:3.2} and (\ref{kappa}) that the set of values that $f^{(k)}_{E_{k+1}}$ takes as $v_{\bar{y}}$ varies is spaced apart by at least
$(\gamma^2/N)\lvert a^{(\text{r})}_1(\bar{y})\rvert^2 \gg \veps_{k+1}$. Hence there is at most one value of $v_{\bar{y}}$ such that
 $f^{(k)}_{E_{k+1}}$ lies in $I_{2\veps_{k+1}}(E_{k+1})$,
 and consequently $\hat{n}_f <\hat{n}$ for all but one value of $v_{\bar{y}}$.

For $\hat{n}>1$, we use a basic fact about spreads of Hermitian matrices. The spread of a Hermitian matrix is defined as the difference between its largest and smallest eigenvalues. By Weyl's inequality, the spread of 
$M_1+M_2$ is at least $\lvert \text{spread}(M_1)-\text{spread}(M_2)\rvert$. 
Observe that 
$M_0\equiv\mathbf{a}^{(\text{r})}(\bar{y})\mathbf{a}^{(\text{r})}(\bar{y})^{\text{tr}}$ is a rank-one 
matrix, so its spread is equal to its nonzero eigenvalue 
$\lvert \mathbf{a}^{(\text{r})}(\bar{y}) \rvert ^2 \ge \gamma^{3.1L_k} \gg \veps_{k+1}$.

From Lemma \ref{prop:3.2} we have that
\begin{equation}\label{psiprime}
f_{E_{k+1}}^{(k)}= f_\lambda^{(k-1)}-\frac{\gamma^2}{v_{\bar{y}}+2d\gamma-\lambda}
M_0- \calC
-\calR(v_{\bar{y}}),
\end{equation}
with $\|\calC\| \le \gamma\veps_k$, $\|\calR (v_{\bar{y}})\| \le \gamma^{2.5}\lvert \mathbf{a}^{(\text{r})}(\bar{y}) \rvert ^2$.
Let $v_{\bar{y},1}$ and $v_{\bar{y},2}$ be two different values of $v_{\bar{y}}$, and put
\begin{align}\label{M1M2}
M_1 &\equiv f_\lambda^{(k-1)}-\frac{\gamma^2}{v_{\bar{y},1}+2d\gamma-\lambda}
M_0- \calC,\nonumber\\
M_2 &\equiv \bigg(\frac{\gamma^2}{v_{\bar{y},1}+2d\gamma-\lambda}-\frac{\gamma^2}{v_{\bar{y},2}+2d\gamma-\lambda}\bigg)M_0. 
\end{align}
Then from (\ref{kappa}) we have that $\text{spread}(M_2) \ge \tfrac{\gamma^2}{N}\lvert \mathbf{a}^{(\text{r})}(\bar{y}) \rvert ^2$.
Suppose that for $v_{\bar{y}} = v_{\bar{y},1}$ the spread of $M_1$ is smaller than 
$\tfrac{\gamma^2}{3N}\lvert \mathbf{a}^{(\text{r})}(\bar{y}) \rvert ^2$.
Then for any other value $v_{\bar{y}} = v_{\bar{y},2}$, 
$\text{spread}(M_1+M_2)\ge \tfrac{2\gamma^2}{3N}\lvert \mathbf{a}^{(\text{r})}(\bar{y}) \rvert ^2$.
Adding in a correction for the $v_{\bar{y}}$-dependent term $\calR(v_{\bar{y}})$, we have shown that
%\begin{equation}
%\left|\text{spread}(M_2) - \text{spread}\bigg(\frac{\gamma^2}{v_{\bar{y}}+2d\gamma-\lambda}M_1\bigg)\right| \le \frac{\gamma^2}{3N}\lvert \mathbf{a}^{(\text{r})}(\bar{y}) \rvert ^2,
%\label{spread}
%\end{equation}
%for at most one value of $v_{\bar{y}}$. 
%(The values of the second spread are sufficiently spaced out, so cancellation to this degree of accuracy can happen at most once.)
%As our bound on $\|\calR(v_{\bar{y}})\|$ is much smaller than this, we see that
\begin{equation}
\text{spread}\big(f_{E_{k+1}}\big) \ge
\Big(\frac{\gamma^2}{3N} - \gamma^{2.5}\Big)\lvert \mathbf{a}^{(\text{r})}(\bar{y}) \rvert ^2 \ge \frac{\gamma^2}{4N}\gamma^{3.1L_k} \gg \veps_{k+1} = \gamma^{3.2L_k},
\label{spreadprime}
\end{equation}
for all but one value of $v_{\bar{y}}$. Hence for all but one value of $v_{\bar{y}}$, at least one eigenvalue of $f_{E_{k+1}}$ must fall outside of $I_{2\veps_{k+1}}(E_{k+1})$. Thus we have demonstrated that
$\hat{n}_f < \hat{n}$ for all but one value of $v_{\bar{y}}$.

Now consider the alternative situation, with $\lvert\tilde{\lambda}-E_{k+1}\rvert > \gamma^2
\lvert\mathbf{a}^{(\text{r})}(\bar{y})\rvert^2\cdot 10N$. This is actually the simpler case, because the separation between $\tilde{\lambda}$ and $E_{k+1}$ pulls spectrum away from $E_{k+1}$. Inserting the lower bound on $\lvert\tilde{\lambda}-E_{k+1}\rvert$ into (\ref{3.29a}), we have that
\begin{equation}\label{*}
\|f_{\tilde{\lambda}}^{(k)\text{c}} - f_{E_{k+1}}^{(k)}\| \le
(\tfrac{5}{10} + 2\gamma)\lvert \tilde{\lambda}-E_{k+1} \rvert \le
\tfrac{2}{3}\lvert \tilde{\lambda}-E_{k+1}\rvert.
\end{equation}
Recall that $\tilde{\lambda} \in \text{spec}\,f_{\tilde{\lambda}}^{(k)\text{c}}$. By (\ref{*}), the shift of the spectrum between $f_{\tilde{\lambda}}^{(k)\text{c}}$ and $f_{E_{k+1}}^{(k)}$ is no more than $\tfrac{2}{3}$ of the distance from $\tilde{\lambda}$ to $E_{k+1}$. 
As $2\veps_{k+1} \ll \gamma^2\lvert\mathbf{a}^{(\text{r})}(\bar{y})\rvert^2 < \tfrac{1}{3}\lvert \tilde{\lambda}-E_{k+1} \rvert$, no more than $\hat{n}-1$ eigenvalues of $f_{E_{k+1}}^{(k)}$ may lie in
$I_{2\veps_{k+1}}(E_{k+1})$. Consequently, $\hat{n}_f < \hat{n}$. 

To sum up, we have obtained that  $\hat{n}_f < \hat{n}$ for all but one value of $v_{\bar{y}}$ in the case where $E_{k+1}$ is close to $\tilde{\lambda}$, and all values of $v_{\bar{y}}$ when it is far from $\tilde{\lambda}$.
%added paragraph above this and modified paragraph below on 3-20-21

Finally, we work on the outer inequalities of the chain $\hat{n}_k(B_k)\le\hat{n}_f < \hat{n} \le\hat{n}_{k-1}(B_{k-1})$. The rightmost inequality is
(\ref{(P8)}). Hence the proof will be complete once we establish that
$\hat{n}_k(B_k) \le \hat{n}_f$.
We may compare the spectrum of $\tilde{F}^{(k)}_{E_{k+1}}$ in
$I_{\veps_{k+1}}(E_{k+1})$ with that of $f^{(k)}_{E_{k+1}}$ in $I_{2\veps_{k+1}}(E_{k+1})$ as in the proof of (\ref{evmove}) in Lemma \ref{fundamental}.
%We may take $\tilde{\veps} = 2\veps_{k+1}$.
Recall the block decomposition $\tilde{F}^{(k)}_{E_{k+1}}(B_{k})= \left( \begin{smallmatrix}\tilde{q}&\tilde{r} \\\tilde{s}&\tilde{t} \end{smallmatrix}\right)$
We have already established that 
$\|(\tilde{t}-E_{k+1})^{-1}\| \le 20/\veps_k$, $\|\tilde{r}\| \le \gamma\veps_k$, 
$\|\tilde{s}\| \le \gamma\veps_k$ (see the discussion following \ref{star5a}). 
These bounds allow us to estimate the product of operators in (\ref{(10')}), and we learn that
for each eigenvalue $\lambda_i$ of
$\tilde{F}^{(k)}_{E_{k+1}}$ in $I_{\veps_{k+1}}(E_{k+1})$
there corresponds an eigenvalue $\tilde{\lambda}_i$ of 
$f^{(k)}_{E_{k+1}}$ such that
\begin{equation}
\label{last}
\lvert \lambda_i - \tilde{\lambda}_i\rvert \le 2\Big(\frac{\gamma\veps_k}{\veps_k/20}\Big)^2\lvert \lambda_i - E_{k+1} \rvert  \le \gamma\lvert \lambda_i - E_{k+1} \rvert \le \gamma\veps_{k+1}.
\end{equation}
Thus the doubling of the spectral window width from $\veps_{k+1}$ to $2\veps_{k+1}$ is sufficient to capture all of the $\tilde{\lambda}_i$, and we obtain that 
$\hat{n}_k(B_k) \le \hat{n}_f$. \qed

\subsection{Percolation Estimates}\label{ssec:3.2}
Here we set up the percolation estimates for the resonant blocks at each step of the procedure. First we consider the fixed-energy procedure, in which the energies $E_k$ are independent of the potentials. Then we discuss how this needs to be modified for the energy-following procedure. Initially, we have a simple site percolation problem. At the start of section \ref{ssec:first}, we gave a bound of $\veps = \tfrac{1}{N-1}$ on the probability that a given site is resonant to $E$, that is, $\lvert v_x + 2d\gamma-E \rvert  < \veps_1$. 
Then we formed connected components by linking sites $x, y$ such that $\lvert x-y \rvert  \le L_1^\alpha$. In subsequent steps, the criterion for a component or block to be resonant is more involved. 
Only the isolated blocks are candidates for removal from $R^{(k-1)}$ when forming $R^{(k)}$. An isolated block in step $k$ must have
$\text{dist}\big(\text{spec}\,\tilde{F}_{E_k}^{(k-1)}(B_{k-1}),E_k\big) \le \veps_k$
 if it is to remain resonant and become part of $R^{(k)}$.

As the step index $k$ increases, there are several processes in play. According to Definition \ref{def:2.1}, a block $B_{k-1}$ that has diameter greater than $L_{k-1}$ is not isolated. 
However, as $k$ increases, this criterion will eventually be satisfied. 
This may be delayed, however, if the block joins up with other components in later steps due to the lengthening of the distance criterion for connectedness.
Once the block $B_{k-1}$ becomes isolated, it may be dropped from $R^{(k)}$ if it is not resonant in step $k$. If the block remains isolated in step $k+1$ without being joined up with other blocks, one of two things will happen: (i) $\hat{n}_k(B_{k})  < \hat{n}_{k-1}(B_{k-1})$, with probability $\ge 1-\tfrac{1}{N-1}$ or (ii) $\hat{n}_k(B_{k})  = \hat{n}_{k-1}(B_{k-1})$ with probability $\le \tfrac{1}{N-1}$. Eventually, if no further joining takes place,  
and if $\bar{B_k}$ does not exhaust $\Lambda$, $\hat{n}_k(B_{k})$ will reach 0, and the block will be dropped from the resonant set.

These processes and probabilities enter into a multiscale percolation problem. We want to control, for example, the probability that a site $x$ is in $R^{(k)}$. This can be written as the sum over $B_k$ containing $x$ of the probability that $B_k$ is a component of $R^{(k)}$, and this probability has the following bound:
\begin{equation}\label{probability}
\veps^{n(B_k)}\mathbb{E}_k\,\mathbf{1}_{\hat{n}_k(B_k)>0}.
\end{equation}
Here we include explicitly the factor $\veps$ for each site of $B_k$, since $B_k \subset R^{(1)}$. (We have, in fact, site percolation with occupation probability $1/N < \veps$, because for each $x \in R^{(1)}$, there is only one choice of $v_x$ that is resonant.)  Note that $B_k$ is structured as a collection of subcomponents $\{B_{j,\beta}\}_{\beta = 1,\ldots,m}$ at each level $j = 1,\ldots,k-1$, and these subcomponents satisfy connectivity and resonance conditions. We condition on the rest of $R^{(1)},\ldots,R^{(k-1)}$, and this determines all of the $\bar{B}_{j,\beta}$ for $j \le k$. 
The expectation $\mathbb{E}_k$ is with respect to the potentials in
$\bar{B}^{(k)}\setminus R^{(1)}$;
these range freely over the $N-1$ nonresonant possibilities allowed at sites $x \notin R^{(1)}$.
Our focus, however, is on the expectations over the potentials $v_{\bar{y}}$; 
these lead to smallness since only one value of $v_{\bar{y}}$ permits a nondecreasing $\hat{n}_j(B_j)$.
(Recall the construction in Section \ref{ssec:3.1}, in which
subcomponents $B_{j-1, \beta}$ are assigned sites $v_{\bar{y}}$ adjacent to $\bar{B}_{j-1, \beta}$, the choice depending only on the potentials in $\bar{B}_{j-1, \beta}$. This allows us to take the expectation over the $v_{\bar{y}}$'s in sequence, \textit{c.f.} (\ref{mathbfE2}) below.)
All bounds will be uniform in the other potentials in $R^{(1)\text{c}}$, so they may be treated as fixed.  
The indicator enforces the condition that $B_k$ be resonant (which implies that any isolated subcomponents $B_{j-1,\beta}$ must be resonant, as required for them to remain in $R^{(j)}$).
Note that this is only an upper bound because we ignore (\textit{i.e.} upper bound by 1) probabilities of other events that need to occur to make $B_k$ a component of $R^{(k)}$, \textit{e.g.} factors $(N-1)/N$ for sites adjacent to $B_k$, and similar factors in later steps.

We make some definitions that keep track of the probability bounds that are generated with this procedure.
Let $n(B_k)$ denote the number of sites in $B_k$, a component of $R^{(k)}$. Then put
\begin{equation}\label{P00}
\tilde{P}^{(k)}(B_k) \equiv \veps^{n(B_k)} = 
\prod_\beta \tilde{P}^{(k-1)}(B_{k-1,\beta}).
\end{equation}
Here $\{B_{k-1,\beta}\}$ are the subcomponents of $B_k$ on scale $k-1$, \textit{i.e.}
each $B_{k-1,\beta}$ is a connected component of $R^{(k-1)}$, based on connections with range $L_{k-1}
^\alpha$. 
Next, we define a weighted sum of the probability bounds for each $k \ge 1$:
\begin{equation}\label{P0}
Q^{(k)}_{x} \equiv \sum_{B_k\text{ containing }x} \tilde{P}^{(k)}(B_k)
\veps^{-q_kn(B_k)}\mathbb{E}_k\,\mathbf{1}_{\hat{n}_k(B_k)>0}\,\veps^{-q_0\hat{n}_k(B_k)}
\big(\text{diam}(B_k)\vee L_{k-1}\big)^p,
\end{equation}
where $q_0 = \tfrac{1}{3}$, $q_k = q_{k-1} -\lvert \log\veps \rvert ^{-1}L_{k-2}^{-p/3}$, 
with $L_{-1} \equiv \tfrac{1}{2}L_0$. Recall that our constants are chosen in the order
$p,L_0,N,\gamma$, so for a given $p$ we can choose $L_0$ large enough so that
$q_k\ge\tfrac{1}{4}$ for all $k$.
\begin{theorem}\label{thm:Q}
For any sufficiently large $p$, let $L_0$ be sufficiently large (depending on $p$) and
$\veps = \tfrac{1}{N-1}$ sufficiently small (depending on $L_0$), and take $\gamma \le \veps^{20}$. Then for any $k \ge 1$ such that $5.1L_{k-1} < \mathrm{Diam}(\Lambda)$,
and any $x\in\Lambda$,
\begin{equation}\label{P1}
Q^{(k)}_{x} \le 1.
\end{equation}
\end{theorem}
This theorem demonstrates that the probability that $B_k$ is a component of $R^{(k)}$ decreases exponentially
with the volume $n(B_k)$; it also decreases as a power of the diameter of $B_k$,
with a minimum decay length $L_{k-1}$. The proof shows that the power is due to a small factor $\veps^{1/4}$ per scale, for isolated components that defy the odds defined by Proposition  \ref{prop:3.3} and remain resonant. More complicated scenarios involving multiple nearby components complicate the estimates, but lead to subdominant effects with faster than power law decay.

\textit{Proof.} To facilitate an inductive argument, let us extend the definitions by putting
$\tilde{P}^{(0)}(B_0)\equiv \veps$, where $B_0 = \{x\}$ for some $x \in R^{(1)}$.
Let $\text{diam}(B_0) = 0$, $n(B_0) = \hat{n}_0(B_0)=1$, and $\mathbb{E}_0 = 1$. Then with
\begin{equation}\label{P2}
Q^{(0)}_{x} =\tilde{P}^{(0)}(B_0)\veps^{-q_0n(B_0)}\mathbb{E}_0\,\mathbf{1}_{\hat{n}_0(B_0)>0}\,\veps^{-q_0\hat{n}_0(B_0)}
\big(\text{diam}(B_0)\vee L_{-1}\big)^p,
\end{equation}
the bound (\ref{P1}) holds by taking $\veps^{1/3}\le L_{-1}^{-p}$.

We work on the induction step, assuming (\ref{P1}) for $k-1$.
A block $B_k$ that contains $x$ may be decomposed into its connected components
$B_{k-1,0},B_{k-1,1},\ldots,B_{k-1,m}$ on scale $k-1$.
Here $m+1 \ge 1$ is the number of subcomponents, and $B_{k-1,0}$ is the one containing
$x$. Connectivity on scale $k$ requires that there exists at least one tree graph $T$
on $\{0,1,\ldots,m\}$ such that for each link $\beta\beta' \in T$, $\text{dist}(B_{k-1,\beta}, B_{k-1,\beta'}) \le L_k^{\alpha}$.
Writing
\begin{multline}\label{P3}
Q^{(k)}_{x} = \sum_{B_{k-1,0}\ni x} \sum_T \sum_{\substack{B_{k-1,1},\ldots,B_{k-1,m}\\ \text{consistent with }T }} \tilde{P}^{(k)}(B_k)\veps^{-q_kn(B_k)} 
\\
\cdot\mathbb{E}_k\,\mathbf{1}_{\hat{n}_k(B_k)>0}\,\veps^{-q_0\hat{n}_k(B_k)}
\big(\text{diam}(B_k)\vee L_{k-1}\big)^p,
\end{multline}
we divide the factors on the right-hand side amongst the components
$B_{k-1,1},\ldots,B_{k-1,m}$. From (\ref{P00}), we have
\begin{equation}\label{P4}
\tilde{P}^{(k)}(B_k) = 
\prod_{\beta=0}^m \tilde{P}^{(k-1)}(B_{k-1,\beta}). 
\end{equation}
Clearly,
\begin{equation}\label{P4.5}
n(B_k) = \sum_{\beta = 0}^m n(B_{k-1,\beta}).
\end{equation}
We claim that
\begin{equation}\label{P5}
\hat{n}_k(B_k) \le \sum_{\beta = 0}^m \hat{n}_{k-1}(B_{k-1,\beta}).
\end{equation}
To see this, we need to understand three changes effected between the two sides of this inequality:
\begin{equation}\label{P6}
\tilde{F}^{(k)}_{E_{k+1}}(B_k) \rightarrow F^{(k-1)}_{E_{k+1}}\text{ in }\bar{B}_k
\rightarrow \bigoplus_{\beta = 0}^m \tilde{F}^{(k-1)}_{E_{k+1}}(B_{k-1,\beta})
\rightarrow \bigoplus_{\beta = 0}^m \tilde{F}^{(k-1)}_{E_{k}}(B_{k-1,\beta}).
\end{equation}
The first is merely a change in notation; $\tilde{F}^{(k)}_{E_{k+1}}(B_k)$ is the same
as $F^{(k)}_{E_{k+1}}$ computed in the region $\bar{B}_k$ instead of $\Lambda$.
As $\bar{B}_k$ is contained within a $2.05L_k$-neighborhood of $B_k$ and 
$\text{dist}(B_k,R^{(k)}\setminus B_k) > L_k^\alpha$, the scale index can be shifted to $k-1$ without change. The second change is covered by Corollary \ref{cor:2'}; the norm of the difference is $\le \gamma^{3.3L_{k-1}} < \gamma\veps_k$.
The third change is covered by Theorem \ref{thm:3}; the norm of the difference is
$\le\gamma\lvert E_{k+1}-E_k \rvert \le \gamma\veps_k/3$. Thus the eigenvalues may move by no
more than $2\gamma\veps_k$. The interval $I_{\veps_{k+1}}(E_{k+1})$ is contained within
$I_{\veps_k/2}(E_{k})$, so eigenvalues outside of $I_{\veps_k}(E_{k})$
cannot migrate to it. Thus (\ref{P5}) holds.

We may relate $D \equiv \text{diam}(B_k)\vee L_{k-1}$ to the individual
$d_\beta \equiv \text{diam}(B_{k-1,\beta})\vee L_{k-2}$ as follows.
If $m=0$, we have to consider the possibility that $d_0 < L_{k-1}$. Let $\mathbf{1}_D = 1$
if $m=0$ and $d_0 < D$ (equivalently, $\text{diam}(B_{k-1,0}) < L_{k-1}$); and put
$\mathbf{1}_D = 0$ otherwise. Then we have that
\begin{equation}\label{P8}
D^p \le d_0^p \veps^{-\mathbf{1}_D/4},
\end{equation}
by choosing $N$ large enough so that $\veps^{-1/4}=(N-1)^{1/4} > 2^{p} = L_{k-1}^p/L_{k-2}^p$. If $m\ge1$, we need to estimate a sum by a product.
We have that
\begin{equation}\label{P9}
D \le \sum_{\beta = 0}^m \text{diam}(B_{k-1,\beta}) + mL_k^\alpha
\le \sum_{\beta = 0}^m d_\beta + mL_k^\alpha.
\end{equation}
Note that $\xi_0+\ldots+\xi_{2m} \le (2m+1)\xi_0\cdots \xi_{2m}$
if each $\xi_\beta \ge 1$.
Taking $\xi_\beta = d_\beta/L_{k-2}$ for $0\le\beta \le m$ and
$\xi_\beta = L_k^\alpha/L_{k-2} \le L_{k-2}^{\alpha - 1}$ for
$m+1 \le \beta \le 2m$, we obtain
\begin{equation}\label{P10}
D \le L_{k-2}\sum_{\beta = 0}^{2m} \xi_\beta \le (2m+1)L_{k-2}
\prod_{\beta = 0}^m \frac{d_\beta}{L_{k-2}} \left(L_{k-2}^{\alpha-1}\right)^m
\le \prod_{\beta = 0}^m d_\beta \left(3L_{k-2}^{\alpha-2}\right)^m.
\end{equation}
We have used the fact that $2m+1 \le 3^m$. 

If $\mathbf{1}_D = 0$, then by (\ref{P5}) we have 
\begin{equation}\label{mathbfE}
\mathbb{E}_k\,\mathbf{1}_{\hat{n}_k(B_k)>0}\,\veps^{-q_0\hat{n}_k(B_k)} \le \prod_m \left[
\mathbb{E}_{k-1}\,\mathbf{1}_{\hat{n}_{k-1}(B_{k-1})>0}\,\veps^{-q_0\hat{n}_{k-1}(B_{k-1})}\right].
\end{equation}
Note that if $\hat{n}_k(B_k)>0$ then each of the subcomponents of $B_k$ satisfies $\hat{n}_{k-1}(B_{k-1})>0$.
If $\mathbf{1}_D = 1$ (\textit{i.e.}
$m=0$ and $\text{diam}(B_{k-1,0}) < L_{k-1}$), 
then $B_{k-1,0}$ is isolated, \textit{c.f.} (\ref{isolated}). Hence (\ref{(27)}) implies that
$\text{diam}(\bar{B}_{k-1,0}) \le 5.1L_{k-1}$, which by assumption is less than $\text{Diam}(\Lambda)$. Thus the requirement
$\text{diam}(\bar{B}_{k-1}) < \text{Diam}(\Lambda)$ of Proposition \ref{prop:3.3} is satisfied.
Thus $\hat{n}_k(B_k) = \hat{n}_{k-1}(B_{k-1})$ with probability $\le \tfrac{1}{N-1} = \veps$, and $\hat{n}_k(B_k) < \hat{n}_{k-1}(B_{k-1})$ otherwise. Let $\mathbb{E}_{v_{\bar{y}}}$ denote
the expectation over the single $v_{\bar{y}}$ in $\bar{B}_k\setminus\bar{B}_{k-1}$. Then
\begin{align}\label{mathbfE2}
\mathbb{E}_k\,\mathbf{1}_{\hat{n}_k(B_k)>0}\,\veps^{-q_0\hat{n}_k(B_k)} \nonumber
&= \mathbb{E}_{k-1}\,\mathbb{E}_{v_{\bar{y}}}\,\mathbf{1}_{\hat{n}_k(B_k)>0}\,\veps^{-q_0\hat{n}_k(B_k)} 
\\
&\le
\mathbb{E}_{k-1}\,[\veps + (1-\veps)\veps^{q_0}]\mathbf{1}_{\hat{n}_{k-1}(B_{k-1})>0}\,\veps^{-q_0\hat{n}_{k-1}(B_{k-1})} \nonumber
\\
&\le \veps^{1/4}\mathbb{E}_{k-1}\,\mathbf{1}_{\hat{n}_{k-1}(B_{k-1})>0}\,\veps^{-q_0\hat{n}_{k-1}(B_{k-1})} \nonumber
\\
&=\veps^{\mathbf{1}_D/4}\mathbb{E}_{k-1}\,\mathbf{1}_{\hat{n}_{k-1}(B_{k-1})>0}\,\veps^{-q_0\hat{n}_{k-1}(B_{k-1})}.
\end{align}
The key point here is that the tendency of $\hat{n}_k(B_k)$ to decrease enables us to cancel the factor $\veps^{-\mathbf{1}_D/4}$, and so the power-law decay bound of (\ref{P0})-(\ref{P1}) continues even in cases where
``easier'' sources of smallness (such as large blocks) are not available.
Proposition \ref{prop:3.3} provides the needed boost to the estimate in
precisely the case where no other source of smallness is available.

Inserting (\ref{P4}), (\ref{P4.5}), (\ref{P8}), (\ref{P10})-(\ref{mathbfE2}) into (\ref{P3}),
we obtain
\begin{align}\label{P11}
Q^{(k)}_{x} &= \sum_{B_{k-1,0}\ni x} \sum_T \sum_{\substack{B_{k-1,1},\ldots,B_{k-1,m}\\ \text{consistent with }T }} \left(3L_{k-2}^{\alpha-2}\right)^{mp}\nonumber\\
&\cdot \prod_{\beta = 0}^m
\left[ \tilde{P}^{(k-1)}(B_{k-1,\beta})\veps^{-q_kn(B_{k-1,\beta})}
\mathbb{E}_{k-1}\,\mathbf{1}_{\hat{n}_{k-1}(B_{k-1})>0}\,\veps^{-q_0\hat{n}_{k-1}(B_{k-1,\beta})}
d_\beta^p \right].
\end{align}

We may control the tree-graph sum in (\ref{P11}) with the following construction -- see \cite{Imbrie2016c}. Define for any block $B_{k-1}$,
\begin{multline}\label{P12}
K^{(\rho)}(B_{k-1}) \equiv \sum_{T: \text{ depth}(\bar{T})\le \rho}\,  
\sum_{\substack{B_{k-1,1},\ldots,B_{k-1,\ell}\\ \text{consistent with }T }}\,
\\
\cdot\prod_{\beta=1}^{\ell}\left[ \tilde{P}^{(k-1)}(B_{k-1,\beta}) \veps^{-q_k n(B_{k-1,\beta})}\mathbb{E}_{k-1}\,\mathbf{1}_{\hat{n}_{k-1}(B_{k-1})>0}\,\veps^{-q_0 \hat{n}_{k-1}(B_{k-1,\beta})}
d_\beta^p\cdot3L_{k-2}^{(\alpha-2)p}\right].
\end{multline}
As in (\ref{P11}), $T$ is a tree graph with root at $B_{k-1,0}$, but here we are including the factors in square brackets only for the non-root
vertices. We have used $\text{depth}(T)$ to denote the largest number of links in $T$ that are required to reach any vertex, starting
at the root. Then (\ref{P11}) becomes
\begin{multline}\label{P13}
Q^{(k)}_{x} = \sum_{B_{k-1,0}\ni x}  \tilde{P}^{(k-1)}(B_{k-1,0})
\veps^{-q_kn(B_{k-1,0})}
\\
\cdot\mathbb{E}_{k-1}\,\mathbf{1}_{\hat{n}_{k-1}(B_{k-1,0})>0}\,\veps^{-q_0\hat{n}_{k-1}(B_{k-1,0})}
d_0^p K^{(\infty)}(B_{k-1,0}).
\end{multline}
\begin{lemma}\label{lem:3.5}
Under the same assumptions as Theorem \ref{thm:Q},
\begin{equation}\label{P14}
K^{(\rho)}(B_{k-1}) \le \exp\big(L_{k-2}^{-p/3}n(B_{k-1})\big).
\end{equation}
\end{lemma}
\textit{Proof.} If we take $\rho = 0$, then (\ref{P14}) becomes
$1 \le \exp\big(L_{k-2}^{-p/3}\big)$. Working inductively,
we have a recursion
\begin{multline}\label{P15}
K^{(\rho)}(B_{k-1}) \le \sum_{r=0}^\infty\frac{1}{r!}
 \prod_{\beta=1}^{r}
\Bigg[ 
\sum_{B_{k-1,\beta} \text{ linked to }B_{k-1}}
\tilde{P}^{(k-1)}(B_{k-1,\beta}) 
\\
\cdot
\veps^{-q_k n(B_{k-1,\beta})} \mathbb{E}_{k-1}\,\mathbf{1}_{\hat{n}_{k-1}(B_{k-1})>0}\,\veps^{-q_0 \hat{n}_{k-1}(B_{k-1,\beta})}
d_\beta^p  K^{(\rho-1)}(B_{k-1,\beta})  \cdot3L_{k-2}^{(\alpha-2)p}
\Bigg].
\end{multline}
This inequality results from ignoring any consistency conditions amongst the $r$ sums in (\ref{P15}).
Applying (\ref{P14}) and the relation $q_k = q_{k-1}-\lvert \log\veps \rvert ^{-1}
L_{k-2}^{-p/3}$, we may replace $q_k \rightarrow q_{k-1}$ in each factor. The sum over $B_{k-1,\beta}$ reduces to a sum over $x$ and a sum over
$B_{k-1,\beta}$ containing $x$. There are no more than
$n(B_{k-1})(2L_k^{\alpha }+1)^d$ choices for $x$. The sum over
$B_{k-1,\beta}$ containing $x$ reproduces $Q_x^{(k-1)}$ as in
(\ref{P0}), and it is bounded by 1 by (\ref{P1}).
Since $\alpha = \tfrac{3}{2}$, we may take $p$ large enough so that
$3L_{k-2}^{(\alpha -2)p}(2L_k^{\alpha }+1)^d \le L_{k-2}^{-p/3}$, and we obtain (\ref{P14}). \qed

With the lemma in hand, we find again in (\ref{P13}) that the factor
$K^{(\infty)}(B_{k-1,0})$ leads to the replacement $q_k \rightarrow
q_{k-1}$, and we obtain a bound by $Q_x^{(k-1)}\le1$. This
completes the proof of Theorem \ref{thm:Q}. \qed

Definition (\ref{P00}) is intended for the fixed energy procedure, with
$E_k$ independent of the potentials. We will need a modified version of (\ref{P00})
for the energy-following procedure. In this procedure, we fix a site $x$,
and put $E_1 = v_x$. Let $B_{x,k}$ denote the component of $R^{(k)}$
containing $x$. We choose $E_{k+1}$ close to a solution to
$\lambda \in \text{spec}\,\tilde{F}_\lambda^{(k)}(B_{x,k})$ in
$I_{\veps_k/3}(E_k)$. This ensures that $x$ remains in $R^{(k)}$ for 
all $k$; thus $B_{x,k}$ always exists. For the energy-following procedure, we replace (\ref{P00}) with the following:
\begin{align}\label{Q1}
\hat{P}^{(k)}(B_k) &=  \tilde{P}^{(k)}(B_k) = \veps^{n(B_k)}, \quad \text{for } B_k \not\ni x. \text{ Otherwise:}
\\ \label{Q1'}
 \hat{P}^{(k)}(B_k) &= \veps^{n(B_k) - 1} = \prod_\beta \hat{P}^{(k-1)}(B_{k-1,\beta}).
\end{align}
We may understand this definition by considering various cases.
Initially, $x$ is resonant to $E_1$ by construction, so the
probability that $x$ is resonant is 1, not $\tfrac{1}{N} < \veps$. 
However, the
probability that \textit{other} sites are resonant to $E_1$ is $< \veps$ as in
(\ref{P00}). 

The subsequent accumulation of small probability factors works as in
(\ref{mathbfE2}), except in the case where $B_{k-1}$ contains $x$ and
$\hat{n}_{k-1}(B_{k-1}) = 1$. In this case, the randomness-induced
movement of the eigenvalues (demonstrated in Proposition \ref{prop:3.3}) 
does not lead to smallness as in the factor $\veps^{\mathbf{1}_D/4}$ of (\ref{mathbfE2}), because $E_{k+1}$ can follow
an eigenvalue. Blocks not containing $x$ are unaffected, because the
potentials used to produce eigenvalue movement are disjoint from
the ones that determine $E_{k+1}$. By case 2 of Proposition \ref{prop:3.3}, when $\hat{n}_{k-1}(B_{k-1}) > 1$,
the spread argument works even for $B_{x,k}$; although $E_{k+1}$
is chosen near one of the eigenvalues, the demonstrated
lower bound on the spread pushes at least one eigenvalue out of
$I_{\veps_{k+1}}(E_{k+1})$, with probability $1-\tfrac{1}{N-1}$. 
Thus if $\hat{n}_{k-1}(B_{k-1})>1$, a small power of 
$\veps$ is available as in (\ref{mathbfE2}).

We use (\ref{Q1})-(\ref{Q1'}) to define an associated weighted sum
\begin{equation}\label{Q2}
\hat{Q}^{(k)}_{x} \equiv \sum_{B_{x,k} \ne \{x\}}
\hat{P}^{(k)}(B_{x,k})
\veps^{-\hat{q}_kn(B_{x,k})}\mathbb{E}_k\,\mathbf{1}_{\hat{n}_{k}(B_{k})>0}\,\veps^{-\hat{q}_0\hat{n}_k(B_{x,k})}
\big(\text{diam}(B_{x,k})\vee L_{\hat{k}-1}\big)^{\hat{p}}.
\end{equation}
Here $\hat{p} = p/2$, $\hat{q}_0 = \tfrac{1}{6}$, 
$\hat{q}_k = \hat{q}_{k-1} -\lvert \log\veps \rvert ^{-1}L_{k-2}^{-\hat{p}/3}$, and 
then we have $\hat{q}_k \ge \tfrac{1}{8}$ for all $k$.
Roughly speaking, the loss of some small factors when $\hat{n}_k(B_{x,k}) = 1$ is
compensated by the halving of the exponents in (\ref{Q2}).
Also, in (\ref{Q2}) we are using $\hat{k} = \hat{k}(B_{x,k})$ to denote 
the maximum of all the $j \in [1,k]$
such that $B_{x,j} \setminus B_{x,j-1} \ne \varnothing$ or
$\hat{n}_{j-1}(B_{x,j-1}) >1$.
(As before, $B_{x,0} \equiv \{x\}$ and $\hat{n}_0(B_{x,0}) = n(B_{x,0}) = 1$.)
There must be some $j\in[1,k]$ satisfying the condition, because 
otherwise we would have the trivial case 
$B_{x,j} = \{x\}$ for all $j$, which is not included in (\ref{Q2}).
Thus $\hat{k}$ represents the last scale at which smallness is 
produced, either through joining of blocks, or because
$\hat{n}_{\hat{k}-1}(B_{x,\hat{k}-1})>1$. 
For $j \in (\hat{k},k]$, we have 
$B_{x,j} = B_{x,k}$ and $\hat{n}_{j-1}(B_{j-1}) = 1$.
\begin{theorem}\label{thm:Qhat}
Under the same assumptions as Theorem \ref{thm:Q},
\begin{equation}\label{P1a}
\hat{Q}^{(k)}_{x} \le 1-2^{-k}.
\end{equation}
\end{theorem}
\textit{Proof.} We modify as needed the arguments in the proof of 
Theorem \ref{thm:Q}. Instead of (\ref{P2}), we define 
$\hat{Q}_x^{(0)}=1$, as there is no initial factor of $\veps$ to work with.
As mentioned above, the condition $B_{x,k} \ne \{x\}$ implies that 
at some scale $j \in[1,k]$, $B_{x,j-1} = \{x\}$ was joined with other block(s) to form $B_{x,j}$. 
At this point, $\hat{P}^{(j-1)}(B_{x,j-1}) = \veps^{n(B_{x,j-1})-1} = 1$ -- see (\ref{Q1'}).
We may make up for the missing factor of $\veps$ when $\beta = 0$ by replacing (\ref{P4.5}), (\ref{P5}) with
\begin{align}\label{QQ}
n(B_j)/2 &\le \sum_{\beta = 1}^m n(B_{j-1,\beta}),
 \\ \label{QQQ}
\hat{n}_j(B_j)/2 &\le \sum_{\beta = 1}^m \hat{n}_{j-1}(B_{j-1,\beta}).
\end{align}
These follow from the simple fact that $(1+n)/2 \le n$ if $n \ge 1$ -- with $n$ representing the right-hand side of (\ref{QQ}) or (\ref{QQQ}) and the 1 representing the $\beta=0$ contribution to the left-hand side.
In subsequent steps, we use (\ref{P4.5}), (\ref{P5}) as before.

As in the fixed-energy case, we need to relate
$\hat{D} \equiv \text{diam}(B_{x,k})\vee L_{\hat{k}-1}$ to the individual
$d_\beta \equiv \text{diam}(B_{k-1,\beta})\vee L_{k-2}$. Let us put
\begin{equation}\label{d00}
\hat{d}_0 \equiv \begin{cases}
1, &\text{if } B_{x,k-1} = \{x\}; \\
\text{diam}(B_{x,k-1})\vee L_{\widehat{k-1} -1}, &\text{otherwise.}
\end{cases}
\end{equation}
Let us consider the case $m=0$. We have the possibility that $\hat{d}_0 < \hat{D}$,
so let us put $\mathbf{1}_{\hat{D}} = 1$ if $m=0$ and $\hat{d}_0 < \hat{D}$; and 0
otherwise.
Note that if $\hat{k} = k$ and $m=0$, then $\widehat{k-1} = k-1$ also. (Since
$\hat{n}_{k-1}(B_{x,k-1})>1$, there must have been either (1) blocks joining
to form $B_{x,k-1}$ or (2) $B_{x,k-1}$ consisting of a single subcomponent
$B_{x,k-2}$, in which case $\hat{n}_{k-2}(B_{x,k-2})>1$ by (\ref{P5}).)
Thus $\hat{D}/\hat{d}_0 \le 2 = L_{k-1}/L_{k-2}$, and as in (\ref{P8}) we have that
$\hat{D}^p \le \hat{d}_0^p \veps^{-\mathbf{1}_{\hat{D}}/8}$
(anticipating the halving of exponents, $\hat{q}_0 = q_0/2)$.
Suppose $\hat{k} = k$. Since $m=0$, we must have
$\hat{n}_{k-1}(B_{x,k-1})>1$, and case 2 of Proposition \ref{prop:3.3} applies. Hence we pick up a factor $[\veps + (1-\veps)\veps^{\hat{q}_0}] \le \veps^{1/8}$ as in (\ref{mathbfE2}).
On the other hand,
if $\hat{k}<k$, then $\hat{k} = \widehat{k-1}$, so $\hat{d}_0 = \hat{D}$ and 
$\mathbf{1}_{\hat{D}} = 0$. Thus in both cases the factor $\veps^{-\mathbf{1}_{\hat{D}}}$ is cancelled.

For $m \ge 1$ we use (\ref{P10}) to obtain
\begin{equation}\label{Dhat0}
\hat{D} \le D \le \prod_{\beta = 0}^m d_\beta \left(3L_{k-2}^{\alpha-2}\right)^m
\le \hat{d}_0 \prod_{\beta = 1}^m d_\beta^2 \left(3L_{k-2}^{\alpha-2}\right)^m.
\end{equation}
We have used the fact that $d_0d_1 \le \hat{d}_0d_1^2$. (This is a variant of the halving-exponent argument used above: if
$d_0 > \hat{d}_0$, then $\text{diam}(B_{x,k-1}) \le L_{k-2}$, which implies that
$d_0 \le d_1$.) Then we have
\begin{equation}\label{Dhat}
\hat{D}^{\hat{p}} = \hat{D}^{p/2} \le \hat{d}_0^{p/2}\prod_{\beta = 1}^m d_\beta^p \left(3L_{k-2}^{\alpha-2}\right)^{mp/2} =
\hat{d}_0^{\hat{p}}\prod_{\beta = 1}^m d_\beta^p \left(3L_{k-2}^{\alpha-2}\right)^{m\hat{p}}  .
\end{equation}
We show in the next paragraph how the factors $d_\beta^p$ can be absorbed into sums over blocks not containing $x$; 
these have stronger estimates as in Theorem \ref{thm:Q}.

Putting all these estimates together, we obtain a bound analogous to (\ref{P13}):
\begin{multline}\label{QP}
\hat{Q}^{(k)}_{x} \le \sum_{B_{k-1,0}\ni x, \,B_{k-1,0} \ne \{x\}}  \hat{P}^{(k-1)}(B_{k-1,0})
\veps^{-\hat{q}_kn(B_{k-1,0})} 
\\
\cdot\mathbb{E}_{k-1}\,\mathbf{1}_{\hat{n}_{k-1}(B_{k-1})>0}\,\veps^{-\hat{q}_0\hat{n}_{k-1}(B_{k-1,0})}
\hat{d}_0^{\hat{p}} \hat{K}^{(\infty)}(B_{k-1,0})
+\hat{P}^{(k-1)}(\{x\})\big(\hat{K}^{(\infty)}(\{x\})-1\big).
\end{multline}
Here the second term corresponds to the case $B_{k-1,0} = \{x\}$. We have introduced
$\hat{K}^{(\rho)}$ as in (\ref{P12}) but with $L_{k-2}^{(\alpha-2)\hat{p}}$ in the last
factor instead of $L_{k-2}^{(\alpha-2)p}$
(but the other factors remain the same, in particular the coefficient of $n(B_{k-1,\beta})$ is still $q_k$ and the exponent of $d_\beta$ is still $p$).
The proof of Lemma \ref{lem:3.5} still works,
and so $\hat{K}^{(\infty)}(B_{k-1}) \le \exp\big(L_{k-2}^{-\hat{p}/3}\big)$. 
This bound on $\hat{K}$ 
allows us to absorb the factors  $\veps^{-q_k n(B_{k-1,\beta})}$ and $d_\beta^p$
associated with the ``makeup'' bounds (\ref{QQ}), (\ref{Dhat}), which transfer the burden to the terms $\beta \ge 1$. 
Note that
$q_k > \hat{q}_k$ for all $k$.
The second term
of (\ref{QP}) has no $\veps^{-1}$ factors because (\ref{QQ}) moves them into $\hat{K}$;
we have 
$\hat{d}_0 = 1$ by (\ref{d00}). In fact $\hat{P}^{(k-1)}(\{x\}) = 1$ as well, see (\ref{Q1'}).
The term $r=0$ in (\ref{P15}) is not present if $B_{k-1,0} = \{x\}$, because it would lead
to $B_k = \{x\}$, which is not included in $\hat{Q}^{(k)}_x$, see (\ref{Q2}).
Then we have 
\begin{equation}\label{QP1}
\hat{K}^{(\infty)}(\{x\})-1 \le \exp\big(L_{k-2}^{-\hat{p}/3}\big) -1 \le
2L_{k-2}^{-\hat{p}/3} \le 2^{-k}.
\end{equation}
The bound on $\hat{K}^{(\infty)}(B_{k-1,0})$ leads to the
reduction $\hat{q}_k \rightarrow \hat{q}_{k-1}$ in the first term of (\ref{QP}), and 
it becomes $\hat{Q}_x^{(k-1)}$. This is bounded by $1-2^{-(k-1)}$, by induction,
and hence $\hat{Q}_x^{(k)}\le 1-2^{-(k-1)} + 2^{-k} = 1- 2^{-k}$. \qed

\section{Results}\label{sec:results}
Here we use the percolation estimates from Section \ref{ssec:3.2} and the random-walk
estimates from Section \ref{ssec:random} to obtain our main theorems.
\subsection{Density of States}\label{ssec:4.1}
\textit{Proof of Theorem \ref{thm:1.1}.} We wish to prove that
$\mathbb{E}\,\calN\big(I_\delta(E)\big) \le \lvert \Lambda \rvert(\log_\gamma\delta)^{-p}$, 
for $\delta \in [\gamma^{\text{Diam}(\Lambda)/2},1]$. We may assume
that $\delta<\gamma$, because the total number of eigenvalues is $\lvert \Lambda \rvert$, and if $\delta \ge \gamma$, the inequality (\ref{(5)}) is automatically satisfied.
Let $k$ be defined by the inequality $\veps_{k+1}/3 < \delta \le \veps_k/3$; observe that
$\veps_1 = \tfrac{1}{N-1} > \gamma$.
Let us take the case $k \ge 2$. We have that
\begin{equation}\label{R0}
\gamma^{3.2L_{k-1}} > \gamma^{3.2L_{k-1}}/3 = \veps_k/3 \ge \delta \ge \gamma^{\text{Diam}(\Lambda)/2}.
\end{equation}
Therefore, $3.2L_{k-1} < \text{Diam}(\Lambda)/2$, which implies that the limitation
$5.1L_{k-1} < \text{Diam}(\Lambda)$ in Theorem \ref{thm:Q} is satisfied.

As explained at the start of Section \ref{ssec:3.2}, $\tilde{P}^{(k)}(B_k)\mathbb{E}_k\,\mathbf{1}_{\hat{n}_{k}(B_{k})>0}$ is a bound
for the probability that $B_k$ is a component of $R^{(k)}$. 
The probability that $x \in R^{(k)}$ is therefore bounded by
\begin{equation}\label{R1}
\sum_{B_k \text{ containing } x} \tilde{P}^{(k)}(B_k)\mathbb{E}_k\,\mathbf{1}_{\hat{n}_{k}(B_{k})>0} \le Q^{(k)}_x \veps^{1/4}L_{k-1}^{-p} \le \veps^{1/4}L_{k-1}^{-p}.
\end{equation}
Here we use (\ref{P0}) and Theorem \ref{thm:Q}, noting also that $q_k \ge \tfrac{1}{4}$ and $n(B_k)\ge 1$. We claim that the number of
eigenvalues in $I_{\veps_k/3}(E)$ is bounded by $\lvert R^{(k)} \rvert $, the dimension of the space
on which $F_E^{(k)}$ acts. This holds because repeated application of Lemma \ref{fundamental}
guarantees that all of the spectrum of $H$ in $I_{\veps_k/3}(E)$ is captured by $F_E^{(k)}$.

In detail, we note that in each step $j \le k$ 
Theorem \ref{thm:3} provides the requisite Lipschitz continuity as in Lemma \ref{fundamental}(ii), and so as in the proof of that lemma, we
conclude that all of the eigenvalues 
of $F_E^{(j-1)}$ in $I_{\veps_j/3}(E)$
are in close agreement with those of $F_E^{(j)}$.
Thus by induction, $\calN\big(I_{\veps_k/3}(E)\big) \le \lvert R^{(k)} \rvert $, which verifies the claim.

As a consequence, we may use (\ref{R1}) to obtain
\begin{align}\label{R2}
\mathbb{E}\,\calN\big(I_\delta(E)\big) &\le \mathbb{E}\,\calN\big(I_{\veps_k/3}(E)\big) \le \sum_x P\big(x \in R^{(k)}\big) \notag\\
&\le \lvert \Lambda \rvert\veps^{1/4}L_{k-1}^{-p}
 = \lvert \Lambda \rvert\veps^{1/4}\big(\tfrac{1}{6.4}\log_\gamma\veps_{k+1}\big)^{-p} \le \lvert \Lambda \rvert(\log_\gamma\delta)^{-p}.
\end{align}
We have used the fact that $\gamma \ge \delta \ge \tfrac{1}{3}\veps_{k+1} = \tfrac{1}{3}\gamma^{6.4L_{k-1}}$. 

If $k=1$, then $\delta > \veps_2/3$, so $\log_\gamma\delta < 3.2L_1 +1$. Choosing 
$N$ large enough, depending on $p$ and $L_0$, we obtain the desired conclusion directly, using
\begin{equation}\label{R00}
\mathbb{E}\,\calN\big(I_{\veps_1/3}(E)\big) \le \mathbb{E}\,\lvert R^{(1)} \rvert  = \lvert \Lambda \rvert P\big(x \in R^{(1)} \big)
\le \tfrac{1}{N} \lvert \Lambda \rvert
\le \lvert \Lambda \rvert(\log_\gamma\delta)^{-p}.
\end{equation}
This completes the proof. \qed

\subsection{Energy-Following Procedure}\label{ssec:4.2}
Here we lay out a procedure for constructing all of the eigenfunctions and eigenvalues based on local data.
Starting at some site $x$, we produce a sequence of approximate eigenvalues $E_1, E_2, \ldots$. The associated Schur complements determine resonant blocks, as has been described already. Recall that $B_{x,k}$ denotes the block containing $x$ in step $k$;
it is used to construct the next approximate eigenvalue $E_{k+1}$.
This brings in the effect of the potential in the region $\bar{B}_{x,k}$.
If $B_{x,k}$ is isolated in step $k$, the corrections are exponentially small in the diameter of $\bar{B}_{x,k}$. If $B_{x,k}$ is not isolated in step $k$, then its diameter is at least $L_k$ and then by (\ref{Q2}) and Theorem \ref{thm:Qhat} the probability decays as a large power of 
$\text{diam}(\bar{B}_{x,k})$. Thus we may say that the eigenfunctions and eigenvalues are quasilocal functions of the potentials.

To begin the procedure, recall that $H=H_0-\gamma J$ with
$H_0 = \text{diag}\big(\{2d\gamma+v_x\}_{x\in \Lambda}\big)$.
Thus it makes sense to choose one particular $x$ and put $E_1 = 2d\gamma +v_x$. It is evident that $E_1$ depends on $x$, but we suppress the dependence in the notation.
As described in Section \ref{sec:2}, $E_1$ determines a resonant set
$R^{(1)}$, which may be decomposed into blocks $B_1$. The site
$x$ is automatically in $R^{(1)}$, and the block containing $x$ 
is denoted $B_{x,1}$. Once $R^{(1)}$ is determined, we have
for $\lvert \lambda - E_1 \rvert  \le \veps_1/2$ the Schur complement $F_\lambda^{(1)}$ and its localized versions $\tilde{F}_\lambda^{(1)}(B_1)$.
 
We continue the process in the $k^{\text{th}}$ step. Assume that a sequence of choices $x$, $E_1, \ldots, E_{k-1}$ has been made in previous steps. There is an associated increasing sequence of blocks 
containing $x$, which we denote by
$B_{x,1},\ldots,B_{x,k-1}$. We seek solutions to the condition
$\lambda \in \text{spec}\,\tilde{F}_\lambda^{(k-1)}(B_{x,k-1})$
in $I_{\veps_{k-1}/3}(E_{k-1})$; they should be good approximations 
to eigenvalues of $H$, which satisfy $\lambda \in \text{spec}\,F_\lambda^{(k-1)}$. Note that Theorem \ref{thm:3} shows that
$\tilde{F}_\lambda^{(k-1)}(B_{x,k-1})$ depends weakly on $\lambda$; it satisfies a Lipschitz condition with constant $\gamma$. 
By Weyl's inequality, the same is true of the eigenvalues. Thus we may 
sweep $\lambda$ through $I_{\veps_{k-1}/3}(E_{k-1})$, and for each
solution to $\lambda \in \text{spec}\,\tilde{F}_\lambda^{(k-1)}(B_{x,k-1})$
we choose $E_k$ to be the closest element of 
$\tfrac{1}{2}\veps_k \mathbb{Z}$. 
(If a solution happens to be equidistant between two multiples of
$\veps_k/2$, we take the smaller one.)
It may happen that more than one solution leads to the same choice of $E_k$; this avoids unnecessary proliferation of such choices.
It should be clear that every solution in 
$I_{\veps_{k-1}/3}(E_{k-1})$ is no farther than $\veps_k/4$ from some choice of $E_k$. 
Each resulting choice of $E_k$ is then used as the central energy for the next Schur complement
$F_{E_k}^{(k)}$, and the procedure continues. Note that when we shift
$E_{k-1}\rightarrow E_k$, we shift $F_{E_{k-1}}^{(j)} \rightarrow F_{E_k}^{(j)}$
for the random-walk expansions at level $j<k$ as well. We have the flexibility to do this
because we never leave the ``safe'' zone $\lvert \lambda - E_j \rvert  \le \veps_j/2$ covered by
Theorems \ref{thm:2}--\ref{thm:3}.
(Here we use the condition $\lvert E_k - E_{k-1} \rvert  \le \veps_{k-1}/3$, which implies that
$\lvert E_k-E_j \rvert  \le \veps_j/2$ for $j<k$, since the sum of shifts $\veps_i/3$
for $j \le i < k$ is less than $\veps_j/2$.)

Let $\bar{k}$ denote the smallest integer such that $5.1 L_{\bar{k}-1} \ge \text{Diam}(\Lambda)$. There can be no more than one block $B_{\bar{k} -1}$, because of the
minimum separation distance $L_{\bar{k}-1}^\alpha$. Thus we take $\bar{B}_{\bar{k} -1} =\Lambda$ and so
$\tilde{F}_\lambda^{(\bar{k}-1)}(B_{x,\bar{k}-1})=F_\lambda^{(\bar{k} -1)}$.
Then we choose $E_{\bar{k}}$ from one of the solutions to 
$\lambda \in \text{spec}\,\tilde{F}_\lambda^{(\bar{k}-1)}(B_{x,\bar{k}-1})$ in
$I_{\veps_{\bar{k}-1}/3}(E_{\bar{k}-1})$. Thus each $E_{\bar{k}}$ is an eigenvalue of
$H$, by repeated application of Lemma \ref{fundamental}, as in the proof of Theorem \ref{thm:1.1}. 

We will control the sum over the choices of $x, E_1, E_2, \ldots, E_{\bar{k}}$ in the next section. We conclude this section by stating
a proposition guaranteeing that every eigenvalue of $H$ can be obtained through this approximation scheme. The proof is deferred to Appendix \ref{A}.
\begin{proposition}\label{prop:4.1}
Let $L_0$ be sufficiently large. Take $\veps = \tfrac{1}{N-1}$ to be sufficiently small, depending on $L_0$, and take $\gamma \le \veps^{20}$. 
Let $\lambda_0$ be an eigenvalue of $H$. Then there is at least one set of choices
$x, E_1, \ldots E_{\bar{k}}$ for the energy-following procedure such that $E_{\bar{k}} = \lambda_0$ and such that
\begin{equation}\label{efp}
\lvert E_k - \lambda_0 \rvert  \le .31 \veps_k \text{ and } \lvert E_k - E_{k-1} \rvert  \le \veps_{k-1}/3, \text{ for }k \le \bar{k}.
\end{equation}
\end{proposition}
\subsection{Eigenfunction Correlator}\label{ssec:4.3}
We work toward a proof of Theorem \ref{thm:1.2}, in particular the bound (\ref{(6)}) giving power law decay of 
$\mathbb{E} \,\sum_\beta \lvert \varphi_\beta(x)\varphi_\beta(y) \rvert $.
As a preliminary step, we control the energy-following procedure (EFP) that was used in the previous section to construct all of the eigenfunctions.

Let $N_{x,y,z}$ denote the number of eigenvalues of $H$ that can be reached via the EFP as in Proposition \ref{prop:4.1}, with starting point $x$, and with a resonant region $B_{x,\bar{k}-1}$
that includes $y$ and $z$. Recall that $B_{x,k}$ is the component of 
$R^{(k)}$ containing $x$, and $\bar{k}$ is the smallest integer such that
$5.1 L_{k-1} \ge \text{Diam}(\Lambda)$. 
The block $B_{x,\bar{k}-1}$ is the final block in the EFP, as the procedure terminates with a choice of $E_{\bar{k}}$, an eigenvalue of
$H$.
\begin{proposition}\label{prop:4.3}
For any sufficiently large $p$, let $L_0$ be sufficiently large (depending on p),  $\veps = \tfrac{1}{N-1}$ sufficiently small (depending on $L_0$), and take $\gamma \le \veps^{20}$, $\hat{p} = p/2$. Then
\begin{equation}\label{(4.9)}
\mathbb{E}\,N_{x,y,z} \le \veps^{1/6}\big(\mathrm{diam}(\{x,y,z\}) \vee 1\big)^{-(\hat{p} - 1)} + \mathbf{1}_{\{x = y = z\}}.
\end{equation}
\end{proposition}
\textit{Proof.}
In the EFP we start at $x$, and take $E_1 = v_x + 2d\gamma$. Then
$B_{x,1}$ is determined, and $E_2$ is chosen close to one of the solutions to $\lambda \in \text{spec}\,\tilde{F}^{(1)}_\lambda(B_{x,1})$
in $I_{\veps_1/3}(E_1)$. Then $E_2$ determines $B_{x,2}$, and so on.
The choices of $B_{x,k}$ will be controlled by Theorem \ref{thm:Qhat},
so we focus now on counting the choices for $E_2, E_3,\ldots,E_{\bar{k}}$. 
We have a sequence of sums, so
let $\rho_j$ index
the sum over the choices of $E_j$; these choices depend on all 
previous choices in the EFP. 
We look for convenient positive combinatoric factors
$c_{\rho_j}$ satisfying $\sum_\rho c_{\rho_j}^{-1} \le 1$.
Then we have that
\begin{equation}\label{(4.10)}
\sum_{\rho_2, \ldots,\rho_{\bar{k}}} T_{\rho_2, \ldots,\rho_{\bar{k}}}
\le \sup_{\rho_2, \ldots,\rho_{\bar{k}}} c_{\rho_2}\cdots c_{\rho_{\bar{k}}}
T_{\rho_2, \ldots,\rho_{\bar{k}}}.
\end{equation}
We need to ensure that the product $c_{\rho_2}\cdots c_{\rho_{\hat{k}}}$ remains under control, relative to the smallness implicit in (\ref{Q2})
and Theorem \ref{thm:Qhat}; in particular we will obtain a bound by an
exponential in $\hat{k}+n(B_{x,k})$.
(Recall that $\hat{k} = \hat{k}(B_{x,k})$ was introduced after (\ref{Q2}); in step $k$ it represents the last scale at which smallness is 
produced, either through joining of blocks, or because
$\hat{n}_{\hat{k}-1}(B_{x,\hat{k}-1})>1$.)

Let us define for $2\le k\le \bar{k}$
\begin{equation}\label{(4.11)}
\hat{m}_{k,\rho_k} = \text{ the number of solutions to }\lambda \in
\text{spec}\,\tilde{F}_\lambda^{(k-1)}(B_{x,k-1}) \text{ in } I_{\veps_k/2}(E_{k,\rho_k}).
\end{equation}
This counts the number of solutions assigned to a particular choice of $E_k$. (Solutions are counted with multiplicity, in case of degeneracies in the spectrum.) Recall that we choose
$E_k \in \tfrac{1}{2}\veps_k \mathbb{Z}$ and so $E_{k,\rho_k}$
has to stand in for all solutions in $I_{\veps_k/2}(E_{k,\rho_k})$.
There is double coverage, which leads to a doubling of the 
combinatoric factors $c_{\rho_k}$, but we will see that they remain under control.

Next, we define combinatoric factors. For $k =2$ we put
\begin{equation}\label{(4.12)}
c_{\rho_2} = 2n(B_{x,1})/\hat{m}_{2,\rho_2},
\end{equation}
and for $2 < k \le \bar{k}$ we put
\begin{equation}\label{(4.13)}
c_{\rho_k} = 
\begin{cases}
\hat{m}_{k-1,\rho_{k-1}}, &\text{if }k = \bar{k}; \\
1, &\text{if }k>\hat{k}; \\
2\left(\hat{m}_{k-1,\rho_{k-1}} + n(B_{x,k}) - n(B_{x,k-1})\right)/\hat{m}_{k,\rho_k}, &\text{otherwise}.
\end{cases}
\end{equation}
Observe that the total number of solutions to $\lambda \in \text{spec}\,\tilde{F}_\lambda^{(1)}(B_{x,1})$ in $I_{\veps_1/2}(E_1)$ is no greater than $n(B_{x,1})$, the dimension
of the matrix $\tilde{F}_\lambda^{(1)}(B_{x,1})$. 
(This is evident because $\tilde{F}_\lambda^{(1)}(B_{x,1})$ is the same as 
$F^{(1)}_\lambda$ in $\bar{B}_{x,1}$, so
Lemma \ref{fundamental} implies that each such solution maps to an eigenvalue of $\tilde{F}_{E_1}^{(1)}(B_{x,1})$.)
Allowing for double counting when these solutions are assigned to each $E_{2,\rho_2}$ and
tallied in $\hat{m}_{2,\rho_2}$, we have that $\sum_{\rho_2} \hat{m}_{2,\rho_2} \le 2n(B_{x,1})$. In particular, we have that $\sum_{\rho_2} c_{\rho_2}^{-1} \le 1$.

When the same calculation is performed in later steps, we need to bound
\begin{equation}\label{(4.14)}
\sum_{\rho_k} \hat{m}_{k,\rho_k} \le 2\left(\hat{m}_{k-1,\rho_{k-1}} + n(B_{x,k-1}) - n(B_{x,k-2})\right).
\end{equation}
On the left, we are counting (possibly twice) all the solutions to $\lambda \in
\text{spec}\,\tilde{F}_\lambda^{(k-1)}(B_{x,k-1})$ in $I_{\veps_{k-1}/3}(E_{k-1,\rho_{k-1}})$ (recall that $\lvert E_{k,\rho_k} - E_{k-1,\rho_{k-1}}\rvert \le \veps_{k-1}/3$).
Thus we need to show that all such solutions are counted in the right-hand side of (\ref{(4.13)}).
To see this, decompose $B_{x,k-1}$ into its subblocks $\{B_{k-2,\beta}\}_{\beta = 0}^{m}$,
with $B_{k-2,0} \equiv B_{x,k-2}$.
Working in $\bar{B}_{x,k-1}$, we may replace $F_\lambda^{(k-1)}$ with 
$F_\lambda^{(k-2)}$ since, as explained in the proof of Theorem \ref{thm:Q}, there is no
difference (because separation conditions keep blocks $B_{k-2}$ out of $\bar{B}_{x,k-1}$.)
Corollary \ref{cor:2'} allows for a further replacement with
$\oplus_\beta \tilde{F}^{(k-2)}_\lambda(B_{k-2,\beta})$, making an error with norm
$\ll \veps_{k-1}$.
Hence by Weyl's inequality, the solutions to 
$\lambda \in \text{spec}\,\tilde{F}_\lambda^{(k-1)}(B_{x,k-1})$ in $I_{\veps_{k-1}/3}\big(E_{k-1,\rho_{k-1}}\big)$ can be tallied with a total
no greater than the sum over $\beta$ of the number of solutions to
$\lambda \in \text{spec}\,\tilde{F}_\lambda^{(k-2)}(B_{k-2,\beta})$ in $I_{\veps_{k-1}/2}\big(E_{k-1,\rho_{k-1}}\big)$.
The number of such solutions for $\beta = 0$ is $\hat{m}_{k-1,\rho_{k-1}}$. 
The blocks for $\beta>0$ represent newly attached blocks, and the corresponding increase in solution count is bounded by the site counts of the blocks.
So for each $\beta \in [1,m]$, we  work in $\bar{B}_{k-2,\beta}$, 
and then
as in the proof of Theorem \ref{thm:1.1}, 
repeated applications of Lemma \ref{fundamental} gives a bound on
the number of solutions  by
$n(B_{k-2,\beta}) = 
\lvert R^{(k-2)} \cap \bar{B}_{k-2,\beta}\rvert$.
We have that
$\sum_{\beta = 1}^m n(B_{k-2,\beta}) = n(B_{x,k-1}) - n(B_{x,k-2})$.
Allowing for a factor of 2 from the double counting, we obtain (\ref{(4.14)}).
An immediate consequence is that $\sum_{\rho_k} c_{\rho_k}^{-1} \le 1$,
which validates the use of $c_{\rho_k}$ as a combinatoric factor.
In this argument, one can see that 
the migration of eigenvalues from one scale to the next led to the use of slightly larger, overlapping intervals for counting them.

To handle the case $k > \hat{k}$, recall from the discussion after
(\ref{Q2}) that $\hat{n}_{j-1}(B_{x,j-1}) = 1$ for $j \in (\hat{k},k]$.
Hence $\hat{n}_{k-1}(B_{x,k-1}) = 1$. Furthermore, a comparison of the definitions (\ref{(4.11)}) and (\ref{(P6)}) for $\hat{m}_{k,\rho_k}$ and
$\hat{n}_{k-1}(B_{x,k-1})$, respectively, shows that the latter uses a wider interval. With an application of Theorem \ref{thm:3}, we see that fixed-point solutions in (\ref{(4.11)}) are close to the eigenvalues counted in (\ref{(P6)}), and hence 
$\hat{m}_{k,\rho_k} \le \hat{n}_{k-1}(B_{x,k-1}) = 1$. Thus there is 
no more than one solution to 
$\lambda \in \text{spec}\,\tilde{F}_\lambda^{(k-1)}(B_{x,k-1})$
in $I_{\veps_{k-1}/3}(E_{k-1})$, and hence no more than one choice for $E_k$. This validates the choice $c_{\rho_k} = 1$ when $k>\hat{k}$ in (\ref{(4.11)}).

In the final step, $k = \bar{k}$, there are evidently no more than $\hat{m}_{\bar{k}-1,\rho_{\bar{k}-1}}$ choices for $E_{\bar{k}}$, because the
capture interval in (\ref{(4.11)}) is wider than $\veps_{\bar{k}-1}/3$. Thus
$\sum_{\rho_{\bar{k}}} c_{\rho_{\bar{k}}}^{-1} \le 1$, and so we conclude that (\ref{(4.12)}) and (\ref{(4.13)}) define valid combinatoric factors in all cases.

We claim that products of combinatoric factors satisfy
\begin{equation}\label{(c)}
\prod_{i=2}^j c_{\rho_i} \le 2^{j-2} 2^{n(B_{x,j-1})}/\hat{m}_{j,\rho_j},
\end{equation}
for $2 \le j \le \hat{k}$. This holds for $j=2$ by (\ref{(4.12)}) and the inequality
$2n \le 2^n$ for positive integers $n$. From (\ref{(4.13)}), we may obtain a bound
\begin{equation}\label{(d)}
c_{\rho_j} \le 2 \hat{m}_{j-1,\rho_{j-1}} 2^{n(B_{x,j-1}) - n(B_{x,j-2})}/\hat{m}_{j,\rho_j},
\end{equation}
by letting $a = \hat{m}_{j-1,\rho_{j-1}}$, $b = n(B_{x,j-1}) - n(B_{x,j-2})$ and using
$2(a+b) \le 4ab \le 2a2^b$ (valid for positive integers $a$, $b$).
Multiplying (\ref{(d)}) by the $j-1$ version of (\ref{(c)}), we obtain the $j$ version.

If $\hat{k} < \bar{k} -1$, then $c_{\rho_j} = 1$, $\hat{m}_{j,\rho_j} = 1$, and $B_{x,j-1} = B_{x,\hat{k}-1}$ for $j > \hat{k}$, so
\begin{equation}\label{(f)}
\prod_{j=2}^{\bar{k}} c_{\rho_j} \le 2^{\hat{k}-2} 2^{n(B_{x,\bar{k}-1})}.
\end{equation}
This holds also if $\hat{k} = \bar{k} - 1$, since in that case the final denominator 
$\hat{m}_{\bar{k}-1,\rho_{\bar{k} - 1}}$ is cancelled by $c_{\rho_{\bar{k}}}$.

The estimate (\ref{(f)}) on combinatoric factors allows us to bound $N_{x,y,z}$ by taking
the supremum over $\rho_2,\ldots,\rho_{\bar{k}}$
and including an additional factor $2^{\hat{k}-2}2^{n({B_{x,\bar{k} -1}})}$. Then
\begin{equation}\label{(g)}
\mathbb{E}\,N_{x,y,z} \le \sum_{
B_{x,\bar{k} - 1} \text{ containing }y,z}
2^{\hat{k}-2}2^{n(B_{x,\bar{k} -1})}
\hat{P}^{(\bar{k}-1)}(B_{x,\bar{k} - 1})\mathbb{E}_{\bar{k}-1}\,\mathbf{1}_{\hat{n}_{\bar{k}-1}(B_{x,\bar{k} - 1})>0},
\end{equation}
using the last two factors as a bound for the probability that $B_{x,\bar{k}-1}$ is a component of $R^{(\bar{k}-1)}$.
We have that $2^{\hat{k}-2} \le L_{\hat{k} - 1}$ and $\{x,y,z\} \subseteq B_{x,\bar{k}-1}$, so
\begin{equation}\label{(h)}
\big(\text{diam}(\{x,y,z\})\vee 1\big)^{\hat{p}-1}2^{\hat{k}-2} \le \big(\text{diam}(B_{x,\bar{k}-1})\vee L_{\hat{k}-1}\big)^{\hat{p}}.
\end{equation}
Furthermore, 
\begin{equation}\label{(h')}
\veps^{-1/6} \le \veps^{-q_0 \hat{n}_{\bar{k}-1}(B_{x,\bar{k} - 1})} \text{ and }
2^{n(B_{x,\hat{k}-1})} \le \veps^{-n(B_{x,\bar{k} - 1})/8}\le \veps^{-q_{\bar{k}-1}n(B_{x,\bar{k} - 1})}.
\end{equation}
Therefore, as long as $\{x,y,z\} \ne \{x\}$ so that $B_{x,\bar{k}-1} \ne \{x\}$, 
(\ref{Q2}) and Theorem \ref{thm:Qhat} imply that
\begin{equation}\label{(i)}
\mathbb{E}\, N_{x,y,z} \big(\text{diam}(\{x,y,z\}\big)\vee 1)^{\hat{p}-1}\veps^{-1/6}\le 1,
\end{equation}
which is the same as (\ref{(4.9)}) in this case. If $x = y= z$, then we need to add in the 
case $B_{x,\bar{k}-1} = \{x\}$, which leads to a single eigenvalue, hence the term
$\mathbf{1}_{\{x = y = z\}}$ in (\ref{(4.9)}). \qed

The next corollary simplifies Proposition \ref{prop:4.3} by summing over the starting point $x$. Define $N_{y,z} = \sum_x N_{x,y,z}$.
\begin{corollary}\label{cor:4.4}
Under the same assumptions as Proposition \ref{prop:4.3}, 
\begin{equation}\label{(omega)}
\mathbb{E}\,N_{y,z} \le \veps^{1/6}c_d(\lvert y-z \rvert \vee 1)^{-(\hat{p} - d-1)} + \mathbf{1}_{\{y = z\}},
\end{equation}
where $c_d$ is a constant that depends only on the dimension $d$.
\end{corollary}
\textit{Proof.}
We claim that
\begin{equation}\label{(nu)}
\sum_x \big(\text{diam}(\{x,y,z\})\vee 1\big)^{-(d+1)} \le c_d (\lvert y-z \rvert \vee 1)^{-1}.
\end{equation}
This can be obtained by (1) summing over $x$ such that
$2^m(\lvert y-z \rvert \vee 1) \le \text{diam}(\{x,y,z\}) < 2^{m+1}(\lvert y-z \rvert \vee 1)$, obtaining
a bound $c_d [2^m(\lvert y-z \rvert \vee 1)]^{d - (d+1)}$; and (2) summing this bound over $m$.
The bound (\ref{(omega)}) then follows from (\ref{(4.9)}) and (\ref{(nu)}). \qed

\textit{Proof of Theorem \ref{thm:1.2}}.
Instead of counting eigenvalues as in Proposition \ref{prop:4.3} and Corollary \ref{cor:4.4}, we weight each term with $\lvert \vphi_{\beta}(y_1)\vphi_{\beta}(y_2) \rvert $.
If we work in the last step $\bar{k}$, in the energy-following procedure starting from some $x$, then the block $B_{x,\bar{k}-1}$ has reached its maximum extent.
From (\ref{(1.52)}), each eigenfunction reachable in the EFP starting at $x$ can be 
written as $G_{E_{\bar{k}}}^{(\bar{k}-1)}\vphi^{(\bar{k} -1)}$, for some eigenvector
$\vphi^{(\bar{k} -1)}$ of $F_{E_{\bar{k}}}^{(\bar{k}-1)}$ (here $E_{\bar{k}}$ is the
corresponding eigenvalue). Proposition \ref{prop:4.1} assures us that every eigenvector
of $H$ can be constructed via the EFP starting at some $x$. Thus we may write
\begin{equation}\label{(4.24)}
\sum_\beta \lvert \vphi_{\beta}(y_1)\vphi_{\beta}(y_2) \rvert  \le
\sum_x \sum_{\beta \text{ reachable from }x}
\big|\big(G_{E_{\bar{k}}}^{(\bar{k}-1)}\vphi_{\beta}^{(\bar{k} -1)}\big)(y_1)\big(G_{E_{\bar{k}}}^{(\bar{k}-1)}\vphi_{\beta}^{(\bar{k} -1)}\big)(y_2)\big|.
\end{equation}
Note that the sum over $\beta$ can be taken as the sum over the choices in the EFP, starting at $x$, and these choices determine $B_{x,\bar{k} - 1}$ as well.
(Each $\beta$ may be counted more than once, but this is not a problem, as the sums in the EFP are under control, as demonstrated in the proof of Proposition \ref{prop:4.3}.)
If we take $\vphi_{\beta}^{(\bar{k} -1)}$ to have norm 1, then by (\ref{(1.52)}) $\vphi_\beta$
will have norm at least 1, so our bound will apply also to the eigenfunction correlator (which uses normalized eigenvectors). Thus we may bound $ \lvert \vphi_\beta^{(\bar{k}-1)}(z) \rvert  $ by 1
for $z \in B_{x,\bar{k}-1}$.
Theorem \ref{thm:G} ensures that the eigenfunction-generating kernel 
$G_{E_{\bar{k}},yz}^{(\bar{k}-1)}$ is bounded by $\gamma^{.85\lvert y-z \rvert }$ 
for $y \in \Lambda \setminus B_{x,\bar{k}-1}$, $z \in B_{x,\bar{k}-1}$. When $y, z$ are both in $B_{x,\bar{k}-1}$, 
$G_{E_{\bar{k}},yz}^{(\bar{k}-1)} = \delta_{yz}$.
Thus
\begin{align}\label{(xi)}
\sum_\beta \lvert \vphi_{\beta}(y_1)\vphi_{\beta}(y_2) \rvert  &\le
\sum_{x,z_1,z_2}\big(\delta_{y_1z_1} + \gamma^{.85\lvert y_1-z_1 \rvert }\big)
\big(\delta_{y_2z_2} + \gamma^{.85\lvert y_2-z_2 \rvert }\big)N_{x,z_1,z_2}\notag\\
&=\sum_{z_1,z_2}\big(\delta_{y_1z_1} + \gamma^{.85\lvert y_1-z_1 \rvert }\big)
\big(\delta_{y_2z_2} + \gamma^{.85\lvert y_2-z_2 \rvert }\big)N_{z_1,z_2}.
\end{align}
Take the expectation and apply Corollary \ref{cor:4.4}. We obtain
decay from $y_1$ to $y_2$ via intermediate points $z_1, z_2$. The
resulting bound is governed by the factor with slowest decay. Hence
for $y_1 \ne y_2$,
\begin{equation}\label{(eta)}
\mathbb{E}\,\sum_\beta \lvert \vphi_{\beta}(y_1)\vphi_{\beta}(y_2) \rvert  \le
\big(\lvert y_1-y_2 \rvert \vee 1\big)^{-(\hat{p} - d -1)},
\end{equation}
which is (\ref{(6')}). 
When $y_1=y_2$, we have that $\sum_\beta \lvert\vphi_\beta(y_1)\rvert^2 = 1$ by orthognality. Thus we obtain the first part of the theorem.

To obtain the second part, let $\lvert x-y \rvert  \ge R$ and define
\begin{equation}\label{(4.27)}
X(x,y) = \sum_\beta \lvert \vphi_\beta(x)\vphi_\beta(y) \rvert \gamma^{-\lvert x-y \rvert /5},
\end{equation}
and as in (\ref{(xi)}) we have
\begin{equation}\label{(4.28)}
X(x,y) \le \sum_{z_1,z_2}\big(\delta_{xz_1} + \gamma^{.85\lvert x-z_1\rvert}\big)
\big(\delta_{yz_2} + \gamma^{.85\lvert y-z_2\rvert}\big)N_{z_1,z_2}
\gamma^{-\lvert x-y \rvert /5}.
\end{equation}
Put $X(x,y) = X^{\text{near}}(x,y) + X^{\text{far}}(x,y)$, where $X^{\text{near}}$ contains the terms of (\ref{(4.28)}) with 
$\lvert z_1 -x\rvert \le \lvert x-y \rvert /4$, $\lvert z_2 -y\rvert \le \lvert x-y \rvert /4$, and $X^{\text{far}}$
contains the rest. Then
\begin{equation}\label{(4.29)}
P(X(x,y)> 1) \le P(X^{\text{near}}(x,y) > \tfrac{1}{2}) +
P(X^{\text{far}}(x,y) > \tfrac{1}{2}).
\end{equation}
The terms contributing to $X^{\text{far}}$ satisfy
$\lvert x-z_1 \rvert +\lvert y-z_2 \rvert  \ge \lvert x-y \rvert /4$, so
\begin{equation}\label{(4.30)}
\gamma^{.85(\lvert x-z_1 \rvert  + \lvert y-z_2 \rvert )} \gamma^{-\lvert x-y \rvert /5} \le \gamma^{.05((\lvert x-z_1 \rvert  + \lvert y-z_2 \rvert )}.
\end{equation}
Then we may bound $\mathbb{E}\,X^{\text{far}}(x,y) \le \veps^{1/6}c_d\lvert x-y \rvert ^{-(\hat{p} - d - 1)}$ as in the proof of (\ref{(eta)}). (Here, we are assuming $\lvert x-y \rvert \ge 4$, so all terms have at least a factor $\veps^{1/6}c_d$, as in (\ref{(omega)})).
Hence $P(X^{\text{far}}(x,y) > \tfrac{1}{2}) \le 2\veps^{1/6}c_d\lvert x-y \rvert ^{-(\hat{p} - d - 1)}$.
We may estimate
\begin{align}\label{(4.31)}
P(X^{\text{near}}(x,y) > \tfrac{1}{2}) &\le
\sum_{z_1,z_2}P(N_{z_1,z_2}>0) \le \sum_{z_1,z_2}\mathbb{E}\,N_{z_1,z_2}\\
&\le  \lvert x-y \rvert ^{2d}\veps^{1/6}c_d(\lvert x-y \rvert /2)^{-(\hat{p} - d - 1)}
\le \veps^{1/6}c^{\prime}_d\lvert x-y \rvert ^{-(\hat{p} - 3d - 1)},\notag
\end{align}
using (\ref{(omega)}), the fact that $\lvert z_1-z_2 \rvert \ge\lvert x-y \rvert /2$, and a bound of
$(\lvert x-y \rvert /2 +1)^{2d}<\lvert x-y \rvert ^{2d}$ on the number of choices for $z_1, z_2$.
Combining these results, we obtain that 
\begin{equation}\label{(4.31a)}
P(X(x,y) > 1) \le \veps^{1/6}(2c_d+c^{\prime}_d)\lvert x-y \rvert ^{-(\hat{p} -3d-1)}.
\end{equation}
Summing this bound over $y$ such that $\lvert y-x \rvert \ge R$, we obtain a bound 
$R^{-(\hat{p} -4d-1)}$, which gives the desired result, (\ref{(7)}). \qed

\subsection{Level Spacing}\label{ssec:4.4}
We now prove Theorem \ref{thm:1.3}. As in the proof of Theorem \ref{thm:1.2}, we construct every eigenfunction via the EFP. Instead of counting all eigenvalues in an interval, we count only the ones with additional spectrum within a $\delta$-neighborhood. To this end, we
define $N_x(\delta)$ to be the number of eigenvalues $\lambda_0$ of 
$H$ that can be reached via the EFP as in Proposition \ref{prop:4.3}, 
starting at $x$, and which have another eigenvalue in $I_\delta(\lambda_0)$. Since
\begin{equation}\label{(4.32)}
P\Big(\min_{\beta \ne \tilde{\beta}} \lvert E_\beta  - E_{\tilde{\beta}} \rvert  < \delta\Big) \le \tfrac{1}{2}\sum_x \mathbb{E}\,N_x(\delta),
\end{equation}
Theorem \ref{thm:1.3} will follow from the estimate
\begin{equation}\label{(ab)}
 \mathbb{E}\,N_x(\delta) \le 2\lvert \Lambda \rvert(\log_\gamma \delta)^{-(p/2-1)},
\end{equation}
for $\delta \in [\gamma^{\text{Diam}(\Lambda)},\gamma]$.

Let us define $k $ by the inequality
\begin{equation}\label{(4.34)}
 \veps_{k+1}/4 < \delta \le \veps_k/4.
\end{equation}
Recall that $\bar{k}$ is the smallest integer such that $5.1 L_{\bar{k}-1} \ge \text{Diam}(\Lambda)$, we have that
\begin{equation}\label{(4.35)}
 \delta \ge \gamma^{\text{Diam}(\Lambda)} \ge \gamma^{5.1 L_{\bar{k}-1}} >\gamma^{6.4 L_{\bar{k}-1}} =\veps_{\bar{k}+1} > \veps_{\bar{k}+1}/4,
\end{equation}
which implies that $k \le \bar{k}$. We can assume that
$\delta \le \gamma \ll \veps_1/4$ because otherwise $\log_\gamma \delta < 1$, in which case (\ref{(8)}) is automatic.
In fact, we can assume $k \ge 2$ because the case $k = 1$ with 
$\delta > \veps_2/4$ can be handled by a direct appeal to Weyl's
inequality when all off-diagonal entries of $H$ are turned off. When this is done, we see that all eigenvalues of $H$ are within $O(\gamma)$ 
of their unperturbed values $2d\gamma +v_x$. As
$\veps_1 = \tfrac{1}{3(N-1)}$ is one-third the spacing of allowed
values of $v_x$, we obtain that
\begin{equation}\label{(lambda)}
P\Big(\min_{\beta \ne \tilde{\beta}} \lvert E_\beta  - E_{\tilde{\beta}} \rvert  < \delta\Big) \le \tfrac{1}{N}\tfrac{1}{2}\lvert \Lambda \rvert(\lvert \Lambda \rvert-1) \le \tfrac{1}{N}\lvert \Lambda \rvert^2.
\end{equation}
Furthermore, $N$ is chosen after $p, L_0$, so 
$\tfrac{1}{N} < (\log_\gamma \veps_2/4)^{-(p/2-1)}$.
(Recall that $\veps_2 = \gamma^{1.6L_2} = \gamma^{6.4L_0}$.)
Thus (\ref{(8)}) holds when $k=1$.

We proceed to estimate $\mathbb{E}\,N_x(\delta)$, assuming that
$2\le k\le \bar{k}$. Consider the EFP at the point where $E_k$ is
chosen in $\tfrac{1}{2}\veps_k \mathbb{Z}$ within $\tfrac{1}{4}\veps_k$
of a solution to $\lambda \in \text{spec}\,\tilde{F}^{(k-1)}_\lambda(B_{x,k-1})$. Recall that $\hat{k}(B_{x,k})$ is
the maximum of all the 
$j \in [1,k]$
such that $B_{x,j} \setminus B_{x,j-1} \ne \varnothing$ or
$\hat{n}_{j-1}(B_{x,j-1}) >1$. Consider two cases. For case 1, we assume that 
$\hat{k}(B_{x,k})\ge k$.
Note that in subsequent steps, $\hat{k}$ can only increase.
Hence we can estimate all case 1 terms by ignoring the condition that 
there is another eigenvalue within $\delta$ and requiring instead that
$\hat{k}(B_{x,\bar{k}-1})\ge k$. Thus $\mathbb{E}\,N_x^{(1)}(\delta)$, 
the expected number of eigenvalues in case 1, may be bounded as in 
(\ref{(g)}):
\begin{equation}\label{(4.37)}
\mathbb{E}\, N_{x}^{(1)}(\delta) \le \sum_{
B_{x,\bar{k} - 1}:\,\hat{k} \ge k}
2^{\hat{k}-2}2^{n(B_{x,\bar{k} -1})}
\hat{P}^{(\bar{k}-1)}(B_{x,\bar{k} - 1})\mathbb{E}_{\bar{k}-1}\,\mathbf{1}_{\hat{n}_{\bar{k}-1}(B_{x,\bar{k} - 1})>0}.
\end{equation}
As in (\ref{(h)})-(\ref{(i)}), we use (\ref{Q2}) and Theorem \ref{thm:Qhat}
to reap the smallness entailed in the condition $\hat{k} \ge k$. We have that
\begin{equation}\label{(4.38)}
L_{k - 1}^{\hat{p}-1}2^{\hat{k}-2}\le 
L_{\hat{k} - 1}^{\hat{p}-1}2^{\hat{k}-2} \le \big(\text{diam}(B_{x,\bar{k}-1})\vee L_{\hat{k}-1}\big)^{\hat{p}}.
\end{equation}
and $B_{x,\bar{k}-1}\ne \{x\}$ 
(which must be the case if $\hat{k} \ge 2$).
Then using (\ref{(h')}) as before, we obtain
\begin{equation}\label{(4.39)}
\mathbb{E}\, N_{x}^{(1)}(\delta) \le L_{k-1}^{-(\hat{p}-1)} \veps^{1/6}.
\end{equation}

Now consider case 2: $\hat{k}(B_{x,k}) < k$. This implies that
$B_{x,k} \setminus B_{x,k-1} = \varnothing$ and 
$\hat{n}_{k-1}(B_{x,k-1}) =1$.
As argued in the proof of Theorem \ref{thm:1.1}, all of the eigenvalues of $H$ in $I_{\veps_{k-1}/3}(E_{k-1})$ are in close agreement
with those of
$\oplus_\beta \tilde{F}^{(k-1)}_{E_{k-1}}(B_{k-1,\beta})$,
with differences of size $\gamma^{3.3L_{k-1}} \ll \veps_k$.
Let $\lambda_0$ be the eigenvalue of $H$ that is reached in the EFP.
Then $\lvert \lambda_0-E_{k-1} \rvert  \le .31 \veps_{k-1}$, and there is a corresponding nearby eigenvalue of 
$\tilde{F}^{(k-1)}_{E_{k-1}}(B_{x,k-1})$.
This implies that there is a corresponding solution $\tilde{\lambda}_0$
to $\lambda \in \text{spec}\,\tilde{F}^{(k-1)}_{\lambda}(B_{x,k-1})$ in
$I_{\veps_{k-1}/3}(E_{k-1})$ with $\lvert \lambda_0-\tilde{\lambda}_0 \rvert  \le 
2\gamma^{3.3L_{k-1}} \ll \veps_k$, using Theorem \ref{thm:3}.
We also have that 
$\lvert \lambda_0 - E_k \rvert  \le .31\veps_k$, 
so $\lvert \tilde{\lambda}_0 - E_k\rvert \le \veps_k/3$. But since
$\hat{n}_{k-1}(B_{x,k-1}) =1$, there are no other 
eigenvalues of $\tilde{F}^{(k-1)}_{E_{k-1}}(B_{x,k-1})$ in
$I_{\veps_k}(E_k)$, and hence no other
solutions to
$\lambda \in \text{spec}\,\tilde{F}^{(k-1)}_{\lambda}(B_{x,k-1})$ in
$I_{.9\veps_{k}}(E_{k})$.
However, we have the condition that there is another eigenvalue
$\lambda'$ of $H$ with $\lvert\lambda' - \lambda_0\rvert \le \veps_k/4$.
Then $\lvert\lambda'-E_k\rvert \le .56 \veps_k$, 
$\lvert\lambda' - E_{k-1}\rvert \le \veps_k/4 + .31 \veps_{k-1} \le .32 \veps_{k-1}$,
so by the same reasoning there must be a solution
$\tilde{\lambda}'$ other than $\tilde{\lambda}_0$ to 
$\lambda \in \text{spec}\,\tilde{F}^{(k-1)}_{\lambda}(B_{k-1,\beta})$ for some $\beta$, and it satisfies $\lvert\lambda'-\tilde{\lambda}'\rvert \le 
2\gamma^{3.3L_{k-1}} \ll \veps_k$.
Thus $\lvert\tilde{\lambda}'-E_k\rvert \le .6 \veps_k$, and as $\tilde{\lambda}_0$ is
the only solution in $I_{.9\veps_{k}}(E_{k})$ for $B_{x,k-1} \equiv B_{k-1,0}$, there must be a second block $B_{k-1,1}$ in $R^{(k-1)}$.
Furthermore, it is clear that
\begin{equation}\label{(4.42)}
\text{dist}\big(\text{spec}\,\tilde{F}^{(k-1)}_{E_{k}}(B_{k-1,1}),E_{k}\big)
\le \lvert\tilde{\lambda}' - E_{k}\rvert(1+2\gamma) + 2\gamma^{3.3L_{k-1}}
\le  .7\veps_{k},
\end{equation}
by Theorem \ref{thm:3} and the abovementioned bound on 
$\lvert\tilde{\lambda}' - E_{k}\rvert$. Consequently, $B_{k-1,1}$ is resonant
in step $k$ -- see (\ref{(23)}) -- and it survives to $R^{(k)}$.
Note that since $B_{x,k} \setminus B_{x,k-1} = \varnothing$,
so there must be a second component of $R^{(k)}$. Let us call it
$B'_k$. At its core, this argument for a second component of 
$R^{(k)}$ is a generalization of the one given above for (\ref{(lambda)})
in the case $k = 1$.

As explained at the start of Section \ref{ssec:3.2}, we can bound
the probability that both $B_{x,k}$ and $B'_k$ are blocks of $R^{(k)}$
by the product of $\hat{P}^{(k)}(B_{x,k})\mathbb{E}_k\,\mathbf{1}_{\hat{n}_k(B_{x,k})>0}$ and $\tilde{P}^{(k)}(B'_k)\mathbb{E}_k\,\mathbf{1}_{\hat{n}_k(B'_k)>0}$. We may sum
the latter over all possibilities for $B'_k$ for each
$B_{x,k}$ that arises in the EFP. Using (\ref{R1}), this may be 
bounded by
\begin{equation}\label{(4.43)}
\sum_{y \in \Lambda \setminus B_{x,k}} \,\sum_{B'_k \text{ containing }y}
\tilde{P}^{(k)}(B'_k) \le (\lvert \Lambda \rvert - 1)L_{k-1}^{-p}\veps^{1/4}.
\end{equation}
We may insert this bound in place of the condition $\hat{k}<k$
that defines the second case, and then the EFP sums can be
controlled as in the proof of (\ref{(i)}). We obtain
$(\lvert \Lambda \rvert - 1)L_{k-1}^{-p}\veps^{1/4}$ times the same bound as
we would have obtained for $\mathbb{E}\,N_{x,x,x}$, which is (\ref{(i)})
plus the case $B_{x,\bar{k}-1} = \{x\}$. Thus
\begin{equation}\label{(4.44)}
\mathbb{E}\, N_{x}^{(2)}(\delta) \le 
(\lvert \Lambda \rvert - 1)L_{k-1}^{-p}\veps^{1/4}(1+\veps^{1/6}).
\end{equation}
Combining this with (\ref{(4.39)}), we obtain
\begin{equation}\label{(4.45)}
\mathbb{E}\, N_{x}(\delta) \le 
L_{k-1}^{-(\hat{p}-1)}\veps^{1/6} + (\lvert \Lambda \rvert - 1)L_{k-1}^{-p}\veps^{1/4}(1 + \veps^{1/6})
\le 2\lvert \Lambda \rvert L_{k-1}^{-(\hat{p}-1)}\veps^{1/6}.
\end{equation}

Note that 
\begin{equation}\label{(4.40)}
\log_\gamma \delta \le \log_\gamma\frac{\veps_{k+1}}{4} = 1.6L_{k+1} - \log_\gamma 4 = 6.4L_{k-1}-\log_\gamma 4 \le 7L_{k-1}.
\end{equation}
Hence
\begin{equation}\label{(4.41)}
\mathbb{E}\, N_{x}(\delta) \le 2\lvert \Lambda \rvert(\tfrac{1}{7}\log_\gamma \delta)^{-(\hat{p}-1)} \veps^{1/6}
\le 2\lvert \Lambda \rvert(\log_\gamma \delta)^{-(\hat{p}-1)}.
\end{equation}
Recalling that $\hat{p} = p/2$, we sum this over $x \in \Lambda$ to
obtain (\ref{(ab)}) and complete the proof of Theorem \ref{thm:1.3}. \qed

\appendix
\section{Completeness of the Energy-Following Procedure}\label{A}
Here we prove Proposition \ref{prop:4.1}.
We make a comparison with a ``reference'' fixed-energy procedure (FEP) with 
$E_k = \lambda_0$. For the reference procedure, we halve the energy windows that are used in (\ref{(2.1n)}) and Definition \ref{def:resonant} to define resonant blocks. 
Thus we replace $\veps_k$ with $\veps_k/2$ in (\ref{(2.1n)}), (\ref{(23)}) and denote
the resulting set of blocks $b_k$ (to distinguish them from the ones generated by the EFP).
We will see that the $b_k$'s are necessarily contained in the EFP blocks $B_k$ for at
least one set of choices $x, E_1, \ldots , E_{\bar{k}}$; this will allow us to demonstrate 
convergence to $\lambda_0$ in the sense of (\ref{efp}).
In this way, we are able to deal with the dependence of $R^{(k)}$ (and its components $B_k$) on the sequence $E_1, \ldots , E_{\bar{k}}$.

We may use the FEP blocks $b_k$ to determine a good starting site $x$ for the EFP.
Start from a large enough scale $k$ so that $\bar{b}_k = \Lambda$ and so
$\tilde{F}^{(k)}_{\lambda_0}(b_k) = F_{\lambda_0}^{(k)}$.
As we proceed downward in scale, we claim that for each $j \le k$, at least one
subblock $b_{j,\beta}$ is strongly resonant with $\lambda_0$, in the sense that
\begin{equation}\label{(a.b)}
\text{dist}\big(\text{spec}\,\tilde{F}^{(j)}_{\lambda_0}(b_{j,\beta}),\lambda_0\big) \le \veps_{j+1}/50.
\end{equation}
This is so because Corollary \ref{cor:2'} allows us to replace $F_{\lambda_0}^{(j)}$ with
$\oplus_\beta \tilde{F}^{(j)}_{\lambda_0}(b_{j,\beta})$ with error $\ll \veps_{j+1}$, and
(\ref{(a.b)}) follows by Weyl's inequality.

Continuing down to $j=1$, we obtain a $b_1$ with
$\text{dist}\big(\text{spec}\,\tilde{F}^{(1)}_{\lambda_0}(b_{1}),\lambda_0\big) \le \veps_{2}/50$.
Recalling that $\tilde{F}^{(1)}_{\lambda_0}(b_1)$ is the same as $F_{\lambda_0}^{(1)}(\bar{b}_1)$,
we may apply Lemma \ref{fundamental}
to relate its spectrum to that of $H_{\bar{b}_1}$ in $I_{\tilde{\veps}/2}(\lambda_0)$.
Here we can take $\tilde{\veps} = \veps_1/3$, since with $D = H_{\bar{b}_1 \setminus b_1}$,
we have $\|(D-\lambda_0)^{-1}\| \le \tilde{\veps}^{-1}$ (diagonal entries in 
$\bar{b}_1 \setminus b_1$ are farther than $\veps_1/2$ from $\lambda_0$, and the norm
of the off-diagonal matrix $V^{(1)}$ is $\le 2d\gamma$, see (\ref{(18)}).
We conclude that 
$\text{dist}(\text{spec}\,H_{\bar{b}_1},\lambda_0) \le \veps_{2}/50 + 2(2d\gamma/\tilde{\veps})^2 \cdot \tilde{\veps}/2 \le \gamma$.
With another application of Weyl's inequality, we may eliminate the off-diagonal part of
$H_{\bar{b}_1}$ and conclude that at least one site $x$ in $b_1$ has 
$v_x + 2d\gamma \in I_{2d\gamma + \gamma}(\lambda_0)$. (The site $x$ cannot lie in
$\bar{b}_1 \setminus b_1$ since it contains only sites with
$v_x + 2d\gamma \notin I_{\veps_1/2}(\lambda_0)$.)
This confirms the obvious fact that there must be sites with $\lvert v_x - \lambda_0\rvert \le O(\gamma)$, if $\lambda_0 \in \text{spec}\,H$. But we also have $x$ as a base point for a system of blocks $b_{x,j}$, each of which is resonant to $\lambda_0$ to within $\veps_{j+1}/50$.
We now use $x$ to initiate the EFP as we demonstrate convergence to $\lambda_0$.

Let us analyze the relationship between the blocks $b_k$ of the FEP and the blocks $B_k$ of the EFP. The existence of blocks $b_k$ with spectrum close to $\lambda_0$ will be used to make choices in the EFP so that blocks $B_k$ also have spectrum close to $\lambda_0$.
We establish the following result for use in an induction on $k$.
\begin{lemma}\label{lem:4.2}
Under the same assumptions as Proposition \ref{prop:4.1}, 
let $x$ be a base point arising from the FEP as described above, satisfying 
$\mathrm{dist}(\mathrm{spec}\,\tilde{F}^{(j)}_{\lambda_0}(b_{x,j}),\lambda_0) \le \veps_{j+1}/50$
for all $j$, and put $E_1 = v_x + 2d\gamma \in I_{(2d+1)\gamma}(\lambda_0)$.
Let $k \ge 2$ and assume
$E_j$ are chosen for $1 \le j < k$ so that $\lvert E_j - \lambda_0\rvert \le .31\veps_j$. Then
\begin{enumerate}[(i)]
\item
For all $j<k$, each FEP block $b_j$ is contained in some EFP block $B_j$. Furthermore, if one performs the FEP in the region $\bar{B}_j$ (instead of $\Lambda$), then the resulting collection of blocks $\{b_{j,\beta}\}$ are precisely the ones from the $\Lambda$-construction that happen to be contained in $\bar{B}_j$.
\item
For any FEP block $b_{j-1}$, let $B_{j-1}$ denote the EFP block containing it. 
\begin{enumerate}[(a)]
\item
For each $j \le k$, 
$$
\mathrm{dist}\big(\mathrm{spec}\,\tilde{F}^{(j-1)}_{\lambda_0}(b_{j-1}),\lambda_0\big) \le \veps_{j}/50 \Rightarrow \mathrm{dist}\big(\mathrm{spec}\,H_{\bar{B}_{j-1}},\lambda_0\big) \le 3\veps_{j}/50.
$$
\item
For each $j < k$,
$$
\mathrm{dist}\big(\mathrm{spec}\,\tilde{F}^{(j-1)}_{\lambda_0}(b_{j-1}),\lambda_0\big) \le \veps_{j}/2 \Rightarrow
\mathrm{dist}\big(\mathrm{spec}\,\tilde{F}^{(j-1)}_{E_j}(B_{j-1}),E_j\big) \le \veps_{j}.
$$
\end{enumerate}
\item
There is a choice of $E_k$ in the EFP such that
\begin{equation}\label{efp'}
\lvert E_k - \lambda_0 \rvert  \le .31 \veps_k \text{ and } \lvert E_k - E_{k-1} \rvert  \le \veps_{k-1}/3.
\end{equation}
\end{enumerate}
\end{lemma}
\textit{Proof of Lemma \ref{lem:4.2}.} 
Consider first (i) in case $k = 2$, which will serve as input to (ii), (iii). We will consider
the case $k>2$ at the end of the proof. Recall that in the EFP $R^{(1)}$ is the set of sites $y$ such that
$\lvert v_y+2d\gamma - E_1\rvert \le \veps_1$; components $B_1$ are defined using connections up to a distance $L_1^\alpha$. The FEP blocks $b_1$ are obtained from a resonant set 
$R^{(1)}_{\text{FEP}} = \{y:\,\lvert v_y + 2d\gamma - \lambda_0\rvert \le \veps_1/2\}$. 
Since $\lvert \lambda_0 - E_1 \rvert  \le (2d+1)\gamma \ll \veps_1$, the FEP resonance interval is strictly contained in the EFP resonance interval. Hence
$R^{(1)}_{\text{FEP}} \subseteq R^{(1)}$.
As we use the same distance condition for connectedness in both cases, it is clear that each
$b_1$ is contained in some $B_1$. Furthermore, the width of the collar defining $\bar{B}_1$
is much smaller than the distance between components of $R^{(1)}$, so there is no overlap between $\bar{B}_1$'s. Hence the components of $R^{(1)}_{\text{FEP}} \cap \bar{B}_1$ are the same as the components of $R^{(1)}_{\text{FEP}}$ that
happen to be contained in $\bar{B}_1$.

Now consider (ii)(a) for any $j \le k$. Corollary \ref{cor:2'} implies that
$F_{\lambda_0}^{(j-1)}$ in $\bar{B}_{j-1}$ may be approximated in norm by
$\oplus_\beta \tilde{F}^{(j-1)}_{\lambda_0}(b_{j-1,\beta})$, up to an error of size
$\gamma^{3.3L_{j-1}} \ll \veps_j$.
By (i), $\{b_{j-1,\beta}\}$ are the blocks from the $\Lambda$-construction that are contained in $\bar{B}_{j-1}$. One of the terms in the direct sum is
$\tilde{F}^{(j-1)}_{\lambda_0}(b_{j-1})$, which is assumed in (a) to have spectrum within $\veps_j/50$ of $\lambda_0$. Therefore, $F_{\lambda_0}^{(j-1)}$ in $\bar{B}_{j-1}$ has 
spectrum within $\veps_j/25$ of $\lambda_0$.
Applying Theorem \ref{thm:3} and a fixed-point argument, we obtain a solution
to $\lambda \in \text{spec}\,F_\lambda^{(j-1)}$ (and hence a $\lambda \in \text{spec}\,H_{\bar{B}_{j-1}}$) within $3\veps_j/50$ of $\lambda_0$.
For (b), we weaken the assumption to 
$\text{dist}(\text{spec}\,\tilde{F}^{(j-1)}_{\lambda_0}(b_{j-1}),\lambda_0) \le \veps_{j}/2$,
and then the direct sum argument implies that 
$\text{dist}\big(\text{spec}\,\tilde{F}^{(j-1)}_{\lambda_0}(B_{j-1}),\lambda_0\big) \le \veps_j/2 + \veps_j/50$.
As we are taking $j<k$, we 
have by assumption that
$\lvert E_j-\lambda_0 \rvert  \le .31\veps_j$.
Theorem \ref{thm:3} then implies that 
\begin{equation}\label{(4.8)}
\text{dist}\big(\text{spec}\,\tilde{F}^{(j-1)}_{E_j}(B_{j-1}),E_j\big) \le \veps_j/2 + \veps_j/50 + .31\veps_j + \gamma\cdot .31 \veps_j \le \veps_{j}.
\end{equation}

To obtain (iii), observe that the precondition in (ii)(a) was already established -- see (\ref{(a.b)}). 
Using (i) for $j = k-1$, we have that 
$b_{x,k-1}$ is contained in an EFP block $B_{x,k-1}$, 
and we conclude that 
$\text{dist}(\text{spec}\,H_{\bar{B}_{k-1}},\lambda_0) \le 3\veps_{k}/50$.
Note that $\lambda \in \text{spec}\,H_{\bar{B}_{k-1}}$ is equivalent
to $\lambda \in \text{spec}\,\tilde{F}_\lambda^{(k-1)}(B_{k-1})$,
and so in the EFP there is a choice of $E_k$ satisfying
$\lvert E_k - \lambda \rvert  \le \veps_k/4$. Hence
$\lvert E_k - \lambda_0 \rvert  \le \veps_k/4 + 3\veps_k/50 = .31\veps_k$.
From the previous induction step, $\lvert E_{k-1} - \lambda_0 \rvert  \le .31\veps_{k-1}$ (this is true also for $k=2$ because $E_1 \in I_{(2d+1)\gamma}(\lambda_0)$). Thus $\lvert E_k - E_{k-1} \rvert  \le \veps_{k-1}/3$.

It remains for us to verify (i) for $k > 2$. 
Each FEP block $b_{k-1}$ is formed by joining together resonant
blocks $b_{k-2}$ that are within a distance $L^\alpha_{k-2}$.
Here resonant means that
$\text{dist}\big(\text{spec}\,\tilde{F}_{\lambda_0}^{(k-2)}(b_{k-2}), \lambda_0\big) \le \veps_{k-1}/2$. By (i) in the previous step, each 
such block $b_{k-2}$ is contained in an EFP block $B_{k-2}$.
By (ii)(b), $B_{k-2}$ satisfies
$\text{dist}\big(\text{spec}\,\tilde{F}^{(k-2)}_{E_{k-1}}(B_{k-2}),E_{k-1}\big) \le \veps_{k-1}$. Thus all these blocks are resonant in the EFP and become part of $R^{(k-1)}$. The connectivity distance
$L^\alpha_{k-1}$ is the same in both procedures. Therefore, if
two $b_{k-2}$ blocks are joined by the proximity condition, then
so are the corresponding $B_{k-2}$ blocks that contain them. 
Thus each new block $b_{k-1}$ is contained within one of the
new EFP blocks $B_{k-1}$.
Recall from the discussion at the start of Section \ref{ssec:random} that the collared blocks
$\bar{B}_{i,\beta}$ are either well separated from each other or well inside one another.
%The collared blocks $\bar{B}_{k-1}$ satisfy $\text{dist}(\bar{B}_{i,\beta},\bar{B}_{j,\beta'}) > L_i^{\sqrt{\alpha}}$ for $i\le j$. 
Thus the containment of the $b_{k-1}$ blocks inside
$B_{k-1}$ blocks ensures that each $b_j$ block, $j<k$, lies entirely
inside a given $\bar{B}_{k-1}$ or entirely outside of it.
Hence the ones inside $\bar{B}_{k-1}$ cannot be affected by the
situation outside of $\bar{B}_{k-1}$; they are the same as the ones from the construction in $\Lambda$ which happen to lie inside $\bar{B}_{k-1}$. This completes the proof of (i) and the lemma. \qed

\textit{Proof of Proposition \ref{prop:4.1}}.
Lemma \ref{lem:4.2} allows us to run the EFP up to the point at which
$\bar{B}_{x,\bar{k}-1} = \Lambda$ while maintaining
convergence of $E_k$ towards $\lambda_0$ as in (\ref{efp'}).
In the final step, we pick $E_{\bar{k}}$ equal to one of the eigenvalues of $H$ in $I_{\veps_{\bar{k}-1}/3}(E_{\bar{k}-1})$. In particular,
we may take $E_{\bar{k}} = \lambda_0$. Thus we have demonstrated
that every eigenvalue of $H$ can be reached by the EFP. This completes the proof of Proposition \ref{prop:4.1}. \qed

\begin{footnotesize}

\providecommand{\bysame}{\leavevmode\hbox to3em{\hrulefill}\thinspace}
\providecommand{\MR}{\relax\ifhmode\unskip\space\fi MR }
\providecommand{\MRhref}[2]{%
  \href{http://www.ams.org/mathscinet-getitem?mr=#1}{#2}
}
\providecommand{\href}[2]{#2}

\end{footnotesize}

\begin{thebibliography}{BLMS17}

\bibitem[AM93]{Aizenman1993}
M.~Aizenman and S.~Molchanov, \emph{{Localization at large disorder and at
  extreme energies: an elementary derivation}}, Commun. Math. Phys.
  \textbf{157} (1993), 245--278.

\bibitem[And58]{Anderson1958}
P.~Anderson, \emph{Absence of diffusion in certain random lattices}, Phys. Rev.
  \textbf{109} (1958), 1492--1505.

\bibitem[BK05]{bourgain2005localization}
J.~Bourgain and C.~E. Kenig, \emph{On localization in the continuous
  {A}nderson-{B}ernoulli model in higher dimension}, Invent. Math. \textbf{161}
  (2005), 389--426.

\bibitem[BLMS17]{Buhovsky2017}
L.~Buhovsky, A~Logunov, E.~Malinnikova, and M.~Sodin. 
\emph{A discrete harmonic function bounded on a large portion of $\mathbb{Z}^2$ is constant}, arXiv:1712.07902.

\bibitem[Bou12]{bourgain2012furstenbergmeasure}
J.~Bourgain, \emph{On the {F}urstenberg measure and density of states for the
  {A}nderson-{B}ernoulli model at small disorder}, JAMA \textbf{117} (2012),
  273--295.

\bibitem[Bou14]{bourgain2014application}
\bysame, \emph{An application of group expansion to the {A}nderson-{B}ernoulli
  model}, Geom. Funct. Anal. \textbf{24} (2014), 49--62.

\bibitem[CKM87]{carmona1987anderson}
R.~Carmona, A.~Klein, and F.~Martinelli, \emph{{A}nderson localization for
  {B}ernoulli and other singular potentials}, Commun. Math. Phys. \textbf{108}
  (1987), 41--66.

\bibitem[CS83]{craig1983log}
W.~Craig and B.~Simon, \emph{Log {H}{\"o}lder continuity of the integrated
  density of states for stochastic {J}acobi matrices}, Commun. Math. Phys.
  \textbf{90} (1983), 207--218.

\bibitem[DS20]{Ding2019}
J.~Ding and C.~K. Smart, \emph{Localization near the edge for the
  {A}nderson-{B}ernoulli model on the two dimensional lattice}, Invent. Math. \textbf{219}, (2020), 467--506.

\bibitem[DSS02]{damanik2002localization}
D.~Damanik, R.~Sims, and G.~Stolz, \emph{Localization for one-dimensional,
  continuum, {B}ernoulli-{A}nderson models}, Duke Math. J. \textbf{114} (2002),
  59--100.

\bibitem[FS83]{Frohlich1983}
J.~Fr{\"{o}}hlich and T.~Spencer, \emph{{Absence of diffusion in the Anderson
  tight binding model for large disorder or low energy}}, Commun. Math. Phys.
  \textbf{88} (1983), 151--184.

\bibitem[GHK07]{germinet2007localization}
F.~Germinet, P.~Hislop, and A.~Klein, \emph{Localization for the
  {S}chr{\"o}dinger operator with a {P}oisson random potential}, J. Eur. Math.
  Soc. \textbf{9} (2007), 577--607.

\bibitem[GK07]{germinet2007cantor}
F.~Germinet and A.~Klein, \emph{Localization for some {C}antor-{A}nderson
  {S}chr{\"o}dinger operators}, Adventures in Mathematical Physics (F.~Germinet
  and P.~D. Hislop, eds.), Contem. Math., vol. 447, American Mathematical
  Society, Providence, RI, 2007, pp.~103--112.

\bibitem[GK13]{germinet2013comprehensive}
\bysame, \emph{A comprehensive proof of localization for continuous {A}nderson
  models with singular random potentials}, J. Eur. Math. Soc. \textbf{15}
  (2013), 53--143.

\bibitem[IM16]{Imbrie2016c}
J.~Z. Imbrie and R.~Mavi, \emph{Level spacing for non-monotone {A}nderson
  models}, J. Stat. Phys. \textbf{162} (2016), 1451--1484.

\bibitem[Imb16a]{Imbrie2016}
J.~Z. Imbrie, \emph{Multi-scale {J}acobi method for {A}nderson localization},
  Commun. Math. Phys. \textbf{341} (2016), 491--521.

\bibitem[Imb16b]{Imbrie2016a}
\bysame, \emph{On many-body localization for quantum spin chains}, J. Stat.
  Phys. \textbf{163} (2016), 998--1048.

\bibitem[KM06]{Klein2006}
A.~Klein and S.~Molchanov, \emph{{Simplicity of eigenvalues in the Anderson
  model}}, J. Stat. Phys. \textbf{122} (2006), 95--99.

\bibitem[KT16]{klein2016quantitative}
A.~Klein and C.~Tsang, \emph{Quantitative unique continuation principle for
  {S}chr{\"o}dinger operators with singular potentials}, Proc. Am. Math. Soc.
  \textbf{144} (2016), 665--679.

\bibitem[LZ19]{Li2019}
L.~Li and L.~Zhang,
\emph{{A}nderson-{B}ernoulli localization on the 3{D} lattice and discrete unique continuation principle},
arXiv:1906.04350, Duke Math. J., to appear.

\bibitem[Min96]{Minami1996}
N.~Minami, \emph{{Local fluctuation of the spectrum of a multidimensional
  Anderson tight binding model}}, Commun. Math. Phys. \textbf{177} (1996),
  709--725.

\bibitem[SVW98]{shubin1998some}
C.~Shubin, R.~Vakilian, and T.~Wolff, \emph{Some harmonic analysis questions
  suggested by {A}nderson-{B}ernoulli models}, Geom. Funct. Anal. \textbf{8}
  (1998), 932--964.

\bibitem[Weg81]{Wegner1981}
F.~Wegner, \emph{{Bounds on the density of states in disordered systems}}, Z.
  Phys. B \textbf{44} (1981), 9--15.

\end{thebibliography}
\end{document}